%% file: main.tex
\documentclass[10pt,journal,compsoc]{sec/IEEEtran}

\ifCLASSOPTIONcompsoc
  \usepackage[nocompress]{cite}
\else
  \usepackage{cite}
\fi

\usepackage{booktabs} %
\usepackage{graphicx}
\usepackage{epstopdf}
\usepackage{multirow}
\usepackage{subfigure}
\usepackage{array}
\usepackage{multicol}
\usepackage{xcolor}
\usepackage{amsfonts}
\usepackage{amsmath}
\DeclareMathOperator*{\argmax}{arg\,max}

\renewcommand\fbox{\fcolorbox{orange}{white}}
\usepackage[export]{adjustbox}
\usepackage{tikz}
\usetikzlibrary{calc,fit,arrows.meta}
\newlength\aelength

\AppendGraphicsExtensions{.gif}

\graphicspath{
  {high_res/}
}

\usepackage[ruled]{algorithm2e} %

\SetAlFnt{\small}
\SetAlCapFnt{\small}
\SetAlCapNameFnt{\small}
\SetAlCapHSkip{0pt}
\usepackage[numbers]{natbib}

\usepackage{caption}
\newcommand\MYhyperrefoptions{bookmarks=true,bookmarksnumbered=true,
pdfpagemode={UseOutlines},plainpages=false,pdfpagelabels=true,
colorlinks=true,linkcolor={black},citecolor={black},urlcolor={black},
pdftitle={SceneHGN: Hierarchical Graph Networks for 3D Indoor Scene Generation with Fine-Grained Geometry},%
pdfsubject={Computing methodologies~Computer graphics,Computing methodologies~Scene understanding,Computing methodologies~Neural networks,Computing methodologies~Hierarchical representations,Computing methodologies~Learning latent representations,Computing methodologies~Shape modeling},%
pdfauthor={
Lin~Gao,
        Jia-Mu~Sun, 
        Kaichun~Mo, %
        Yu-Kun~Lai, 
        Leonidas~Guibas 
        and~Jie~Yang},
pdfkeywords={3D indoor scene synthesis, deep generative model, recursive neural network, variational autoencoder, graph neural network, relationship graphs, fine-grained mesh generation}}%
\ifCLASSINFOpdf
\usepackage[\MYhyperrefoptions,pdftex]{hyperref}
\else
\usepackage[\MYhyperrefoptions,breaklinks=true,dvips]{hyperref}
\usepackage{breakurl}
\fi
\usepackage[nameinlink,noabbrev]{cleveref}

\newcommand{\yj}[1]{{\color{black}#1}}
\newcommand{\jm}[1]{{\color{black}#1}}

\newcommand{\yjr}[1]{{\color{black}#1}}

\newcommand{\eg}{\textit{e.g. }}

\newcommand{\etal}{\textit{et al.}}

\newcommand{\tabincell}[2]{\begin{tabular}
{@{}#1@{}}#2\end{tabular}}
\newcommand{\titlecap}[2]{\textbf{#1} #2}
\newcommand{\name}{\textsc{SceneHGN}\xspace}

\begin{document}
\title{\name: Hierarchical Graph Networks for 3D Indoor Scene Generation with Fine-Grained Geometry}

\author{
       Lin~Gao,
        Jia-Mu~Sun,
        Kaichun~Mo, %
        Yu-Kun~Lai, 
        Leonidas~J.~Guibas 
        and~Jie~Yang%
\IEEEcompsocitemizethanks{
\IEEEcompsocthanksitem Lin Gao, Jia-Mu Sun, and Jie Yang are with the Beijing Key Laboratory of Mobile Computing and Pervasive Device, 
Institute of Computing Technology, Chinese Academy of Sciences, Beijing, China, and also with the University of Chinese Academy of Sciences, Beijing, China.%
\protect\\
E-mail: \{gaolin, sunjiamu21s, yangjie01\}@ict.ac.cn
\IEEEcompsocthanksitem Kaichun Mo and Leonidas Guibas are with Stanford University.%
\protect\\
E-mail: kaichunm@stanford.edu, guibas@cs.stanford.edu
\IEEEcompsocthanksitem Yu-Kun Lai is with the School of Computer Science \& Informatics, Cardiff University, U.K.
\protect\\
E-mail: LaiY4@cardiff.ac.uk
\IEEEcompsocthanksitem Project webpage: http://geometrylearning.com/scenehgn/.
}%
\thanks{Manuscript received April 19, 2005; revised August 26, 2015.}}

\markboth{IEEE Transactions on Pattern Analysis and Machine Intelligence,~Vol.~xx, No.~xx, June~2021}%
{\name: Hierarchical Graph Networks for 3D Indoor Scene Generation with Fine-Grained Geometry}

\IEEEtitleabstractindextext{%

\input{sec/0Abs}

\begin{IEEEkeywords}
3D indoor scene synthesis, deep generative model, recursive neural network, variational autoencoder, graph neural network, relationship graphs, fine-grained mesh generation
\end{IEEEkeywords}}

\maketitle

\IEEEdisplaynontitleabstractindextext

\IEEEpeerreviewmaketitle

\begin{figure}
\centering
{
    \includegraphics[width=0.95\linewidth]{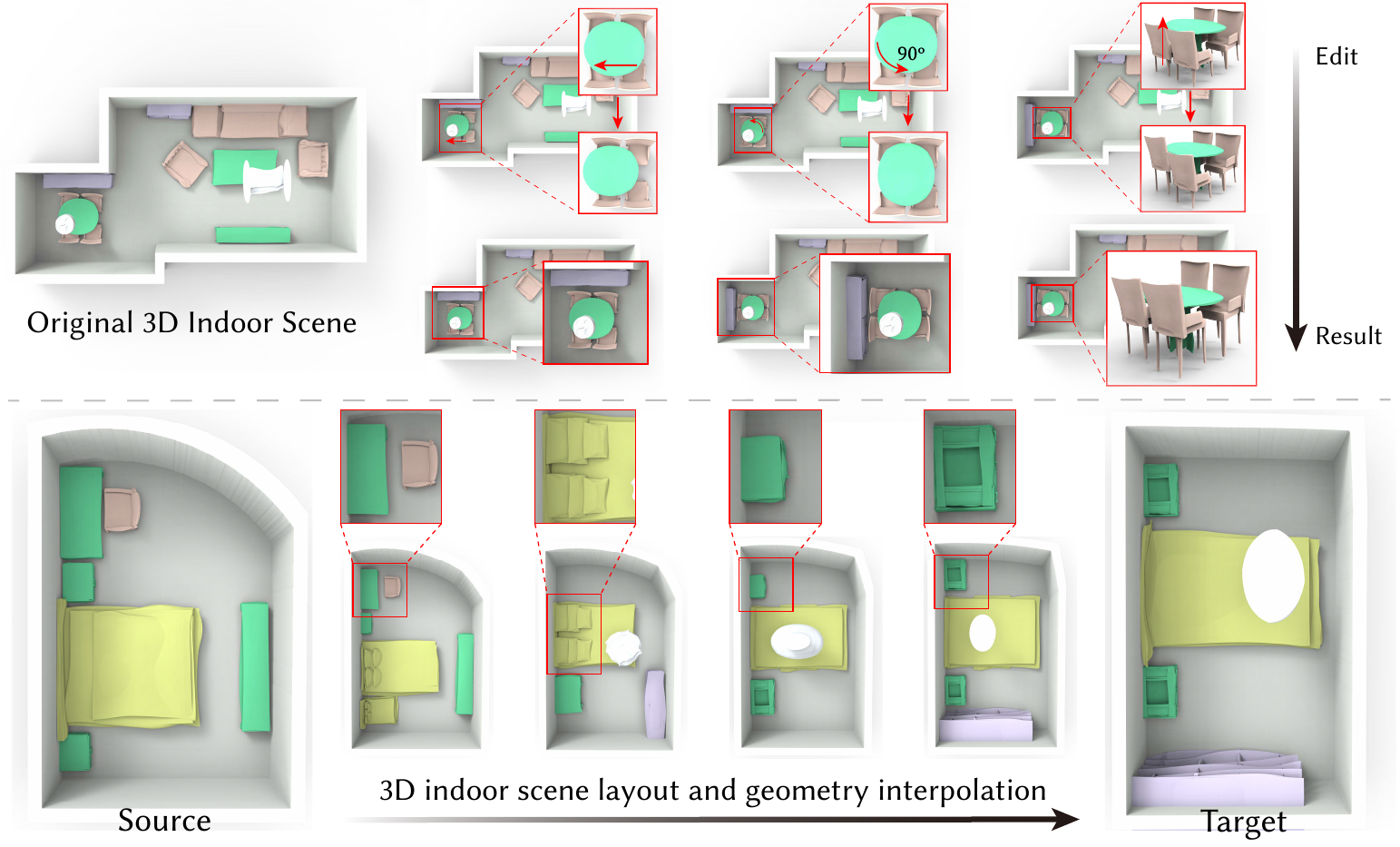}
}\caption{Our deep generative model \name encodes the indoor scene across multiple conceptual levels: the room, functional regions, furniture objects, and even fine-grained object part geometry. We utilize edges, including our proposed hyper-edges to strengthen the relations between objects during decoding. 
This enables some interesting applications, such as room editing with part-level geometry and scene interpolation. Our approach allows the entire 3D room to be represented and synthesized. Based on this, we can achieve part geometry editing (at different scales) in the scene, such as rigid transformation in a functional region and non-rigid deformation at the part level. 
Meanwhile, our network is capable of capturing the smooth latent space near similar scenes for plausible scene interpolation.
}
\label{fig:teaser}
\end{figure}

\input{sec/1Intro}

\input{sec/2Related}

\input{sec/3Overview}

\input{sec/4Rep}

\input{sec/5Method}

\input{sec/6Exps}

\input{sec/7Conclusion}

\section*{Acknowledgment}
This work was supported by the Beijing Municipal Natural Science Foundation for Distinguished Young Scholars (No. JQ21013), the National Natural Science Foundation of China (No. 62061136007) and the Youth Innovation Promotion Association CAS.
Kaichun Mo and Leonidas J. Guibas were supported by the ARL grant W911NF-21-2-0104, a Vannevar Bush Faculty Fellowship, and a gift from the Adobe Corporation.

\appendices
\section{Summary}
This supplementary material accompanies the main paper, which presents the data preparation and its description on the 3D-Front~\cite{fu20203dfront} and PartNet~\cite{mo2019partnet}, more network training and implementation details, and more scene reconstructions/comparisons.

All sections are listed as follows:
\begin{itemize}
    \item Section~\ref{sec:dataset} provides more details on the dataset creation for the 3D-Front and PartNet;
    \item Section~\ref{sec:implementation} provides more training and implementation details for our hierarchical graph networks.
    \item Section~\ref{sec:recon} provides more indoor scene reconstruction evaluation and comparison with the implicit-based approach and GRAINS.
\end{itemize}

\input{sec/supp_1data.tex}

\input{sec/supp_2recon.tex}

\ifCLASSOPTIONcaptionsoff
  \newpage
\fi

{
\bibliographystyle{IEEEtran}
\bibliography{bibliography}
}

\end{document}

%% file: sec/0Abs.tex
\begin{abstract}
3D indoor scenes are widely used in computer graphics, with applications ranging from interior design to gaming to virtual and augmented reality. 
They also contain rich information, including room layout, as well as furniture type, geometry, and placement. 
High-quality 3D indoor scenes are highly demanded while it requires expertise and is time-consuming to design high-quality 3D indoor scenes manually. 
Existing research only addresses partial problems: some works learn to generate room layout, and other works focus on generating detailed structure and geometry of individual furniture objects. 
However, these partial steps are related and should be addressed together for optimal synthesis. 
We propose \textsc{Scene}HGN, a hierarchical graph network for 3D indoor scenes that takes into account the full hierarchy from the room level to the object level, then finally to the object part level. 
Therefore for the first time, our method is able to directly generate plausible 3D room content, including furniture objects with fine-grained geometry, and their layout. 
To address the challenge, we introduce functional regions as intermediate proxies between the room and object levels to make learning more manageable. 
To ensure plausibility, our graph-based representation incorporates both vertical edges connecting child nodes with parent nodes from different levels, and horizontal edges encoding relationships between nodes at the same level. 
Our generation network is a conditional recursive neural network (RvNN) based variational autoencoder (VAE) that learns to generate detailed content with fine-grained geometry for a room, given the room boundary as the condition. 
Extensive experiments demonstrate that our method produces superior generation results, even when   comparing results of partial steps with alternative methods that can only achieve these. 
We also demonstrate that our method is effective for various applications such as part-level room editing, room interpolation, and room generation by arbitrary room boundaries. 
\end{abstract}

%% file: sec/1Intro.tex
\section{Introduction}
\label{sec:intro}

3D indoor scenes are useful for a wide range of applications, such as  smart digital houses, virtual reality/argument reality, robotics, virtual room planning, etc. Therefore, high-quality 3D indoor scenes are in high demand. However, they are compositionally complex and individual furniture objects often contain rich geometric details. Creating high-quality 3D indoor scenes is not only time-consuming but also requires expertise for designers. So research that can automate 3D scene generation would be highly valuable.

Although existing research works have considered some of the problems related to 3D indoor scene generation, they usually focus only on partial steps, rather than the whole process. For example, a large body of work addresses indoor scene layout generation, including traditional data-driven models (\eg~\cite{fisher2012example,henderson2017generative}) and more recent deep generative models (\eg~\cite{wang2019planit,li2019grains}). Although such works can generate diverse and plausible furniture layouts, they do not generate furniture geometry at the same time, and usually only retrieve existing furniture shapes from a repository. However, shape geometry and layout are related and treating these two steps separately may produce suboptimal results. 
Moreover, as furniture can have a large variety in terms of both geometry and structure, retrieving shapes inevitably restricts the diversity of furniture shapes that may appear in the synthesized scenes. Some other works (\eg~\cite{li2017grass,mo2019structurenet,gao2019sdm}) explicitly consider part-aware 3D shape generation, which can be applied to furniture objects to synthesize objects with various structures and/or geometry details, but such works are restricted to individual objects, rather than at the 3D indoor scene level.

While it is possible to apply these individual steps in sequence to synthesize 3D indoor scenes, it is difficult to ensure consistency and compatibility between objects, and if the early stage output is treated as a constraint to the next stage, it may also unnecessarily restricts the diversity of the generated scenes.
\yj{A key observation is that the layout and geometry of furniture are entangled. For example, when a chair is placed underneath a table, its geometry cannot be arbitrary as the chair must fit in the space there. Hence, the geometry must be compatible with its layouts.}
To address these challenges, we propose to \emph{jointly} model the layout and fine-grained geometry, and synthesize an entire 3D room using a single deep generative model. This has significant advantages: by treating it as a joint optimization problem, our approach is able to generate diverse indoor scenes with rich geometric details, while ensuring object room relationships, contextual relationships between objects, and consistency/compatibility of content and style for furniture objects.

However, there are many challenges due to the complexity of data and the problem. While a room is naturally hierarchical: it contains multiple furniture objects, and each furniture object can also be modeled using a part-based approach~\cite{yang2020dsm} for flexible structure and fine-grained geometry, a room may contain a large number of objects (\eg~a room in the 3D-FRONT~\cite{fu20203dfront} dataset contains up to 188 objects), which makes learning difficult. Considering that for larger rooms, smaller groups of objects are more likely to be related to each other (\eg  several chairs surrounding a table), we introduce \emph{functional regions} (\eg dining regions, sofa regions), as intermediate proxies to bridge the gap between the room and objects. When generating indoor scenes, the room shape is usually given. We represent the room shape flexibly by deforming a unit square, and the coding of the deformation is treated as a condition for indoor scene generation. To ensure plausible synthesis, it is also essential to take rich relationships into account. 
These include vertical relationships in the hierarchy: regions must be within the room boundary, objects should be within the region boundary, etc., and also horizontal relationships constraining objects at the same level: \eg symmetry of objects, adjacency between objects,  the symmetry between object parts, etc.  To achieve this, our hierarchical graph network is a recursive neural network (RvNN) based variational autoencoder (VAE) that covers 4 levels, namely: room, regions, objects and parts,  with carefully designed edges between graph nodes to enforce constraints. Training of such a large network can also be challenging, and we propose a multi-stage training strategy to ensure training is stable and effective.

As no publicly available datasets~\cite{fu20203dfront,zheng2020structured3d,li2021openrooms,roberts:2021} contain rich 3D indoor scenes at the part level, we use a hybrid dataset combining 3D-FRONT~\cite{fu20203dfront} data (object-level 3D indoor scenes), with PartNet~\cite{mo2019partnet} data which contains objects with detailed part-level annotation, where each object in the 3D-FRONT dataset is replaced with the most similar PartNet object to make the obtained 3D scene with part-level annotations. Extensive experiments show that our method is superior to baselines (even when comparing the partial steps they are designed for), and allow a range of interesting applications, including reconstruction, generation, completion, and interpolation. 

In summary, the main contributions of our paper are:
\begin{enumerate}
\item To the best of our knowledge, this is the first deep generative model capable of synthesizing an entire room with plausible furniture, including object layout and fine-grained object geometry. 
\item To achieve this, we propose a hierarchical graph network based on an RvNN VAE, that covers 4 levels from a room to objects and object parts. We introduce functional regions and carefully designed both vertical and horizontal edges in the graph, including hyper-edges to represent relationships among multiple objects, to ensure effective learning and plausible generation. We further encode room boundary as a deformed square and incorporate it as the condition for a controlled generation.
\item Extensive experiments show that our method outperforms existing baselines, and supports a range of applications from indoor scene synthesis to editing and completion.
\end{enumerate}

%% file: sec/2Related.tex
\section{Related Work}
\label{sec:related}
In this section, we first briefly review the compositional scene representations and approaches in the literature.
Then, we summarize existing research efforts on leveraging hierarchical graph representations for learning 3D generative models. 

\textbf{Indoor Scene Representations.}
Due to the complexity and diversity of realistic 3D scenes, researchers have pursued compositional scene representations, such as scene graphs that explicitly model the entities in the scene (\eg rooms, objects, walls) and the rich relations among them (\eg adjacency, symmetry).
Scene graphs have been shown to be powerful to generate 2D images~\cite{johnson2018image,ashual2019specifying}.
Recent works~\cite{chang2015text,wang2019planit,luo2020end,Graph2Plan20,Nauata2020HouseGANRG,para2020generative} have been exploring leveraging scene graphs to guide 3D scene generation.
To name a few, Luo~\etal~\cite{luo2020end} learned to generate furniture layout in a room given a scene graph as input.
House-GAN~\cite{Nauata2020HouseGANRG} tackled the floor plan generation problem given room graph user inputs as constraints.
In our method, we use a hierarchical graph scene representation that not only exploits the advantages of scene graph representations for encoding rich relationships among scene entities at the same level, but also leverages hierarchical decomposition to abstract nodes at different levels (\eg regions, objects, object parts).

3D scenes can be hierarchically decomposed into multiple semantic levels of content nodes: regions, objects and object parts.
Some parent-child inclusion constraints, such as objects that should lie within the room boundary, should hold.
Huang~\etal~\cite{huang2018holistic} used holistic scene grammars to parse scenes as hierarchical structures for reconstruction from a single image.
Armeni~\etal~\cite{armeni20193d} introduced a scene hierarchical representation from the entire building to rooms and objects.
GRAINS~\cite{li2019grains} proposed to represent the objects and their relationships in a four-wall room as a hierarchy and employed recursive neural networks~\cite{socher2011parsing,socher2012convolutional} to conduct learning on such a representation.
Shi~\etal~\cite{shi2019hierarchy} explored to predict 3D scene layout using hierarchy denoising recursive autoencoders.
Our work introduces a hierarchical graph scene representation with four compositional levels: the room, functional regions, objects, and object parts, augmented with horizontal edges forming smaller graphs among sibling nodes at each level.

\textbf{Indoor Scene Synthesis.}
There are many works exploring furniture layout generation; \yj{we refer to survey papers~\cite{zhang2019survey,pintore2020state} for a comprehensive discussion. Next, we give a brief review of this task.}

The usual setting of works in this line is to retrieve 3D models from a given database and predict the model positions for the generation of semantically and functionally realistic indoor scenes.
Before deep learning gained its popularity, a substantial body of works~\cite{yu2011make,merrell2011interactive,fisher2012example,yeh2012synthesizing,xu2015wall,kermani2016learning,henderson2017automatic,liang2017automatic,zhang2021fast} has explored this problem and constantly pushes the frontier.
\yjr{For real indoor scenes, these works~\cite{xu2017autonomous,dong2019multi,xu2015autoscanning} perform dense and realistic reconstruction of the real scene via a scanning approach using robots.}
More recently, deep learning-based methods further boost performance.
PlanIT~\cite{wang2019planit} introduced an image-based generative model reasoning over relation graphs.
Ritchie~\etal~\cite{ritchie2019fast} and Wang~\etal~\cite{wang2018deep} proposed to learn image-based deep convolutional generative models.
GRAINS~\cite{li2019grains} leveraged a recursive neural network to layout furniture within a room with four walls.
Zhang~\etal~\cite{zhang2020deep} solved the task by training a generative adversarial network to achieve a free-form generation without any floor constraint via a hybrid representation.
\yjr{SceneFormer~\cite{wang2020sceneformer} achieved faster realistic 3D scene generation with self-attention of transformers, which can predict a sequence of object locations conditioned on the room layout or text descriptions.
ATISS~\cite{paschalidou2021atiss} also uses autoregressive transformers for automatic layout synthesis, scene completion, and object suggestion under some constraints.
}
\yj{For learning the location recommendation, 
Zhou~\etal~\cite{zhou2019scenegraphnet} used neural message passing to predict the probability of newly added objects for learning the spatial and structural relationships between objects within an incomplete indoor scene.
Liu~\etal~\cite{liu2021game} introduced a visual context-aware graph generation network to learn  global implicit relations on the in-game residential home complex.}
\yjr{Furthermore, Sync2Gen~\cite{yang2021scene} uses the learned parametric prior distribution to regularize the unrealistic indoor scenes from feed-forward neural models.}
There are also other works learning to produce furniture layout under language~\cite{ma2018language}, activity~\cite{fu2017adaptive}, human~\cite{qi2018human} and action~\cite{ma2016action} constraints.
Different from these works, our approach learns to generate novel 3D furniture shapes instead of retrieving existing models from a database and also extends to a more fine-grained level of object parts.

\textbf{Hierarchical Graph Networks.}
Designing neural architectures for processing hierarchical data is highly non-trivial.
Recent works have shown promising results using recursive neural networks (RvNN)~\cite{goller1996learning,socher2011parsing,socher2012convolutional} and Tree-LSTM~\cite{tai2015improved} to encode tree-structured natural language sentences and natural scenes.

Recently, we witnessed a surge of success using RvNNs to model 3D shapes and scenes.
GRASS~\cite{li2017grass} first proposed to represent 3D shapes as a binary hierarchy of parts and employed an RvNN variational autoencoder~\cite{kingma2013auto} for a 3D shape generative model.
Follow-up works~\cite{niu2018im2struct,yu2019partnet} extended GRASS for reconstructing the part hierarchy from a single image or 3D point cloud.
Using the PartNet dataset~\cite{mo2019partnet}, which provides large-scale hierarchical shape part segmentation annotations, StructureNet~\cite{mo2020structedit} and its follow-ups~\cite{mo2020structedit,mo2020pt2pc,yang2020dsm} extended the GRASS binary hierarchy into more flexible $n$-ary hierarchies and augmented the representation with adjacent and symmetric part relations forming local graphs among sibling nodes.
In this paper, we further explore using hierarchical graph networks, specifically RvNNs in this paper, for 3D scene generation, where we model each 3D scene as an $n$-ary hierarchy with graphs.

%% file: sec/3Overview.tex
\section{Overview}
\label{sec:overview}

\begin{figure*}[h]
    \centering
    \includegraphics[width=\linewidth]{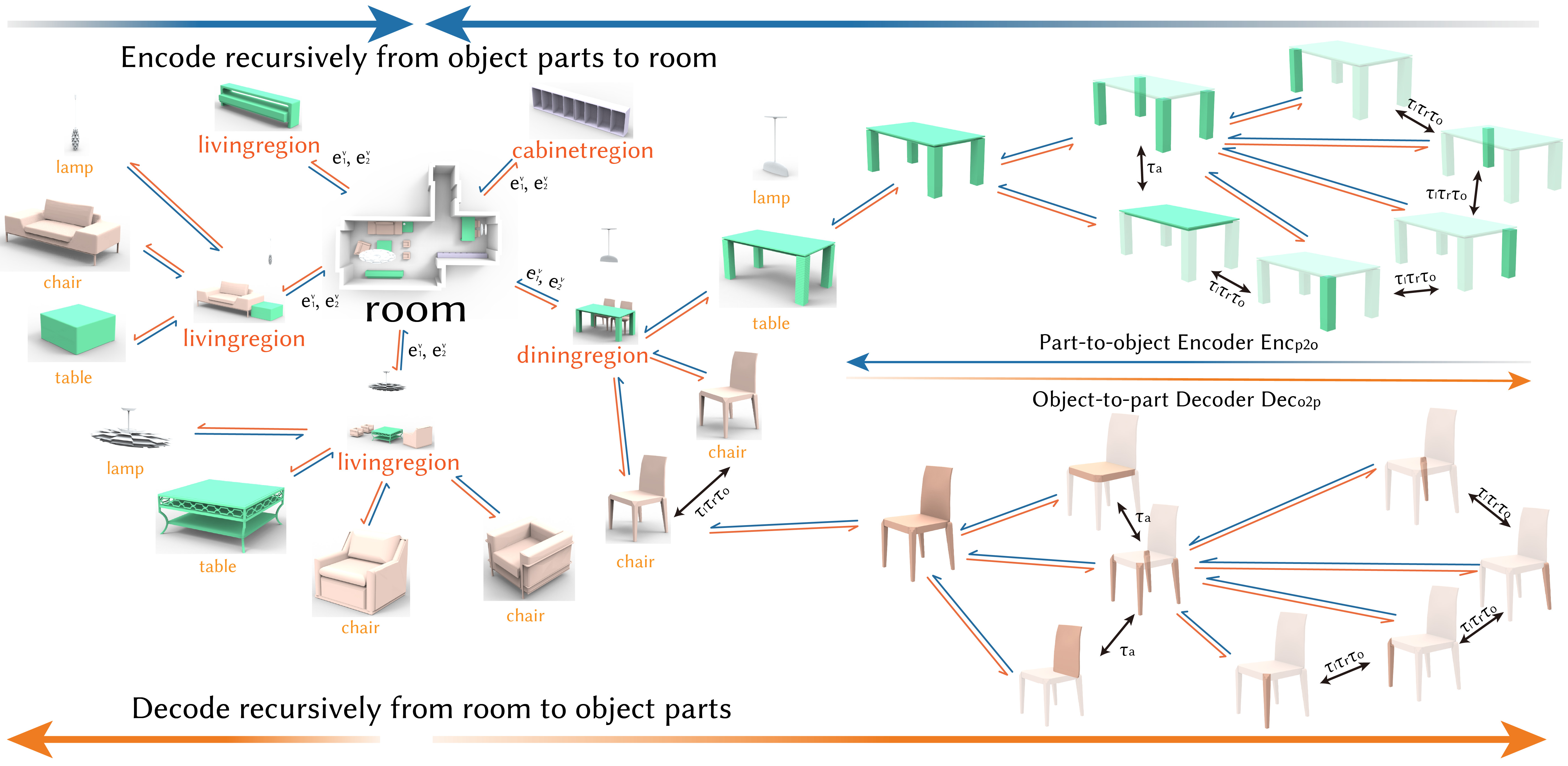}
    \caption{\titlecap{Hierarchical Scene Representation.}{Our scene hierarchy has four conceptual levels: the room root node, functional regions, objects, and object parts. To train the recursive autoencoder, we use an encoder network to summarize the features in a bottom-up fashion and a decoder network that reconstructs the scene hierarchy from the room root node to regions to objects and finally to object parts in a top-down manner. We also model the rich edge relationships at different levels in this process to enforce the validity of the generated scene structures.}}
    \label{fig:hier_simple_viz}
    \vspace{-3mm}
\end{figure*}

We propose \name, a hierarchical graph network for 3D indoor scene mesh generation that is end-to-end differentiable across multiple conceptual levels: rooms, functional regions, furniture objects, and even fine-grained object part geometry. 
Given room boundary layouts as inputs, \name learns smooth and continuous conditional latent spaces for generating diverse and novel indoor room scenes with semantically plausible furniture layouts and shape geometry with part-level details. 
Figure~\ref{fig:hier_simple_viz} presents a high-level architecture overview of our hierarchical conditional variational autoencoder (VAE) for 3D scene generation.

\textbf{Hierarchical Graph Representation.}
We represent every 3D indoor scene as a hierarchical tree with multiple levels: 
the room$\Rightarrow$functional region$\Rightarrow$objects$\Rightarrow$object parts.
We find that modeling the complicated space of highly compositional 3D indoor scenes at reasonable levels is necessary for producing high-quality results.

Aside from the naturally defined object and part levels, we additionally introduce an abstract level of functional regions (\eg dining region, sleeping region) to better organize the furniture objects within big rooms.
Figure~\ref{fig:subregion} presents example regions commonly seen in 3D indoor rooms.
The functional region level serves as an intermediate proxy for the hierarchy and helps group smaller sets of furniture shapes within a big room.
All the objects within a room are thus more consistently organized by an explicit hierarchy: room$\Rightarrow$region$\Rightarrow$objects.

Objects in the scenes also have their own compositional part structures: their constituent parts and part relationships. 
Hence, in our approach, we follow~\cite{mo2019structurenet,yang2020dsm} to organize an object as an $n$-ary hierarchical graph tree with its part geometry and edge relations.
Note that the object nodes in the entire scene hierarchy not only encode the $n$-ary hierarchical graph tree to represent the geometry and structure of the object, but also encode the spatial parameters in the context of the room furniture layout.

In the scene hierarchy, we also encode rich relationships among nodes at different levels.
Besides the parent-child vertical relationships naturally defined in the hierarchical representation, the nodes at the same level also have horizontal edge constraints, such as between shapes and walls (\eg a bed that is well-aligned with the wall), between two objects (\eg two symmetric nightstands), and even among multiple objects (\eg a dining table and four surrounding chairs, two nightstands co-aligned with the bed).
Not only do we adopt the binary edge relationships that are commonly used in previous works~\cite{mo2019structurenet,li2019grains}, we also propose to enforce hyper-edge constraints among multiple objects.

\textbf{Hierarchical Graph Network.}
We train a conditional recursive variational autoencoder to learn a smooth latent space for scene generation. 
Our framework consists of a room layout encoder, a scene hierarchy encoder, and a scene hierarchy decoder.
The room layout encoder takes the deformation gradients of the floor boundary as input and extracts a vector that is used as the floor boundary condition for the decoder.
The scene hierarchy encoder maps the indoor scene hierarchies from the room level to the functional region level, the object level, and finally down to the object part geometry level into a common latent space hierarchically and recursively. 
In contrast, the scene hierarchy decoder performs an inverse mapping which decodes a latent vector and floor boundary condition vector into the furniture layouts and shapes with detailed geometry in a top-down manner.

During the encoding process, furniture objects are firstly encoded as object features with a pre-trained DSG-Net~\cite{yang2020dsm}.
Then, combining the object features and their spatial location information in the regions, the region-level features are extracted.
Finally, a single room-level root feature summarizes all region information together with the region-level spatial arrangement.
The scene hierarchy decoder inversely reconstructs the indoor scene hierarchy from a room-level latent code to more fine-grained levels in a top-down manner.
During the encoding and decoding of the training, several graph message passing operations are performed to capture rich edge relations and constraints between object and object, between room boundary and object, and among multiple objects.
Figure~\ref{fig:edges} visualizes the adopted binary edge relationships for scene generation and our proposed hyper-edge constraints among multiple objects.

\textbf{Paper Organization.}
In the following sections, we first describe the detailed definitions for our indoor scene hierarchical representation in Sec.~\ref{sec:rep}. 
We will introduce our concrete node level designs and rich edge relationships, including our proposed hyper-edges constraining $n$-ary part relationships beyond $n=2$, which are important for indoor scene generation.
Then, we present our hierarchical architecture designs for learning hierarchical 3D indoor scene generation in Sec.~\ref{sec:method}.
Learning to represent the input room boundary layouts as conditions to our hierarchical framework, we introduce FloorNet which learns to encode and decode input room boundary layouts represented as 2D deformation representation in vertex neighborhoods.
We also discuss the key roles of the introduced functional regions and how to model the rich relationships among objects and room boundaries.
Thanks to our learned latent space and decoder, in Sec.~\ref{sec:exps}, we show that our framework enables some interesting applications, such as indoor scene editing, completion, and conditional generation with some constraints (scene generation from 3D box layouts).

%% file: sec/4Rep.tex
\section{Hierarchical Graph Representation}
\label{sec:rep}

Below, we detail our node and edge compositional and relational designs of our proposed hierarchical graph representation.

\subsection{Hierarchical Node Decomposition}
Given a 3D indoor scene, we represent it as a hierarchical graph structure.
There are multiple conceptual levels in the hierarchy: the room, functional regions, furniture objects, and object part geometry. 
Besides the natural concepts of objects and object parts defined in PartNet~\cite{mo2019partnet}, \yjr{we additionally propose a new conceptual level of functional regions to further divide the objects in big rooms into smaller clusters of objects for producing 3D scenes with higher quality via the parameter-free method.}
We describe the detailed node definitions as follows.

\textbf{Regions.}
According to the functionality of furniture shapes, functional regions divide the whole scene into smaller groups of objects for a more consistent and learning-friendly scene hierarchical representation. 
Different types of rooms usually have very disparate functional regions. For example, one may have entertainment and living regions in a living room, and have sleeping and cabinet regions in a bedroom.
Explicitly modeling the region semantics not only provides many meaningful semantic labels as parts of the scene generation results, \yjr{but it is also beneficial for modeling a large number of objects within each room.}
\yjr{Due to the capacity limit of the RvNNs, we cannot practically model a big graph of a huge number of objects in a single room as the children nodes of a single parent node (\eg close to hundreds of objects for the biggest room in the 3D-FRONT dataset~\cite{fu20203dfront}).}

\begin{figure}[t!]
    \centering
    \includegraphics[width=\linewidth]{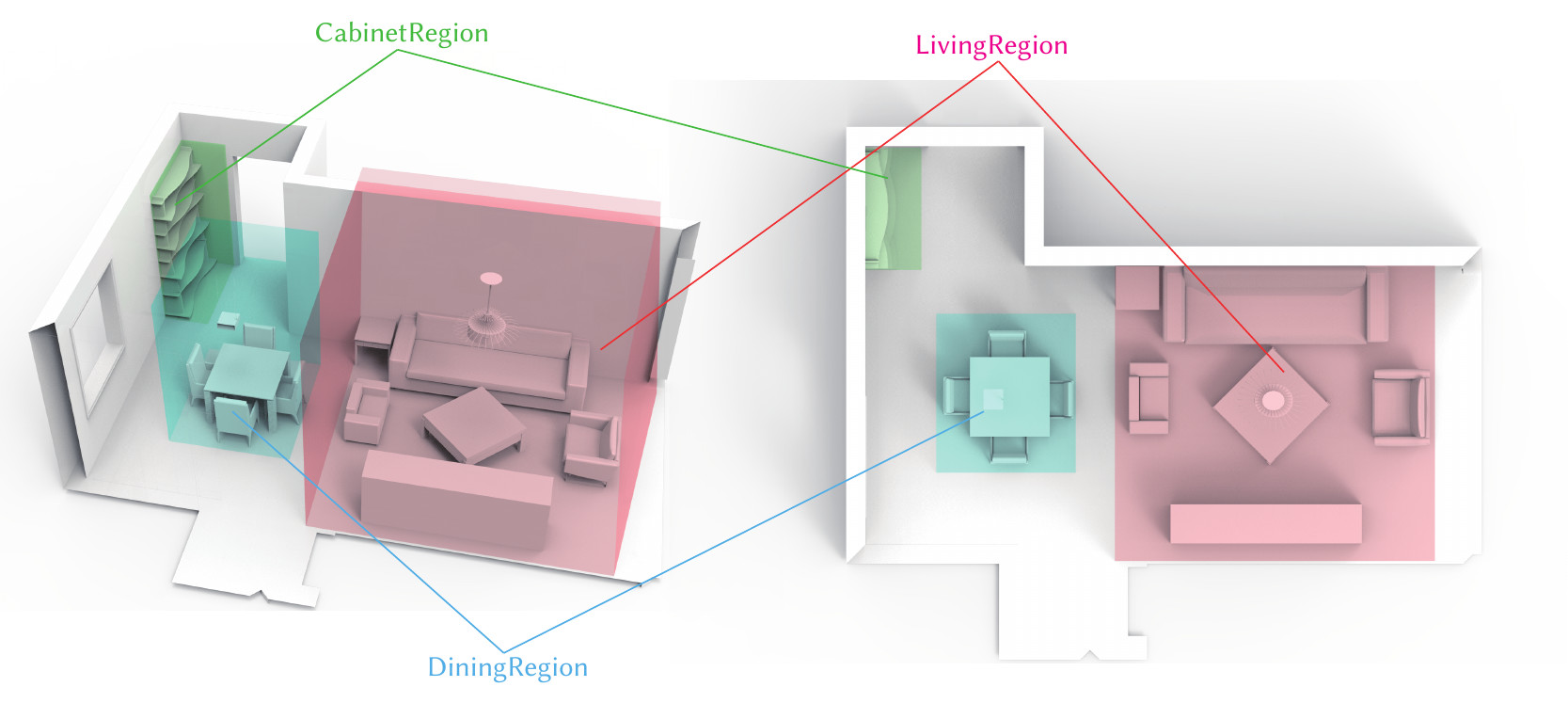}
    \caption{\titlecap{Functional Region Visualization.}
    {In the figure, a whole scene is divided into three functional regions including a Cabinet Region, a Dining Region, and a Living Region, which are highlighted in different colors. 
    The separation is conducted by a spatial clustering algorithm DBSCAN~\cite{DBScan}, which is a density-based and non-parametric clustering algorithm, where the number of clusters is self-adaptive. 
    We can see that an indoor scene can be divided reasonably.}}
    \label{fig:subregion}
    \vspace{-3mm}
\end{figure}

In our implementation, we divide each scene into smaller functional regions by running a spatial clustering algorithm DBSCAN~\cite{DBScan}, which is a density-based and non-parametric clustering algorithm, where the number of clusters is self-adaptive. 
All objects can be grouped and organized according to spatial closeness  by the algorithm, which is often correlated to their functionality.
The functionality of the region is determined by the object with the largest area in the cluster.
Within each detected functional region, many functionally related objects are often clustered together (\eg dining table and its surrounding dining chairs), as illustrated in Figure~\ref{fig:subregion}.
We also detect and correct automatic labeling errors manually for some unreasonable divisions of indoor scene space.

In summary, for a 3D scene $S$, we have $S = \left\{\left\langle g_1,g_2,\cdots,g_K\right\rangle, \mathbf{R}_{region} \right\}$, where $g_i$ means the $i^{\rm th}$ clustered group (a functional region), with $\mathbf{R}_{region}$ means the horizontal relations between these functional regions.
We will describe the region relationships in Sec.~\ref{subsec:edge}.

\textbf{Objects.}
After we divide the scene space into functional regions, there are many furniture objects within each region.
Each shape $O_i \in g_k, 1 \leq k \leq K$ is described by a mesh geometry, along with its semantic object category and its spatial location within the region.

Formally, we define $g_i= \left\{\left\langle O_1,O_2,\cdots,O_M\right\rangle, \mathbf{R}_{object} \right\}$, where $O_i$ means the $i^{\rm th}$ object in the functional region, with $\mathbf{R}_{object}$ denotes the horizontal relations among objects that belong to one functional region.
We will describe the region relationships in Sec.~\ref{subsec:edge}.

\textbf{Object Parts.}
We use the PartNet~\cite{mo2019partnet} shape part hierarchy to decompose every 3D shape into the semantically consistent part hierarchy, organized as an $n$-ary hierarchical graph tree structure that covers different levels of part instances ranging from coarse-grained to fine-grained parts.

Namely, each object $O_i$ is decomposed into some parts $\mathbf{P}_{O_i}^{j}$ organized by a hierarchy structure $\mathbf{H}_{O_i}$ and binary/$n$-ary relations $\mathbf{R}_{O_i}$ between parts. Each part $\mathbf{P}_{O_i}^{j}$ has a pre-defined semantic label and detailed mesh geometry $\mathbf{G}_{O_i}^{j}$.
We follow the exact same set of edge relationships for the part nodes within an object as in StructureNet~\cite{mo2019structurenet}.

\subsection{Edge Relationships}
\label{subsec:edge}
We consider two kinds of node relationships in a scene hierarchy. 
\begin{itemize}
\item vertical edges for the relationships between parent and child nodes;
\item horizontal edges among the nodes at the same level.
\end{itemize}
Extending previous works that only considered binary edges~\cite{mo2019structurenet,li2019grains} for modeling relationships, we propose hyper-edges to model the $n$-ary relationships among $n>2$ parts.
In many indoor scenes, multiple objects may hold one $n$-ary constraint, such as parallel collinearity and $n$-fold rotational symmetry (\eg dining chairs surround a round table). 
We experimentally find that introducing hyper-edges makes our generated indoor scenes more reasonable and realistic.

A notably related work to us is GRAINS~\cite{li2019grains}.
In the GRAINS scene representation, they described the relations between objects and walls with the three relations: supporting, surrounding, and co-occurrence.
Different from GRAINS, we use more accurate relations to describe how to organize these furniture objects, including edges between two objects: adjacency~($\tau_a$), translational symmetry~($\tau_t$), reflective symmetry~($\tau_r$), and rotational symmetry~($\tau_o$); edges among multiple objects: parallel collinearity~($\tau_{np}$) and $n$-fold rotational symmetry~($\tau_{nr}$).
Experiments in Sec.~\ref{sec:exps} show that our proposed set of relationships works better than GRAINS.

\begin{figure}[t!]
    \centering
    \subfigure[room-object edges]{\includegraphics[width=0.49\linewidth]{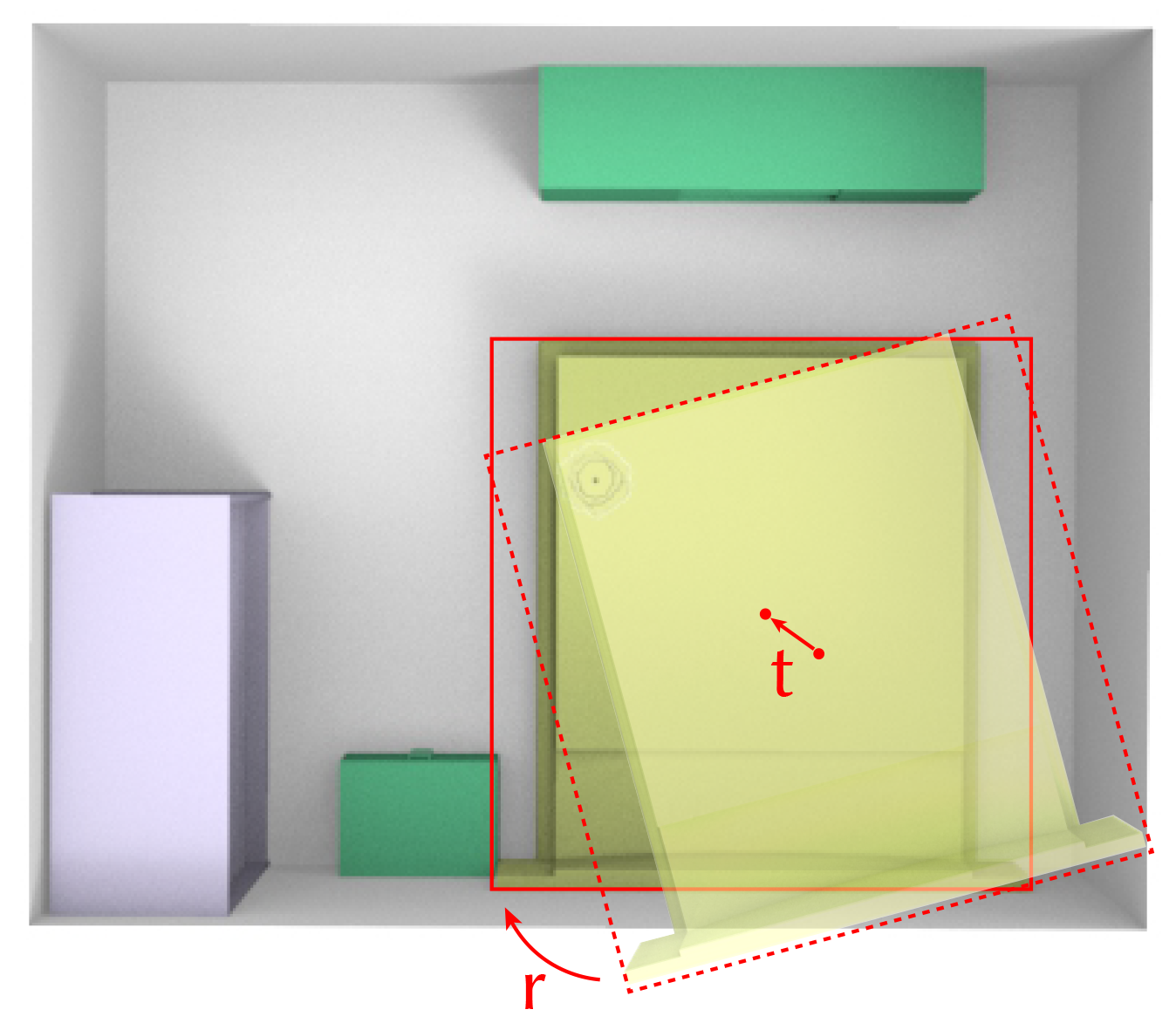}}
    \subfigure[object-object edges]{\includegraphics[width=0.49\linewidth]{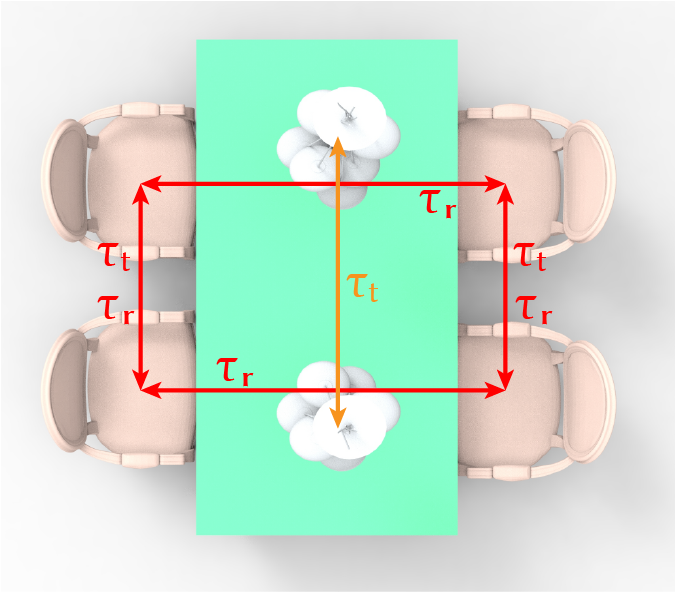}}
    \caption{\titlecap{Two Types of Binary Edges between Objects.}{We illustrate the two types of binary edges at the object level of our hierarchy. In (a), we show a binary edge example of the first kind which is defined between the room wall and an object. It encourages the object to locate within the boundary of the room and align with the room boundary. In (b), another type of binary edge describes the spatial relationship between two objects. For example, any pair of the four chairs have rich symmetry relationships of different kinds.}}
    \label{fig:edges}
    \vspace{-3mm}
\end{figure}

\subsubsection{Vertical Edges}
The vertical edges represent the relationships between parent nodes and children nodes, which are naturally described by the multiple levels of node concepts in the hierarchy: 
\begin{itemize}
    \item the room root node comprises of many functional regions;
    \item every region contains many furniture objects;
    \item each object is further composed of object parts at different granularity.
\end{itemize}

In order to synthesize a realistic and reasonable indoor scene, we must make the generated objects compatible with the room boundary.
So, we propose two additional types of vertical edges connecting the room root node to the object level: 
\begin{itemize}
\item $\mathbf{e}^{v}_1$: the oriented bounding box of an object may have to align with the room boundary in some scenes; 
\item $\mathbf{e}^{v}_2$: \yj{the generated object in a scene %
 must be located within the room boundary walls.}
\end{itemize}
We find that such skip-linked vertical edges help regularize the validity of the generated scene meshes.

\subsubsection{Binary Horizontal Edges}
The horizontal edges are defined on the nodes at the same level, to describe the rich relations and constraints among sibling nodes of a parent node. 
In this work, we consider binary horizontal edges between two objects and between two object parts.
Inspired by \yjr{previous works~\cite{mo2019structurenet, yang2020dsm} on shape generation, we define four types of binary edges, including adjacency~($\tau_a$), translational symmetry~($\tau_t$), reflective symmetry~($\tau_r$), and rotational symmetry~($\tau_o$).
For the adjacency relationship,  we define two parts as adjacent if their smallest distance is below $0.05\times\bar{r}$, where $\bar{r}$ is the average bounding sphere radius of the two parts. 
For the symmetry relationships, we follow the method from~\cite{wang2011symmetry} to detect $\tau_t, \tau_r, \tau_o$.}
Such binary edges are automatically detected from the input training scene data at the object level.
We directly adopt the object part relationships in the previous works~\cite{mo2019structurenet, yang2020dsm}.
Note that there may exist multiple relationships between two nodes.
For example, in Figure~\ref{fig:edges} (b), the two chairs on one side of the table has both the translational symmetry~($\tau_t$) and the reflective symmetry~($\tau_r$).

\begin{figure}[t!]
    \centering
    \includegraphics[width=0.49\linewidth]{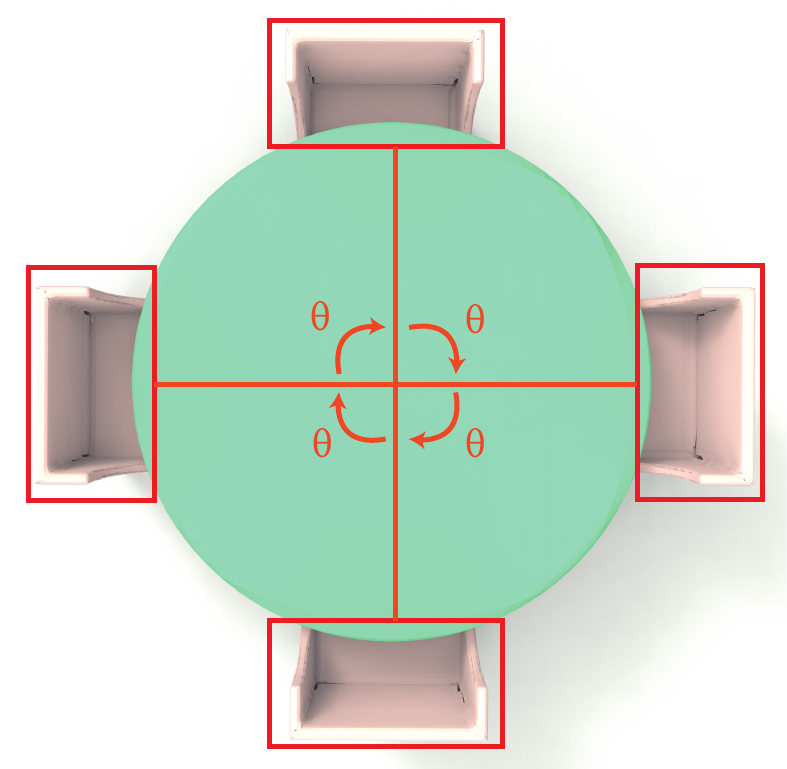}
    \includegraphics[width=0.49\linewidth]{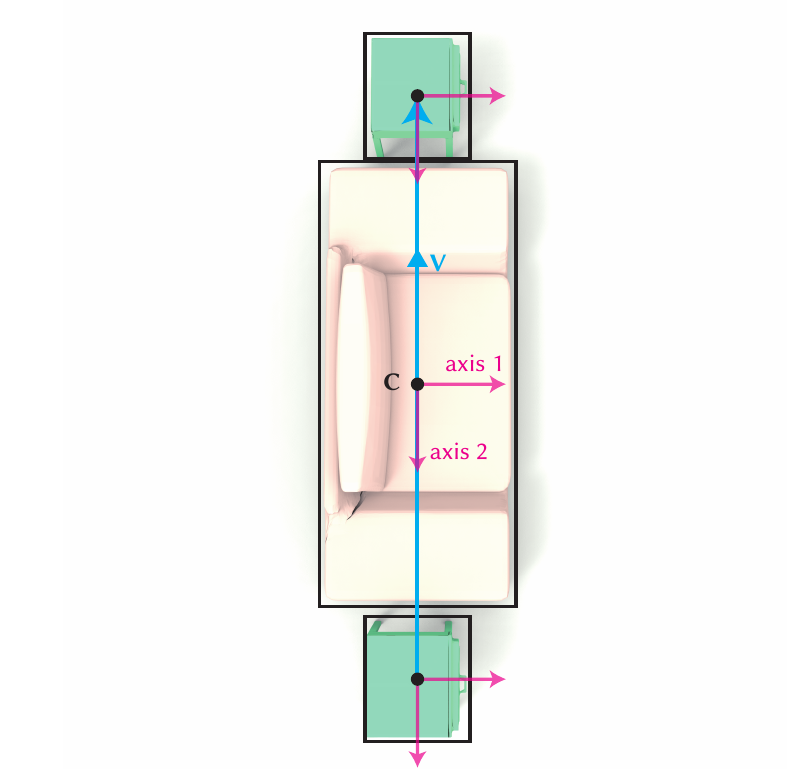}
    \caption{\titlecap{Illustration of Hyper-edges.}{We define two types of hyper-edges that exist across multiple objects: rotation and parallel. A rotation hyper-edge indicates that objects are rotated around a center, and a parallel hyper-edge indicates objects are placed collinearly.}}
    \label{fig:hyperedges}
    \vspace{-3mm}
\end{figure}

\subsubsection{$N$-ary Hyper-edges}
We find that binary relations are not enough to describe the complex object layouts, since some realistic scenes may have more complicated relationships that happen among more than two nodes.
For example, in Figure~\ref{fig:hyperedges} (a), besides the illustrated extensive set of binary relationships, it would be beneficial to consider a $5$-ary hyper-edge relationship constraining that the four chairs surround the central dining table.
Similar hyper-edge relationships may also be helpful, such as the example in Figure~\ref{fig:hyperedges} (b) where the two nightstands and the bed should have their oriented bounding boxes in parallel to each other.
Thus, in this work, we introduce two types of hyper-edge relationships:
\begin{itemize}
    \item $n$-fold rotational symmetry $\mathbf{e}^{hyper}_1$: \eg the dining table is surrounded by some dining chairs;
    \item parallel collinearity $\mathbf{e}^{hyper}_2$: \eg multiple box objects are in parallel to each other and their centers may be collinearly aligned.
\end{itemize}
\yjr{Our hyper-edges are detected at object level and within a functional region, where the number of objects is larger than 2. The relationships between two objects are represented by binary edges.}
For another reason, if we only consider the relations between any two nodes, a dense graph will be constructed for the network learning, which is very hard for training and conducting message-passing operations.

\textbf{$N$-fold Rotational Symmetry.}
Consider $N$ objects $\textbf{O} = \{O_1,\cdots, O_N\}$.
They are in an \textit{N-fold rotational symmetry hyper-edge} if and only if they satisfy:
\begin{equation}
\begin{aligned}
    &\exists p,\ \forall O_i \in \textbf{O}\ {\rm where}\ i \le N - 1,\\ 
   & CD\left(O_{i+1},Rot\left(p, \frac{2\pi}{N} \right) \times O_i\right) \le \epsilon_R
    \label{eqn:hyperedgerotate1}
\end{aligned}
\end{equation}
and
\begin{equation}
    CD\left(O_{1},Rot\left(p, \frac{2\pi}{N} \right) \times O_N\right) \le \epsilon_R 
     \label{eqn:hyperedgerotate2}
\end{equation}
Here, $CD$ indicates the Chamfer Distance between two objects. $Rot\left(p,\theta\right)$ denotes a rotation matrix that rotates an object around an axis that is parallel to the world up-axis and passes through the point $p$ by the angle $\theta$ (in radians). 
$\epsilon_R$ is a constant threshold. Note that the point $p$ in \autoref{eqn:hyperedgerotate1} and \autoref{eqn:hyperedgerotate2} must be the same. 
We show an example of multiple objects satisfying an $N$-fold rotational hyper-edge in \autoref{fig:hyperedges} (a).

\textbf{Parallel Collinearity.}
Given $N$ objects $\textbf{O} = \{O_1,\cdots, O_N\}$, they satisfy a \textit{collinearly parallel hyper-edge} if and only if 
\begin{itemize}
    \item the two main axes of the 2D oriented bounding boxes of all objects are parallel to each other (we exclude the world up-axis here as all objects are placed on the ground floor);
    \item they satisfy the following equation:
\end{itemize}
\begin{equation}
\begin{aligned}
    &\exists \textbf{v},\ \forall O_i \in \textbf{O}\ {\rm where}\ i \le N - 1, \exists d \in \mathbb{R},\\
   & ||C_{i+1} - (d\textbf{v} +  C_i)||^2 \le \epsilon_T
    \label{eqn:hyperedgetranslation}
\end{aligned}
\end{equation}
where $\textbf{v}$ is an arbitrary vector, $d$ is an arbitrary non-negative real number, $C_i$ is the center of the oriented bounding box of $O_i$, and $\epsilon_T$ is a constant threshold. 
\autoref{fig:hyperedges} (b) illustrates an example of collinearly parallel objects.

Different from superstructures (Hub and Spokes, Chains) in PlanIT~\cite{wang2019planit}, since our representation takes the part into consideration, the hyperedges are detected by calculating the object part bounding box instead of the bounding box of the whole shape. 
Further, the orientation of the object is involved in the detection: 1. the orientation of all objects that satisfy the \textit{Parallel Collinearity} must be the same; 2. the orientation of all objects that satisfy the \textit{$N$-fold Rotational Symmetry} must point to the same center object.

%% file: sec/5Method.tex
\section{Hierarchical Graph Network}
\label{sec:method}

Our network is a conditional RvNN~\cite{socher2011parsing,socher2012convolutional} VAE~\cite{kingma2013auto} on the hierarchical scene graph representation defined in the previous section, with many edge losses to encourage more realistic and plausible structural relations in the generated scene.

Our \name takes the scene hierarchy $S$ from the room root node down to object part geometry as inputs, with an additional input room boundary $b_i$ as the condition.
The room boundary $b_i$ is mapped into a conditional feature $f_{b_i}$ by a floor encoder $Enc_{floor}$, while the scene hierarchy $S$ is also mapped into a latent vector $f_{S}$ by the encoder of \name $Enc_{\name}$.
Then, the conditional room boundary feature $f_{b_i}$ and the encoded scene hierarchy feature $f_{S}$ are concatenated together, which is subsequently fed into the decoder of \name $Dec_{\name}$ to reconstruct the input scene hierarchy $S$.
To train the VAE generative model, we add a regularization, using the KL-Divergence, on the latent space to map all scenes onto a standard Gaussian distribution, from which we can smoothly sample novel scenes and interpolate between given scenes.

\begin{figure}[t!]
    \centering
    \includegraphics[width=\linewidth]{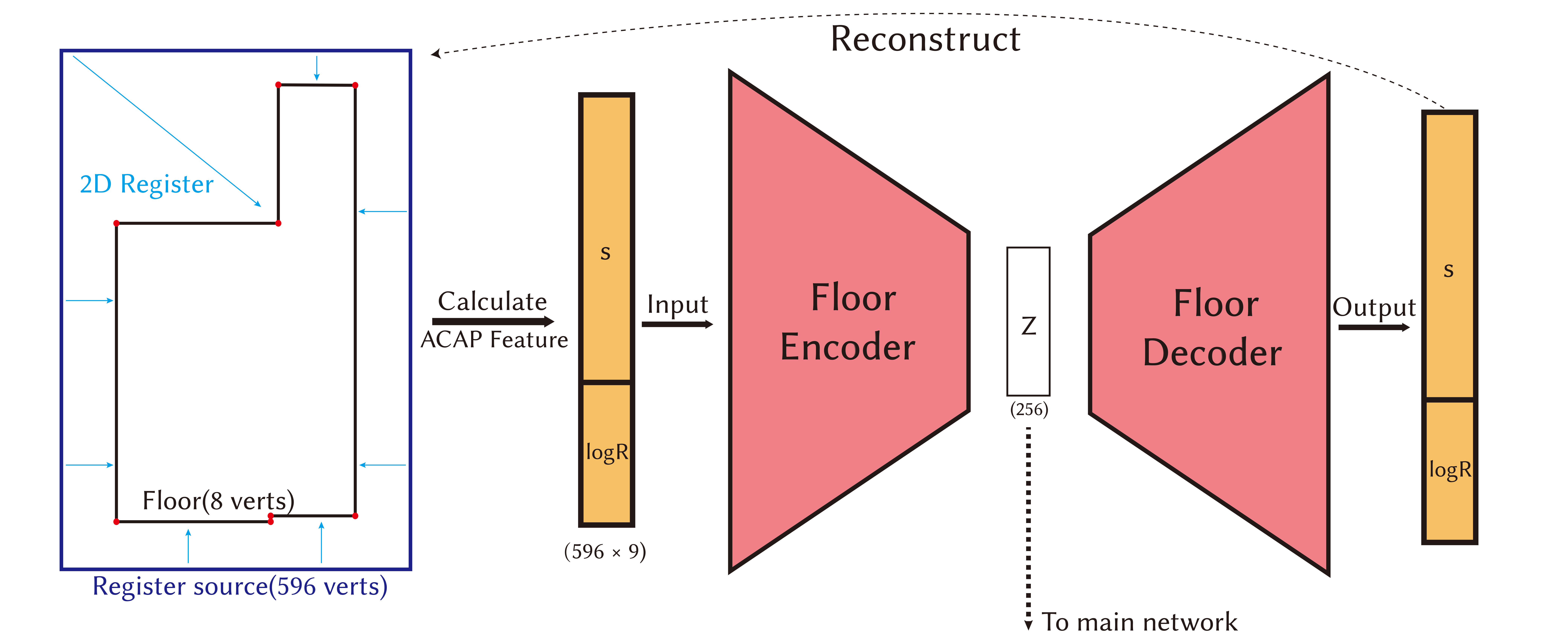}
    \caption{\titlecap{Floor VAE.}{We train a separate Variational AutoEncoder for encoding floor boundaries. Specifically, we register a 2D ring of vertices onto the input floor boundary map, and then calculate the ACAP feature~\cite{Gao2021SparseDD} on the registered ring structure. Finally, the VAE maps the ACAP feature into a latent vector which will serve as a condition for scene generation.}}
    \label{fig:part_floor_vae}
\end{figure}

Since we adopt the object part hierarchy from DSG-Net~\cite{yang2020dsm}, the part geometry encoder $Enc_{PG}$ and part geometry decoder $Dec_{PG}$ are following DSG-Net. 
The $Enc_{PG}$ takes the part geometry information $G_{P_i} = (X_{P_i}, c_{P_i})$, including deformation gradients $X_{P_i}$ of each part of object $O_j$, its center $c_{P_i}$, and its structural information $S_{P_i}$, and maps to a latent embedding feature $f_{P_i} = Enc_{PG}(G_{P_i}, S_{P_i})$. 
Inversely, the $Dec_{PG}$ maps the latent feature $f_{P_i}$ and its structural information $S_{P_i}$ back-into deformation gradient space $\hat{G}_{P_i} = (\hat{X}_{P_i}, \hat{c}_{P_i}) = Dec_{PG}(f_{P_i}, S_{P_i})$.

Below, we focus on introducing our room boundary layout and scene hierarchy VAEs, along with the training strategy and loss terms.

\subsection{Room Boundary Layout VAE}
\yj{Our goal is room generation given an arbitrary room boundary \yjr{which is topologically isomorphic with a ring-shaped boundary}. For generating reasonable indoor rooms and modeling the relationships between the objects and room boundaries, the detailed geometry of the room boundaries needs to be effectively represented. For achieving this goal,}
we train a FloorNet VAE, consisting of an encoder $Enc_{floor}$ and a decoder $Dec_{floor}$, to map the floor boundary of a room to a latent space.
We propose to use the deformation features of a 2D ring of vertices for the representation of the floor boundary geometry. 
Any closed floor boundary can be represented as a deformed 2D unit square boundary. 
In our paper, the unit square boundary consists of 596 vertices and 596 edges.
The architecture of the proposed FloorNet is shown in \autoref{fig:part_floor_vae}. 

The floor boundary in our dataset is a 2D ring represented by 596 vertices and \yj{596} edges.
Just for the illustration purpose, for example, the floor in \autoref{fig:part_floor_vae} (left) consists of 8 vertices and 8 edges.
To efficiently and accurately encode the floor boundary (especially for sharp corners), we apply the 2D non-rigid registration technique~\cite{bouaziz2014dynamic} rather than vertex coordinates to deform a source 2D mesh with 596 vertices to fit the shape of the floor boundary. 
Then, we calculate the ACAP deformation gradients~\cite{Gao2021SparseDD} on the registered 2D mesh. For every vertex, we finally extract a 6-dimensional feature $s \in \mathbb{R}^{6}$ and a 3-dimensional feature $\log R \in \mathbb{R}^{3}$, which represent scaling/shear and rotation respectively.

The $596 \times 9$ feature matrix is then fed into the encoder of our FloorNet. 
The key component of the FloorNet is a Graph Convolutional Network, where we treat the whole registered floor boundary as a cycle graph and perform convolution operation on it to extract 2D mesh features. 
After two convolutional operations, the features pass through an MLP which outputs the latent vectors encoding the room floor boundary.
\yj{We perform 2 iterations for this message to pass to neighboring vertices to learn the angle between adjacent edges for sharp corners.} 
The decoder is basically the inverse of the encoder to map the latent code back into the \yj{ACAP deformation gradients of} room boundary.
We use another Graph Convolutional Network to iteratively decode the vertex locations of the 2D ring structure.

\subsection{Scene Hierarchy Encoder}
Our encoder consists of two parts: a recursive encoder $Enc_{p2o}$ from the part geometry to object and another recursive encoder $Enc_{o2r}$ from the object to the whole room level. 
We directly follow the design of $Enc_{p2o}$ in DSG-Net~\cite{yang2020dsm}.
Below, we focus on describing our $Enc_{o2r}$ design that learns to map objects, through functional regions, and finally to the room root node.

For each object node, there are three types of information stored in it: the structural and geometric information $f_{O_i}$ \yjr{extracted using $Enc_{p2o}$ for the objects}, the placement parameters $Pos_{O_i} \in \mathbb{R}^{7}$ and the one-hot vector of semantic category label $l_{O_i}$.
The $Pos_{O_i}$ parameters include the center position $c_{O_i} \in \mathbb{R}^3$, the scales $s_{O_i} \in \mathbb{R}^3$, and the orientation $r_{O_i} \in \mathbb{R}$ around the world up-axis.
The object feature encoder $Enc_{obj}$ encodes the above information together into an object latent code $f_{O_i}^{obj}$.
\begin{equation}
    f_{O_i}^{obj} = Enc_{obj}([f_{O_i}; Pos_{O_i}; l_{O_i}])
\end{equation}
where the $Enc_{obj}$ is a full-connected layer and [;] means the concatenation operator.

\yjr{
For other non-object node $N_i$, the recursive encoder $Enc_{o2r}$ is used to gather the features of all children and perform message passing along the part relation edges among nodes.
For the hyper-edges among multiple nodes, we first only aggregate the type of hyper-edges into the corresponding nodes to update the node features by an MLP $Enc_{hyper}$, which consists of two fully-connected (FC) layers and a Leaky ReLU activation.
Then, we perform two message passing operations within the sub-graph and aggregate all features of child nodes by an FC layer with LeakyReLU activation.
Finally, the feature $f_{N_i}$ gathers information from the features of all children nodes.

\begin{equation}
\begin{aligned}
    \bar{f}_{O_i}^{obj} &= Enc_{hyper}(f_{O_i}^{obj}, l_{O_i}^{hyper}), s.t.~~O_i \in N_{hyper}\\
    f_{N_i} &= Enc_{o2r}\left(\left\{\bar{f}_{O_j}^{obj}\right\}_{(N_i, O_j) \in \mathbf{H}}, l_{O_j}\right)    
\end{aligned}
\end{equation}
where $(N_i, O_j) \in \mathbf{H}$ denotes that node $O_j$ is a child of $N_i$, $l_{O_i}$ is the one-hot vector of the semantic label, $l_{O_i}^{hyper}$ is the one-hot vector of the hyper edge type for object $O_i$, and $N_{hyper}$ is the set of nodes with the attribute of hyper edges.
If the node does not have any hyperedges associated with it, $l_{O_i}^{hyper}$ is empty (filled with zero).
}

The whole process $Enc_{p2o}$ and $Enc_{o2r}$ are repeated until to the root node. 
A fully connected layer maps the final feature of the root node into the latent space.
We add a regularization term, namely the KL-divergence, to the latent space for encouraging the latent space to be close to the standard Gaussian distribution.

\subsection{Scene Hierarchy Decoder}
The decoding process is conditioned on the feature of floor boundary extracted by the floor encoder $Enc_{floor}$.
The scene hierarchy decoder takes the floor condition and the root node feature outputted by the scene hierarchy encoder as inputs and learns to decode the scene hierarchy down to the part geometry in a recursive manner.
It also includes two parts: one recursive decoder $Dec_{o2p}$ that predicts part geometry from an object feature, and another recursive decoder $Dec_{r2o}$ that consumes the parent node feature $\hat{f}_{N_i}$, along with the conditioned feature $f_{floor}$, and decodes the object root node features.
For the object decoding, the $Dec_{o2p}$ decoder follows the design in DSG-Net. We refer the readers to DSG-Net~\cite{yang2020dsm} for more details.
Below, we focus on introducing the network design for $Dec_{r2o}$.

The recursive decoder $Dec_{r2o}$ takes the room node feature as input and infers the node features of its children functional regions until reaching the object level.
For each decoding step, we assume there are 10 children nodes at most for each parent and learn to predict the probability of node existence likelihood scores using a binary classification network (implemented by an MLP with a final Sigmoid activation function). 
We also predict the children's node semantic information as outputs.
For the object nodes, we train an MLP to predict the placement parameters, categorically semantic labels, and object features, which will then be fed to the $Dec_{o2p}$ for decoding the part hierarchy.
The placement parameter prediction network consists of two fully-connected layers, a Leaky ReLU activation, and a skip-link.
The network predicts the center $c_{O_i} \in \mathbb{R}^3$, scales $s_{O_i} \in \mathbb{R}^3$, and the orientation $r_{O_i} \in \mathbb{R}^1$ around the world up-axis by three individual fully-connected layers respectively.
\yjr{Hence, according to the extracted %
node features, we firstly predict node existence, semantics and geometry. For  object nodes, the placement parameters are also predicted. Then we predict the edges between existing nodes according to their geometry and node features, and the edge type will be compared to the ground truth (GT) edge types for optimizing the hyper-parameters of the network during the recursive decoding for training.
}

For the binary edge predictions, we draw all pairs of existing nodes and predict the edge existence for every edge.
For the hyper-edges, we predict a mask $\mathbf{M}_i$ by an attention mechanism to obtain which nodes share a hyper-edge attribute.
Leveraging the predicted edge connections among nodes, we perform two iterations of message-passing for updating the node features.
Finally, we obtain the node features $\left\{f_{N_{j_1}}, f_{N_{j_2}}, \cdots, f_{N_{k_i}}\right\}$, where $k_i$ denotes the number of existing nodes for parent node $N_i$. 
Above all, we have
\begin{equation}
    \left\{\hat{f}_{N_{j_1}}, \hat{f}_{N_{j_2}}, \cdots, \hat{f}_{N_{j_{k_i}}}, \hat{\mathbf{R}}_i, \mathbf{M}_i\right\} = Dec_{r2o}(\hat{f}_{i})
\end{equation}
where \yjr{$M_i \in \mathbb{R}^{N \times 3}$ is a matrix. Here, $N$ is the number of nodes of the sub-graph for the parent node, and 3 is the number of hyper-edge types. For each row, it predicts the probabilities of different hyper-edge types, and the predicted type is the one with the highest probability (selected by `argmax' operation). 
According to our defined hyper-edges, each object only has one type of hyper-edges, such as none, N-fold Rotation Symmetry, and parallel collinearity.}

The recursive decode process of $Dec_{r2o}$ is repeated until it reaches the object level, which is then the job of $Dec_{o2p}$ to further decode it into object parts.

\subsection{Training and Losses}
We describe our training strategy and loss terms as follows.

\subsubsection{Training Strategy}
Since our scene hierarchy is a very deep tree from the room root node to functional regions, objects, and finally to object parts, it is very difficult to train it effectively together from scratch. 
We thus choose to train the network in two stages. 
We first train the recursive network from object to part geometry and then train the whole network while fine-tuning the pre-trained object-to-part network.
For the floor boundary VAE, we train it separately from our backbone network.
We conduct an ablation study in Sec.~\ref{sec:abla} for evaluating the benefit of such a training strategy.

\subsubsection{Loss Terms}
We define the total training loss $\mathcal{L}$ as the following:
\yjr{
\begin{equation}
    \mathcal{L} = \mathbb{E}_{S \sim \mathfrak{S}}\left[\mathcal{L}_{recon} + \mathcal{L}_{struc} + \gamma\mathcal{L}_{KL}\right]%
\end{equation}%
}%
where \yjr{$\mathfrak{S}$ is the distribution of scenes in the whole dataset}, and the reconstruction loss $\mathcal{L}_{recon}$ includes leaf loss, semantic loss, edges/node existence loss, geometry loss, placement loss, and some edge losses (\eg room-object edge, object-object edge, hyper-edge, and part-part edge). Except for placement loss and edge losses for the room-object edges and the proposed hyper-edges, the other loss terms are following the StructureNet~\cite{mo2019partnet}.
We refer the readers to StructureNet~\cite{mo2019partnet} for these losses.
\yjr{For the structure consistency loss, it is used in StructureNet to ensure the generated structures of objects are reasonable and realistic. However, 
different from StructureNet, we only add the loss on the object hierarchy instead of the whole scene hierarchy.}
And the regularization $\mathcal{L}_{KL}$ aims to make the latent space smoother and easier for downstream applications (scene generation and interpolation). We set $\gamma=0.01$ empirically for our experiments.

We now define placement parameter reconstruction loss and edge losses.

\textbf{Placement Parameter Reconstruction Loss.} 
The center $c_{O_i} \in \mathbb{R}^3$, scales $s_{O_i} \in \mathbb{R}^3$, and the orientation $r_{O_i} \in \mathbb{R}$ around the world up-axis are used to represent each object's location in the indoor scene.
We apply the L2-Loss to the center and scale for encouraging \yjr{the perfect reconstruction by $\mathcal{L}_{locate} = d_{center} + d_{scale} + d_{orient}$, where $d$ is  L2 distance metric, $d_{center} = \rVert c_{obj} - \hat{c}_{obj}\rVert^2_2$, $d_{scale} = \rVert s_{obj} - \hat{s}_{obj}\rVert^2_2$. $\hat{c}_{obj}$ and $\hat{s}_{obj}$ denote the center and scale of object location.}
\yj{
Besides, we observe that furniture is typically located on the floor or ceiling, and the furniture shapes in the room 
are usually in 8 orientations with $45^\circ$ intervals ($Angle = [0^\circ,45^\circ,90^\circ,135^\circ,180^\circ,-45^\circ,-90^\circ,-135^\circ]$) in most cases.
So, we use a discrete representation to encode the orientation of furniture for the prediction of coarse orientation from the candidates above, along with a residual offset to fit the ground truth orientation of the object.}
In summary, we have $d_{orient} = \rVert Angle[ \argmax\limits_k (\rho_k)] + b - o_{obj}\rVert^2_2$,
\yj{where $\rho = (\rho_1,\rho_2,\cdots,\rho_8)$ is a predicted 8-d vector whose entity is the probability of the furniture at every orientation, $b$ is the predicted offset in the range of $[-22.5^\circ,22.5^\circ]$, and $o_{obj}$ is the ground truth orientation of the furniture.}

\textbf{Room-object Binary Edge Loss.}
The main purpose of this loss term is to align the position of a predicted object with the boundary walls of the room as much as possible.
In the original indoor scene data, the room boundary is mostly well aligned with the world $x$-axis and $z$-axis, if we denote the world up-axis as the $y$-axis. 
\yjr{Under this assumption, we encourage the oriented bounding box of object to align with the $x$-axis and $z$-axis. The loss term is only applied to the objects with the attribute (room-objects edge) detected during decoding.}
We add a loss to approximate the distance between normals of room box and object box: $\mathcal{L}_{Ro} = \sum_{\forall O_i \in \mathbf{S}} d_{chs}(T(q_i)\mathbf{N}, \mathbf{N})$, where $\mathbf{S}$ is a set of all predicted objects, 
$d_{chs} = \frac{1}{\left|A_{i}\right|} \sum_{x_{i} \in A_{i}} \min\limits_{x_{j} \in A_{j}}\left\|x_{i}-x_{j}\right\|_{2}^{2}+\frac{1}{\left|A_{j}\right|} \sum_{x_{j} \in A_{j}} \min\limits_{x_{i} \in A_{i}}\left\|x_{j}-x_{i}\right\|_{2}^{2}$ is Chamfer Distance~\cite{fan2017point,barrow1977parametric}, 
and $\mathbf{N}$ is the six unit normal vectors of a unit box, $T(q_i)$ is a transformation matrix rotating the normals to align with orientation $q_i$ of the object box.

\textbf{Hyper-edge Loss.}
These loss terms encourage multiple objects to preserve their $n$-ary hyper-edge relationships. We include two types of losses here corresponding to the two types of hyper-edges.

For objects $\textbf{O} = \{O_1,\cdots,O_N\}$ satisfying the $n$-fold rotational symmetry hyper-edge, we can define the loss:
\begin{equation}
    \mathcal{L}_{\mathbf{e}^{hyper}_1} = \sum\limits_{i = 1}^{N}\left\{\min\limits_{j = 1,\cdots, N,i \neq j}d_{chs}\left[O_i, Rot\left(p,\frac{2\pi}{N}\right) \times O_j\right]\right\}
\end{equation}
where $p = \frac{1}{N}\sum_{i = 1}^{N}C_i$ is the barycenter of all object centers, $Rot\left(p,\theta\right)$ indicates a rotation matrix that rotates an object around an axis that is parallel to the y-axis and passes through point $p$ by $\theta$ in radians. Because the $n$-fold symmetry in \autoref{eqn:hyperedgerotate1} is ordered while our graph decoder is not, we must traverse through all decoded objects and find the one with the minimal Chamfer Distance.

For the collinearly parallel hyper-edges, we define two loss functions. The first one is
\begin{equation}
\mathcal{L}_{hpara_1} = \sum\limits_{i = 1}^{N}\left\{\sum\limits_{j = i+1}^{N}\left[d_{chs}\left(T(q_i)\mathbf{N},T(q_j)\mathbf{N}\right)\right]\right\}
\end{equation}
where $\mathbf{N}$ is the six unit normals of the unit box, $T(q_i)$ is a transformation matrix that rotates the normals to align with orientation $q_i$ of the object box. This loss corresponds to the first condition of a parallel hyper-edge. 
And the second loss is 
\begin{equation}
\mathcal{L}_{hpara_2} = \sum\limits_{i = 1}^{N}\left( dist(C_i,\mathbf{v},p)  \right)
\end{equation}
where $C_i$ is the center of the OBB of $O_i$,  
$p = \frac{1}{N}\sum_{i = 1}^{N}C_i$ is the barycenter of all centers, 
$\mathbf{v} = norm\left(\sum_{i = 1}^{N}\sum_{j = i+1}^{N} C_j - C_i\right)$ is the average of relative position for all pairs of object centers in the hyper-edge, $norm$ indicates normalizing a certain vector. Note that when calculating $\mathbf{v}$, we sort the object centers according to their x-coordinates in an ascending order and then using their y-coordinates in an ascending order to avoid adding vectors pointing at opposite directions.
Finally, we have the total loss function
\begin{equation}
\mathcal{L}_{\mathbf{e}^{hyper}_2} = \mathcal{L}_{hpara_1} + \mathcal{L}_{hpara_2}
\end{equation}

%% file: sec/6Exps.tex
\section{Experiments and Applications}
\label{sec:exps}
We perform extensive experiments evaluating our \name for 3D scene reconstruction, generation, and interpolation, as well as many other applications, such as scene editing, conditional generation from 3D Box layout, and room completion.
\yjr{Since our framework is a generative model, the scene reconstruction evaluations are presented in our supplementary material.}
We compare GRAINS~\cite{li2019grains} and Deep Priors~\cite{wang2018deep} as two state-of-the-art methods in terms of many quantitative metrics and a \yjr{perceptual study}, demonstrating our superior performance.
Ablation studies further validate some of our key module designs.
\yjr{
We use 3D-FRONT dataset~\cite{fu20203dfront} for our training and evaluation. 3D-FRONT is a newly released dataset of 3D indoor scenes which contains 6,815 houses and 51,708 rooms. The room designs are directly sourced from professional creations. In the dataset, each house is divided into several rooms with a room type associated. Among all the rooms, 18,797 rooms are furnished with objects from 3D-FUTURE~\cite{fu20203dfuture}, a dataset of textured 3D furniture models, and each model is labeled with a furniture category. 
We describe more details of dataset preparation in the supplementary material.}

\subsection{Scene Generation}

The generation of 3D indoor scenes is the first and most straightforward application of our network. Our method can do free generation, but in reality, most of the generation tasks require some sort of condition as input (\eg the boundary of a room). So we decide to take a floor boundary (which completely decides the walls) as a condition and input. The generated results are rooms filled with furniture. We show the quantitative and qualitative comparison with GRAINS~\cite{li2019grains}, Deep Priors~\cite{wang2018deep} and ATISS~\cite{paschalidou2021atiss}. Also, the perceptual study is performed for the generative models.

\textbf{Comparison with GRAINS.}
The settings of GRAINS are slightly different from ours. Firstly, GRAINS does not take any condition as input. Secondly, GRAINS uses four walls as `anchors', which it encodes in its hierarchical representation. All the objects in the room need the walls to locate themselves. So GRAINS can only generate rooms with the same shape (in particular, rectangular rooms). But in our settings, the room shape can be arbitrary. 

To reasonably compare the performance of GRAINS with ours, we first use GRAINS to generate 1,000 results for each room type. For an input room boundary and room type, we render the boundary into an image and find its inscribed rectangle with maximum area. With this rectangle, we can find the most similar room shape in the generated results and place the generated furniture into the corresponding rectangle in the input room boundary. This produces the `conditional' generation result. 

GRAINS can only predict the bounding boxes of the objects. With those boxes, it retrieves models from a database (\eg 3D-FRONT, SUNCG) that match the bounding boxes the best. But our method can generate the geometry along with the room layout. In order to compare, we first create a model database using part VAE and graph encoder/decoder in our network. This portion of our network is actually a VAE itself and can perform free generation. We generate 200 objects for all categories for GRAINS to retrieve.

As for training, the original GRAINS VAE uses the SUNCG dataset~\cite{Song2017SemanticSC}. Unfortunately, the SUNCG dataset is unavailable when we do the experiments. So we process the 3D-FRONT dataset and export it in the format of the SUNCG dataset, and train GRAINS on this dataset.

\begin{figure}[t]
    \centering
    \includegraphics[width=0.13\linewidth]{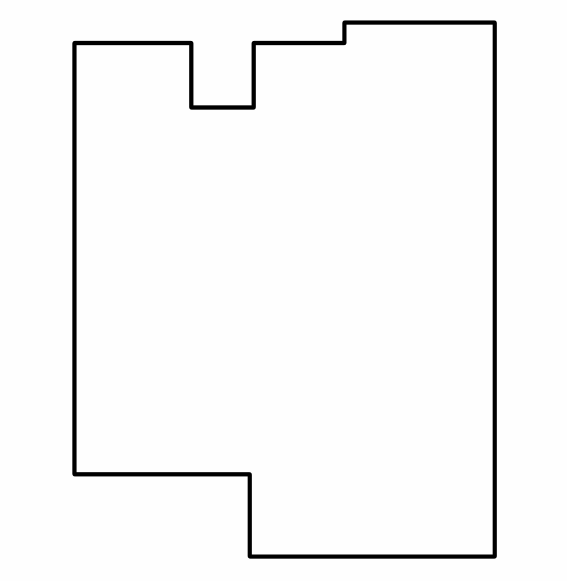}
    \includegraphics[width=0.13\linewidth]{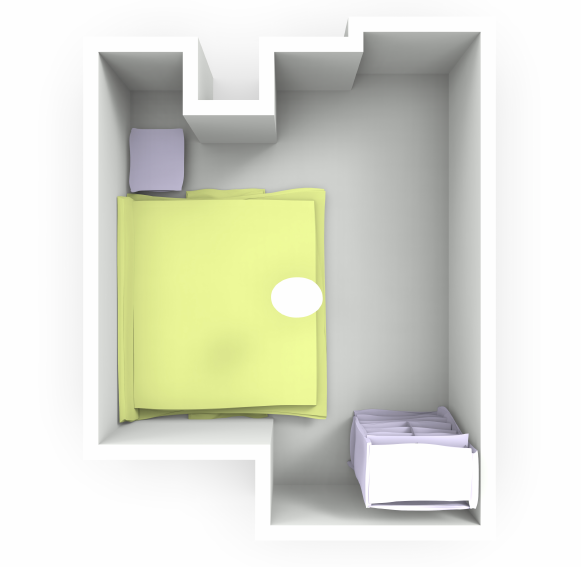}
    \includegraphics[width=0.13\linewidth]{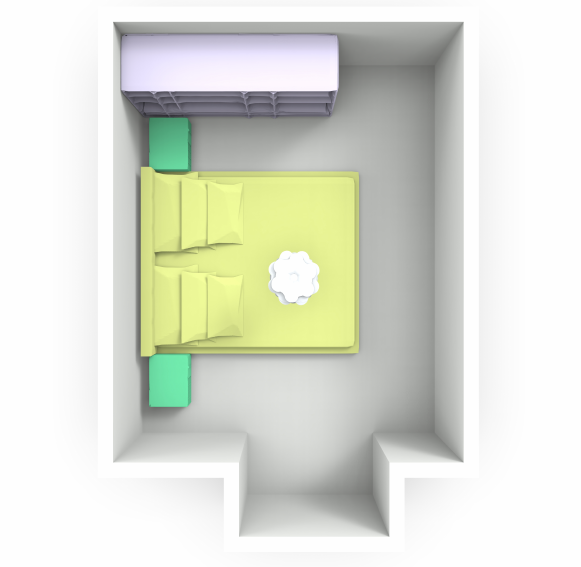}
    \includegraphics[width=0.13\linewidth]{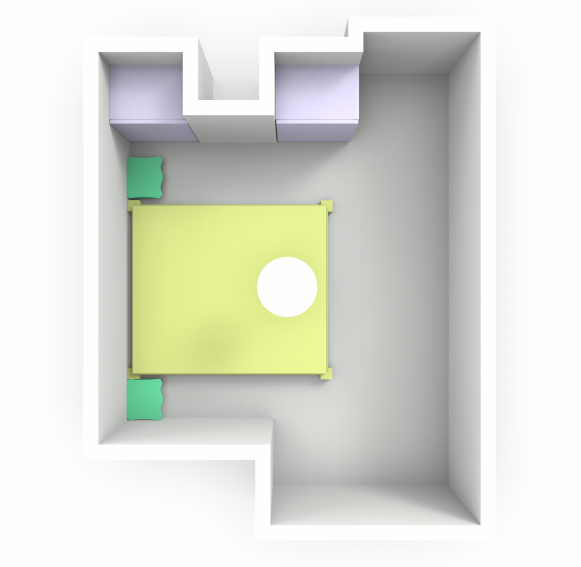}
    \includegraphics[width=0.13\linewidth]{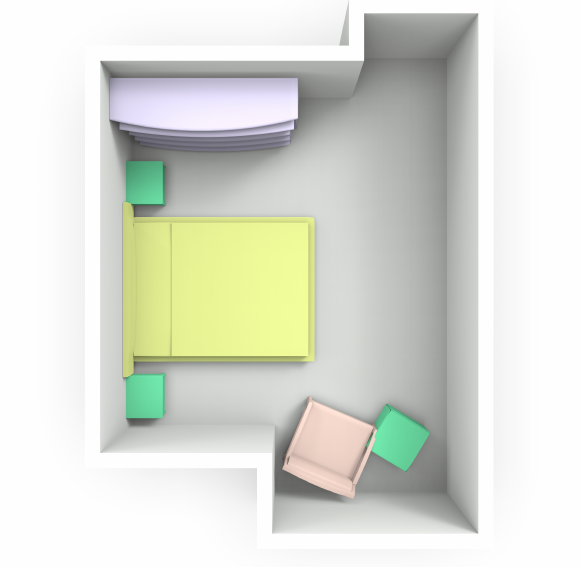}
    \includegraphics[width=0.13\linewidth]{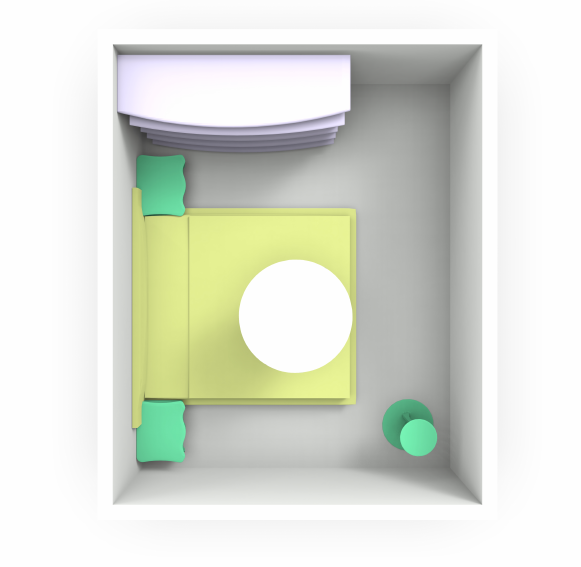}
    \includegraphics[width=0.13\linewidth]{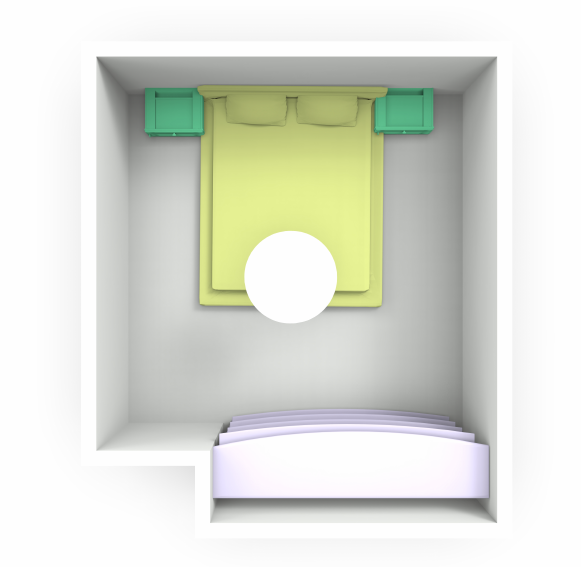}
    \\
    \includegraphics[width=0.13\linewidth]{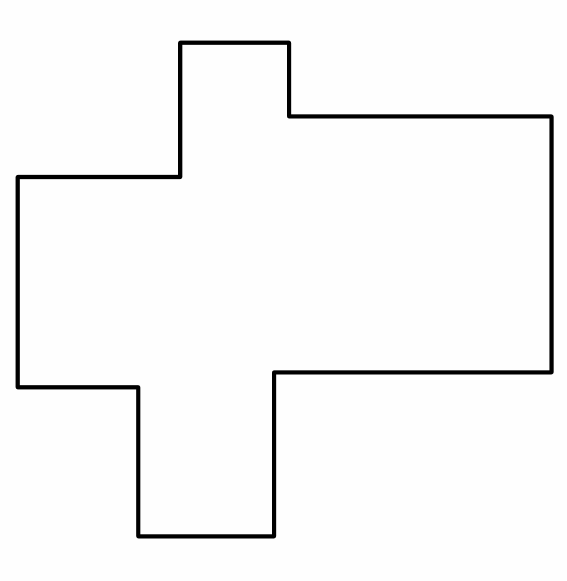}
    \includegraphics[width=0.13\linewidth]{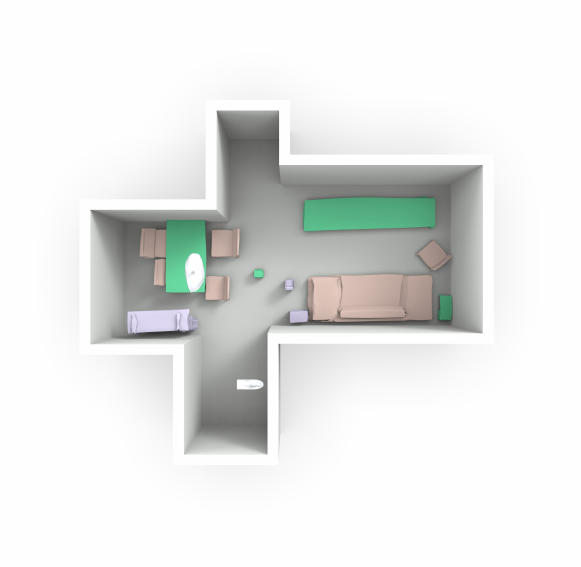}
    \includegraphics[width=0.13\linewidth]{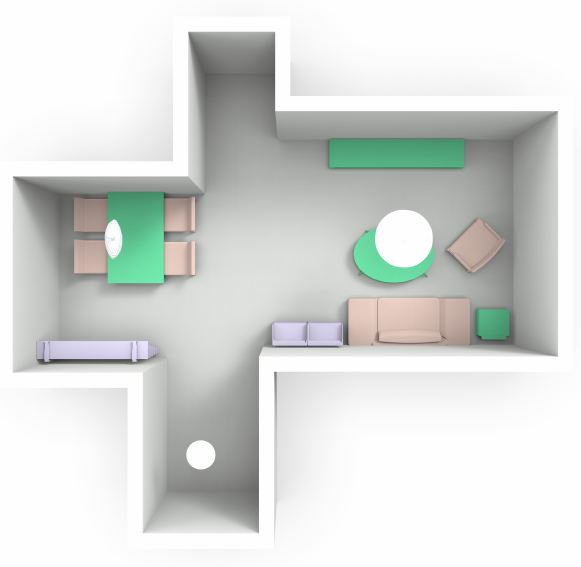}
    \includegraphics[width=0.13\linewidth]{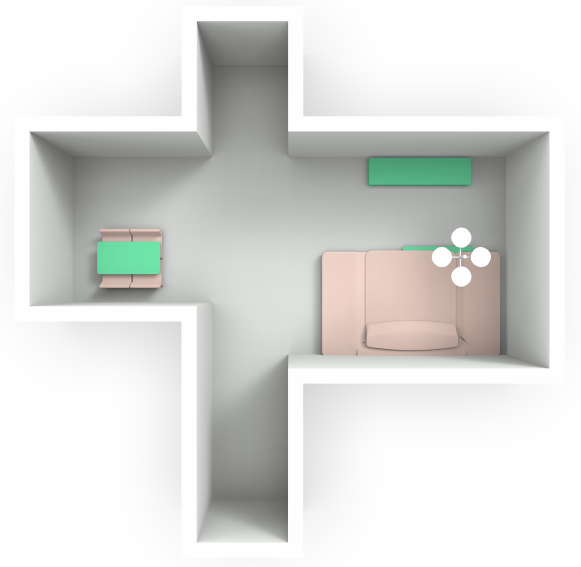}
    \includegraphics[width=0.13\linewidth]{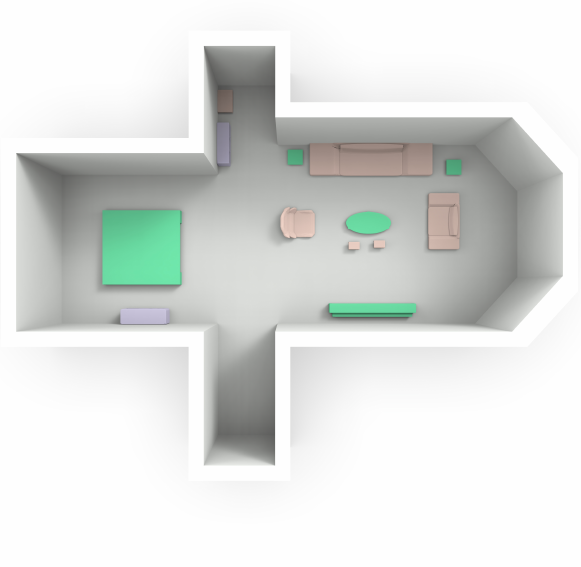}
    \includegraphics[width=0.13\linewidth]{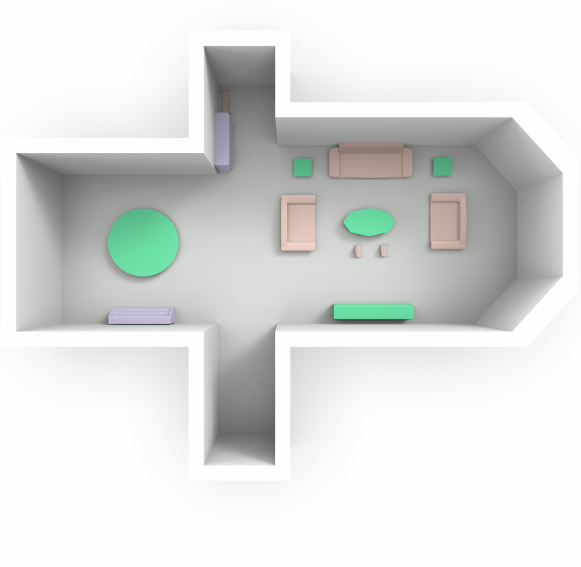}
    \includegraphics[width=0.13\linewidth]{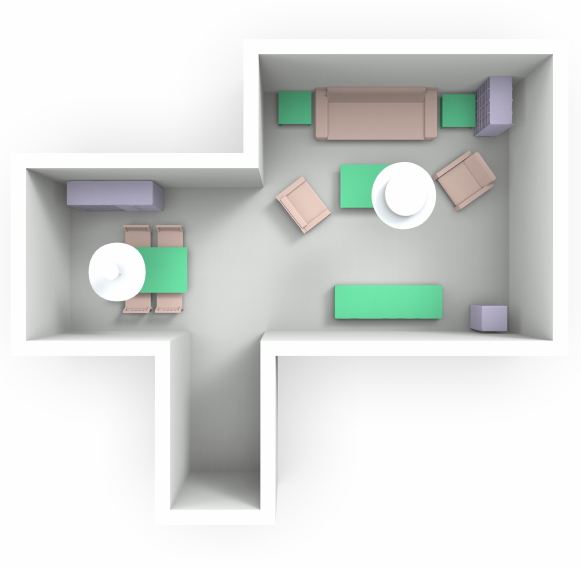}
    \\
    \includegraphics[width=0.13\linewidth]{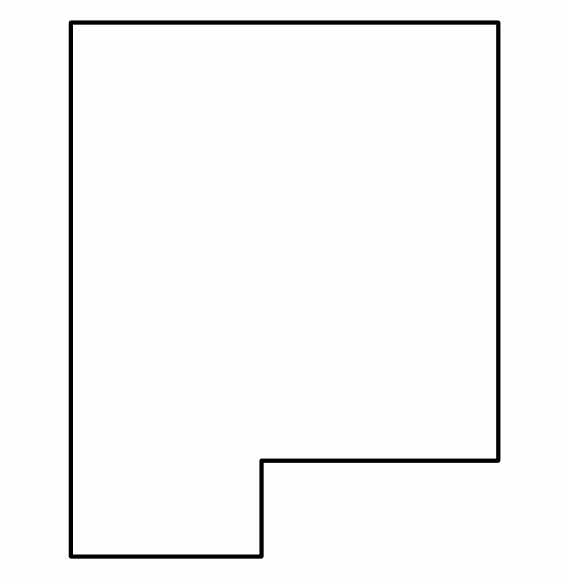}
    \includegraphics[width=0.13\linewidth]{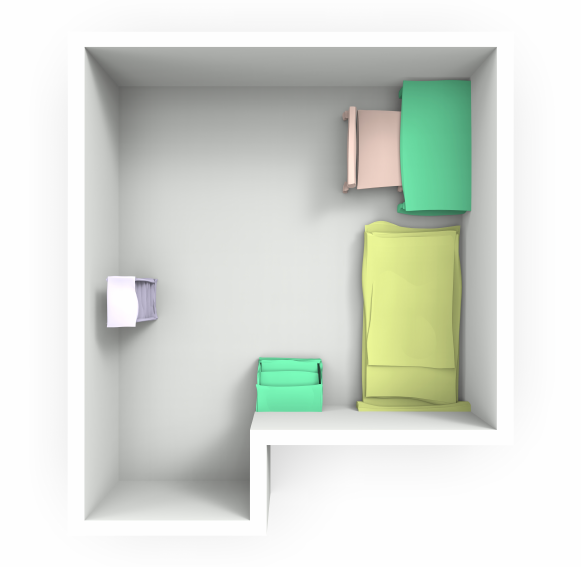}
    \includegraphics[width=0.13\linewidth]{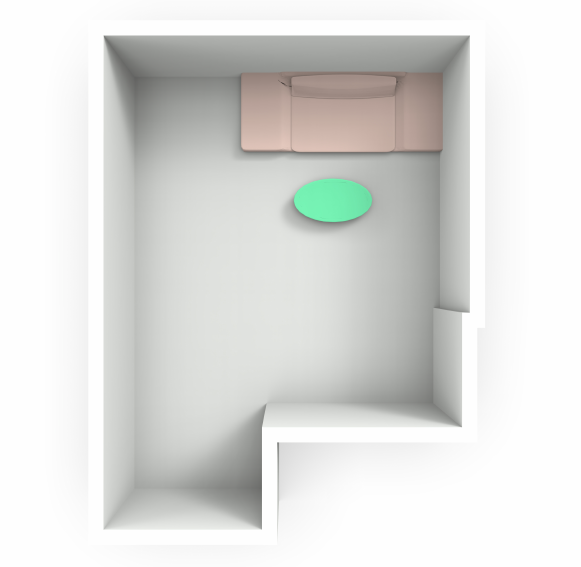}
    \includegraphics[width=0.13\linewidth]{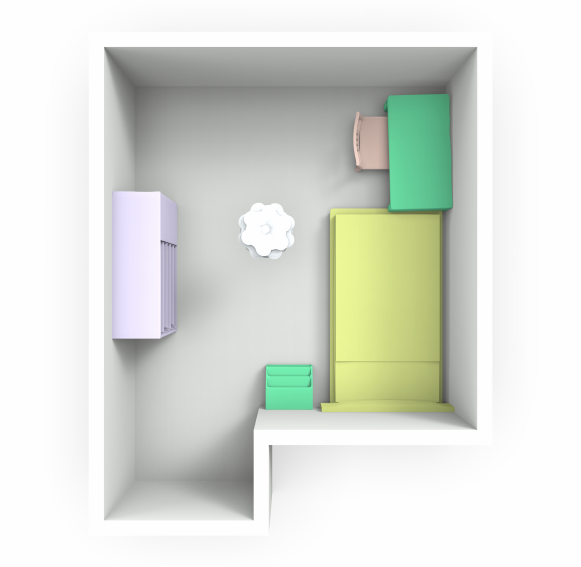}
    \includegraphics[width=0.13\linewidth]{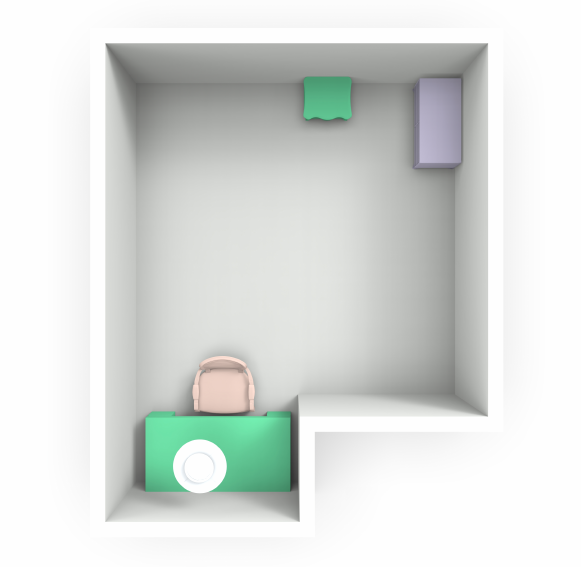}
    \includegraphics[width=0.13\linewidth]{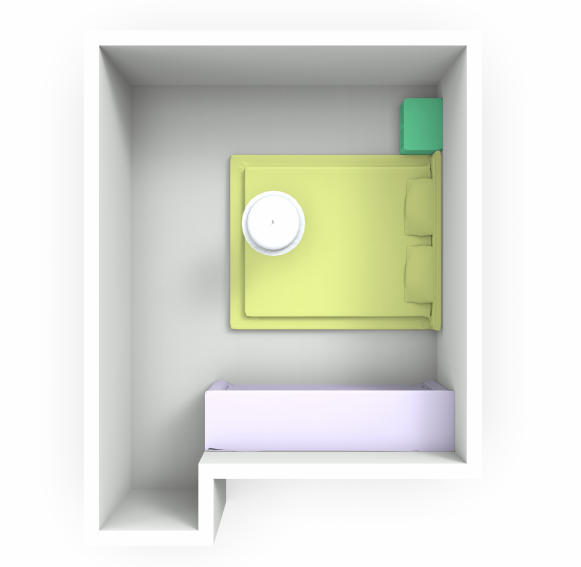}
    \includegraphics[width=0.13\linewidth]{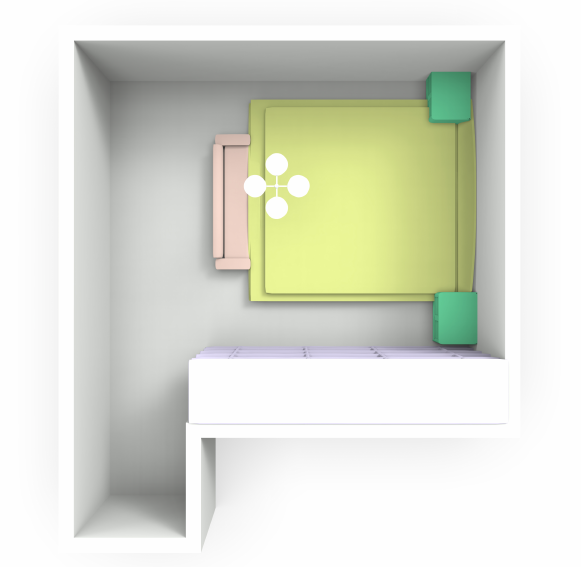}
    \\
    \subfigure[Input]{
    \includegraphics[width=0.125\linewidth]{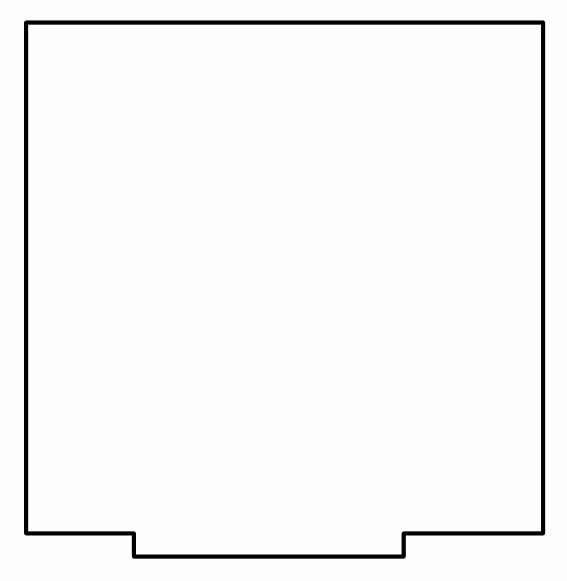}
    }
    \subfigure[Ours]{
    \includegraphics[width=0.125\linewidth]{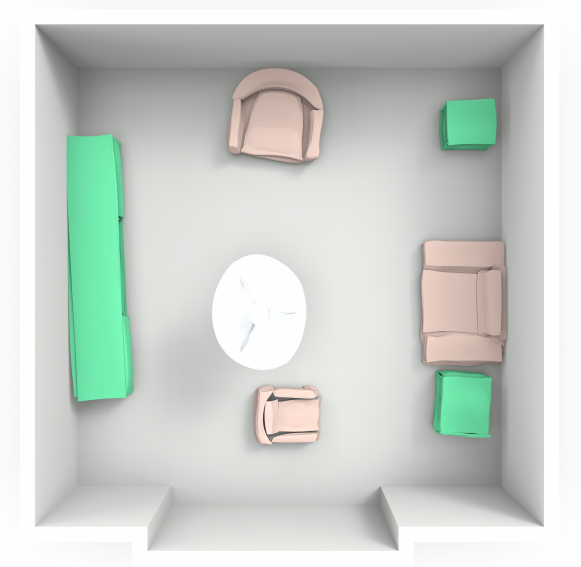}
    }
    \subfigure[Top-5]{
    \includegraphics[width=0.125\linewidth]{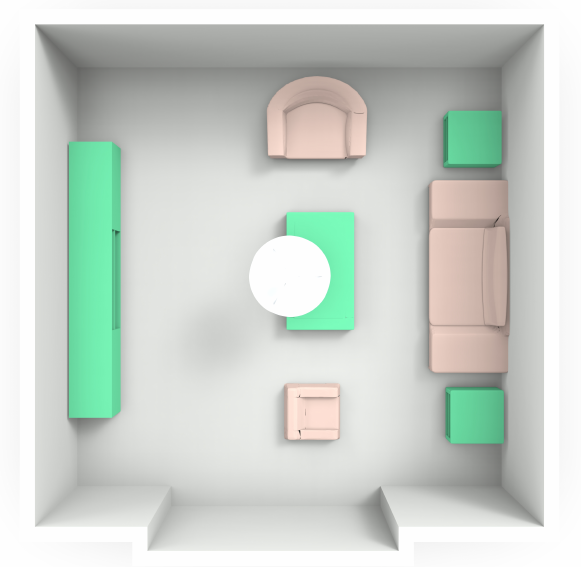}
    \includegraphics[width=0.125\linewidth]{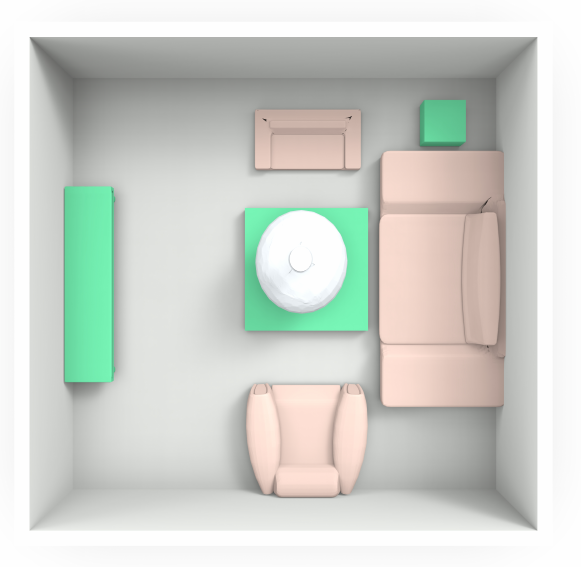}
    \includegraphics[width=0.125\linewidth]{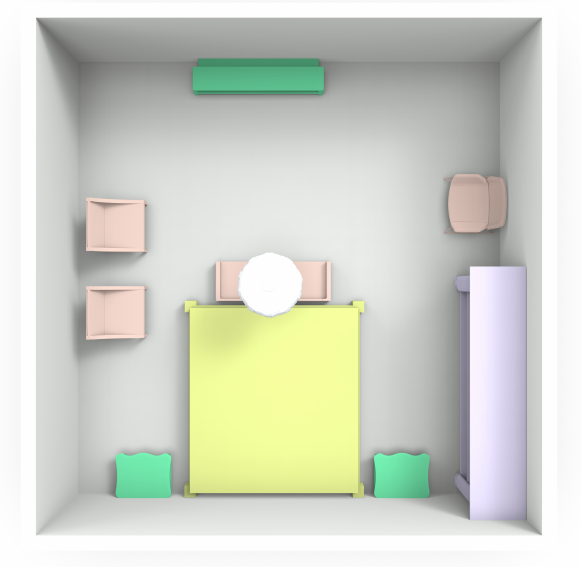}
    \includegraphics[width=0.125\linewidth]{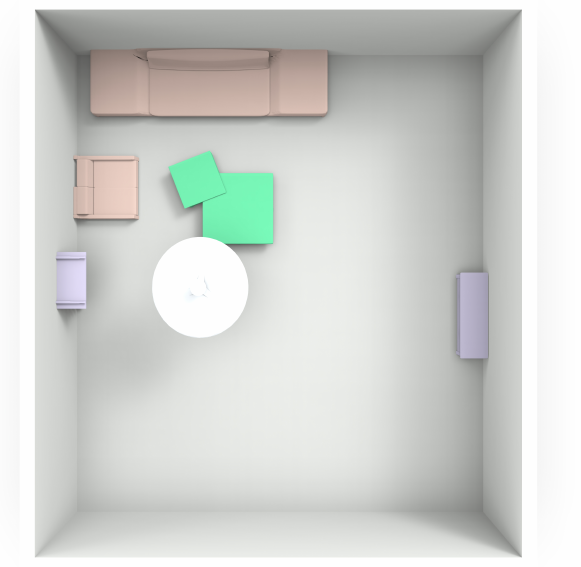}
    \includegraphics[width=0.125\linewidth]{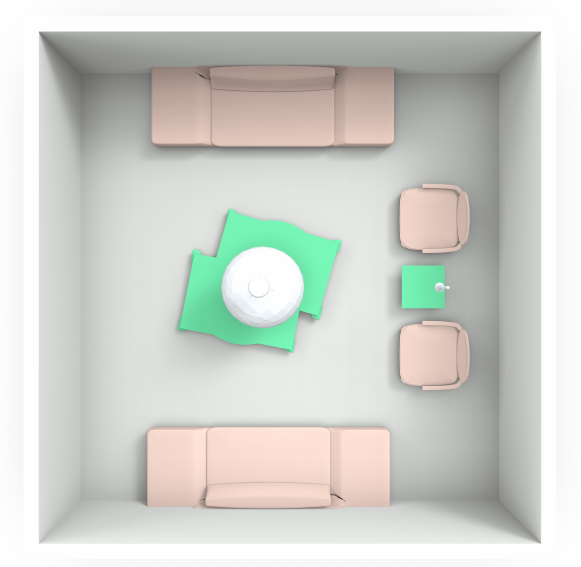}
    }
    \caption{\titlecap{The gallery of shape generation results given the room boundary and top-retrieved rooms in the training set.}{To demonstrate the novelty of room generation, we show the top-5 nearest neighbors in the training set by shape retrieval according to the CD (Chamfer Distance) on the sampled point clouds (with 100,000 points).
    Given the room boundary, we can see that our generated shapes are different from the top-5 retrieved rooms on the object layout and geometric details, which demonstrates the novelty of generated rooms.}
    }
    \label{fig:generation}
\end{figure}

\textbf{Comparison with Deep Priors.}
Deep Priors is a generation pipeline based on top-down view images of the room. Taking a room boundary as input, it iteratively inserts objects into the scene. For each object, it uses multiple networks to decide object location, orientation, and dimension.

Comparison with Deep Priors is straightforward because their settings are almost the same as ours. But there are still some differences. Just like GRAINS, Deep Priors retrieves models from a database. We also use the generated model database for Deep Priors to retrieve models.

The original Deep Priors is trained on the SUNCG dataset which is unavailable, so again the exported 3D-FRONT dataset mentioned above is used instead.

\yjr{
\textbf{Comparison with ATISS.}
ATISS is a transformer-based generative model for a given room boundary. It also iteratively generates the object layouts and retrieves furniture to fill the empty scene. 

ATISS is the state-of-the-art generative model in indoor scene synthesis, which is a very strong baseline in the 3D-Front and the setting is almost the same as ours. But there is also a major difference that is the same as Deep Priors, namely the furniture is retrieved from the original dataset in their method. So for comparison, we use the same generated model database for shape retrieval.
}

\begin{figure}[t]
    \centering
    \includegraphics[width=0.19\linewidth]{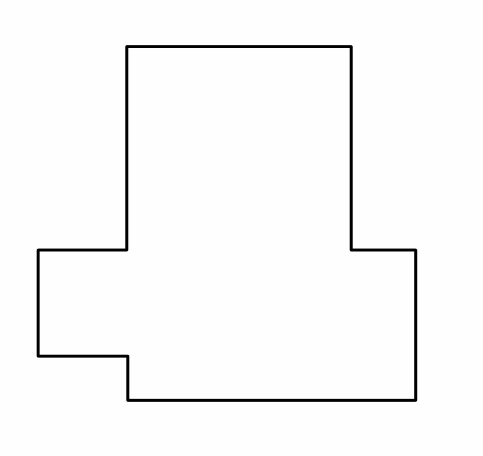}
    \includegraphics[width=0.19\linewidth]{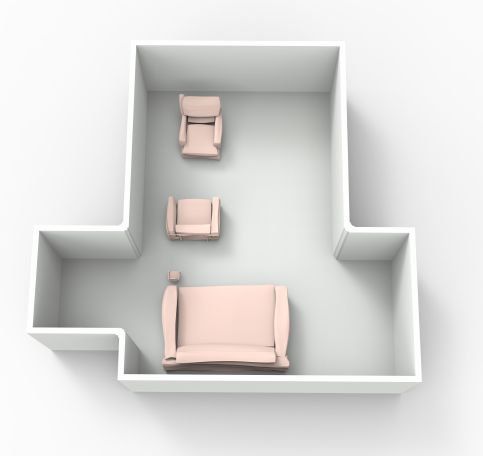}
    \includegraphics[width=0.19\linewidth]{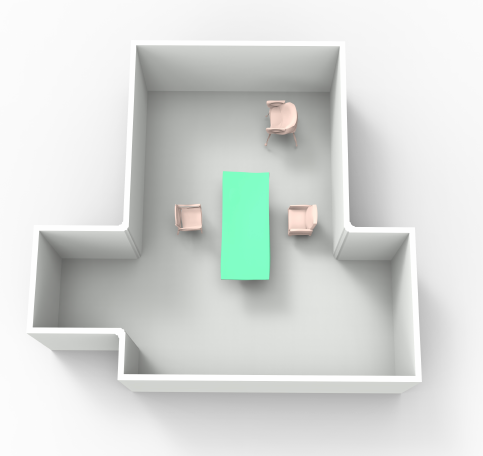}
    \includegraphics[width=0.19\linewidth]{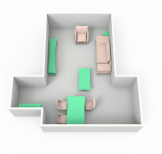}
    \includegraphics[width=0.19\linewidth]{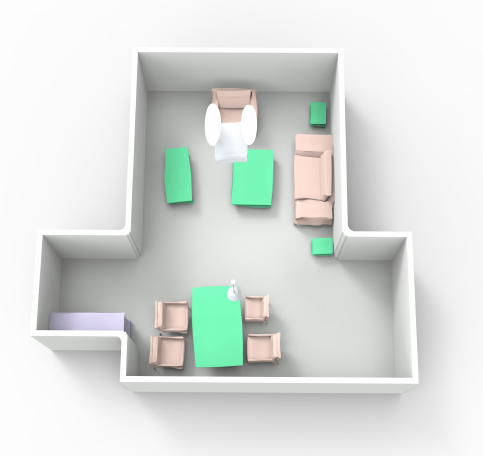}
    \\
    \includegraphics[width=0.19\linewidth]{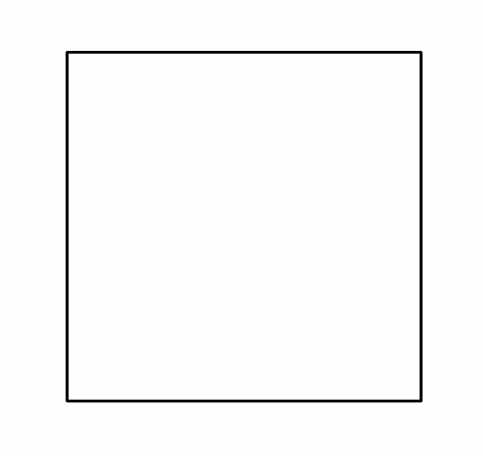}
    \includegraphics[width=0.19\linewidth]{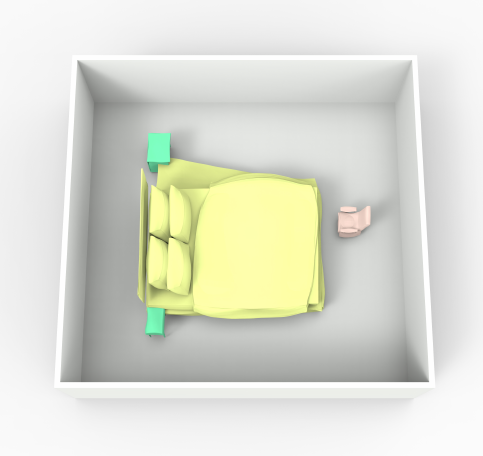}
    \includegraphics[width=0.19\linewidth]{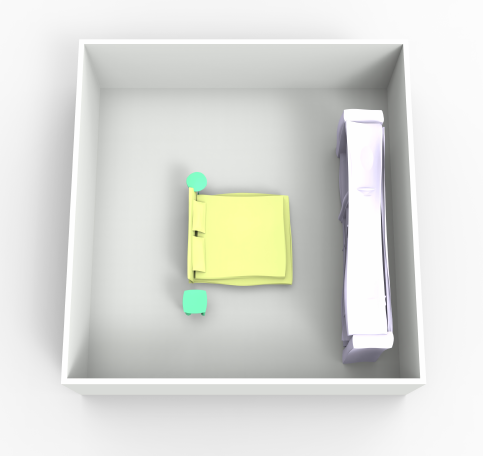}
    \includegraphics[width=0.19\linewidth]{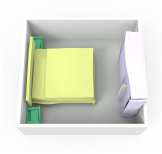}
    \includegraphics[width=0.19\linewidth]{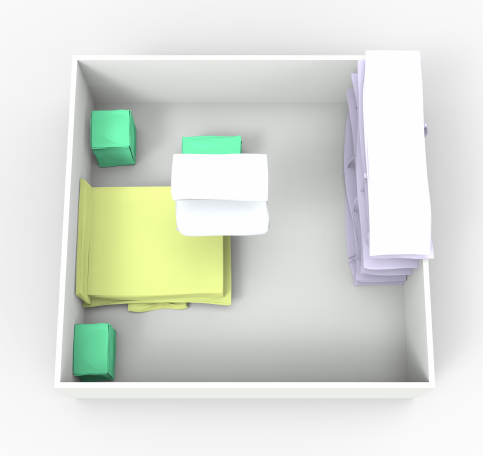}
    \\
    \includegraphics[width=0.19\linewidth]{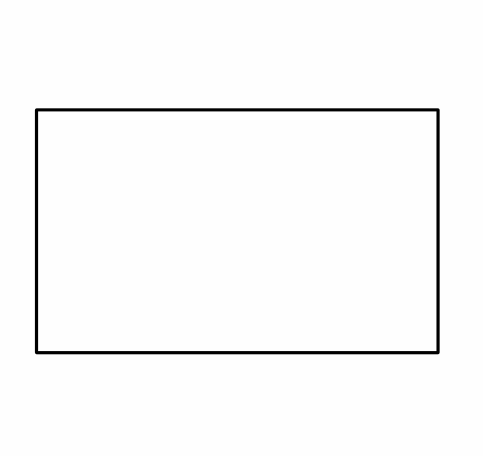}
    \includegraphics[width=0.19\linewidth]{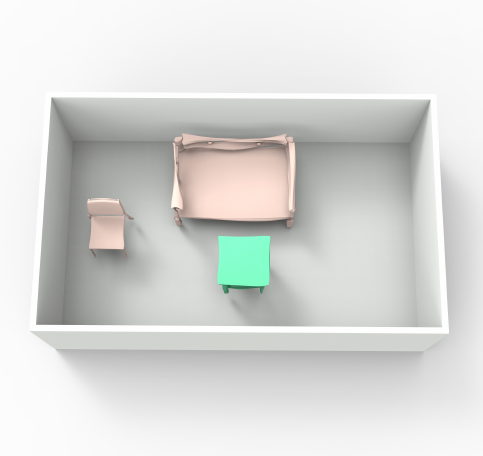}
    \includegraphics[width=0.19\linewidth]{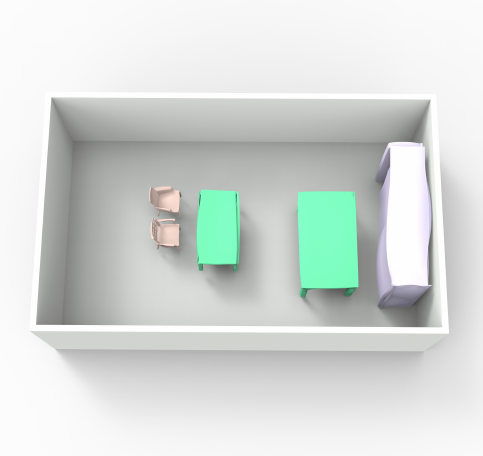}
    \includegraphics[width=0.19\linewidth]{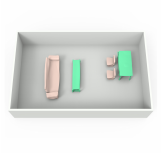}
    \includegraphics[width=0.19\linewidth]{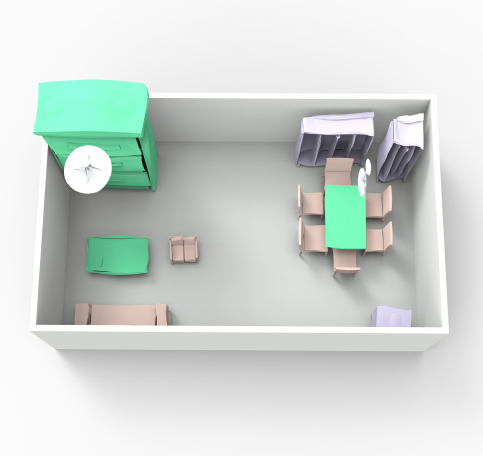}
    \\
    \subfigure[Input]{\includegraphics[width=0.19\linewidth]{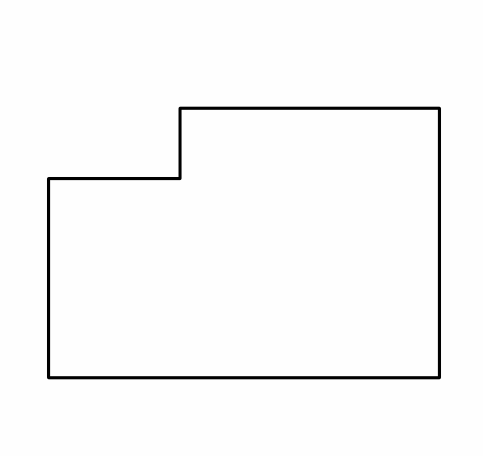}}
    \subfigure[GRAINS]{\includegraphics[width=0.19\linewidth]{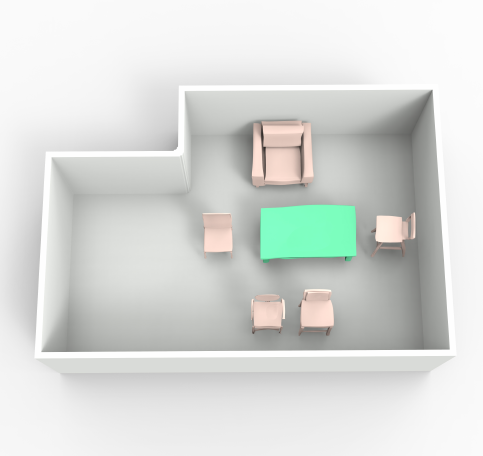}}
    \subfigure[\hspace{-1.5mm}DeepPriors]{\includegraphics[width=0.19\linewidth]{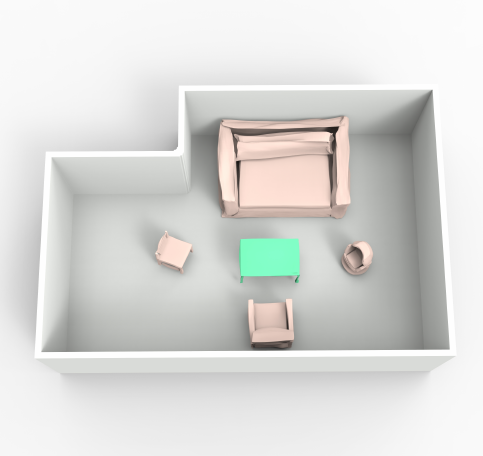}}
    \subfigure[ATISS]{\includegraphics[width=0.19\linewidth]{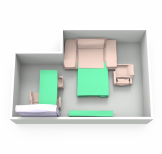}}
    \subfigure[Ours]{\includegraphics[width=0.19\linewidth]{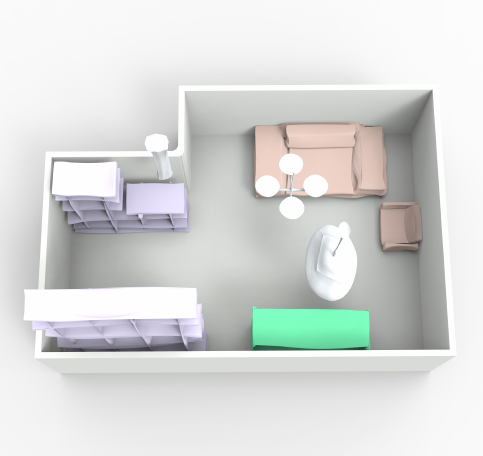}}
    \caption{\titlecap{Comparison on Room Generation.}{\yjr{We show the comparison of generation results of our method, Deep Priors~\cite{wang2018deep}, GRAINS~\cite{li2019grains}, and ATISS~\cite{paschalidou2021atiss}. From the results we can see that our method captures the functional regions (or local gathering of furniture) better. For example, our method can successfully predict four same chairs surrounding a dining table, while the three baseline methods cannot. 
    }}}
    \label{fig:comgeneration}
\end{figure}

\begin{figure}[h]
    \centering
    \includegraphics[width=0.158\linewidth]{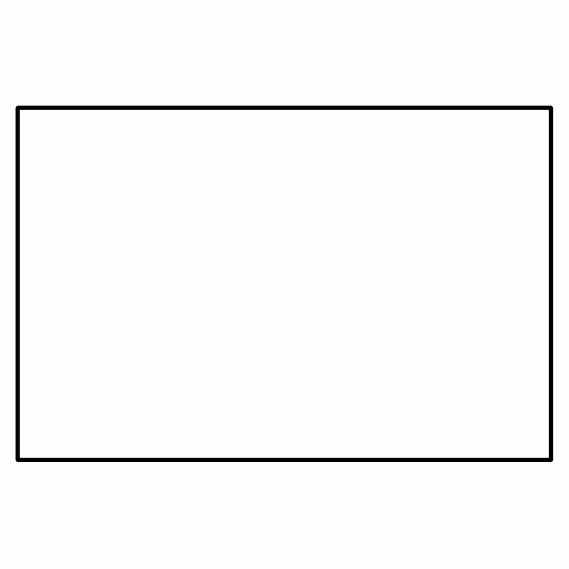}
    \includegraphics[width=0.158\linewidth]{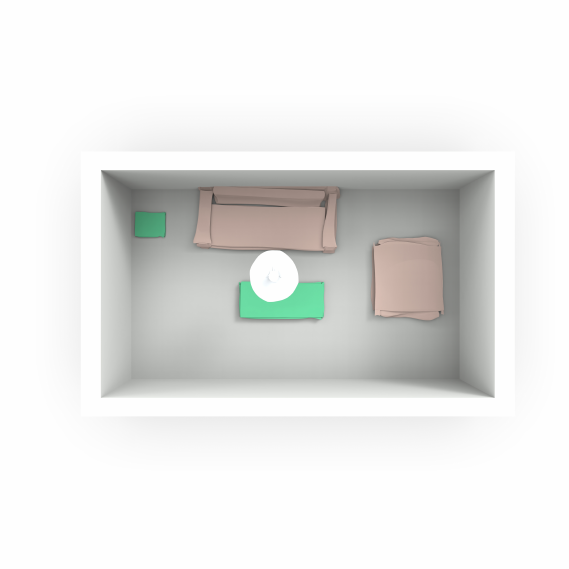}
    \includegraphics[width=0.158\linewidth]{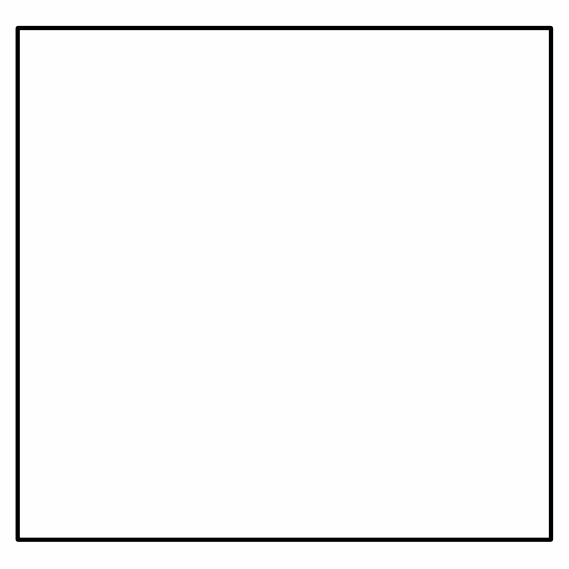}
    \includegraphics[width=0.158\linewidth]{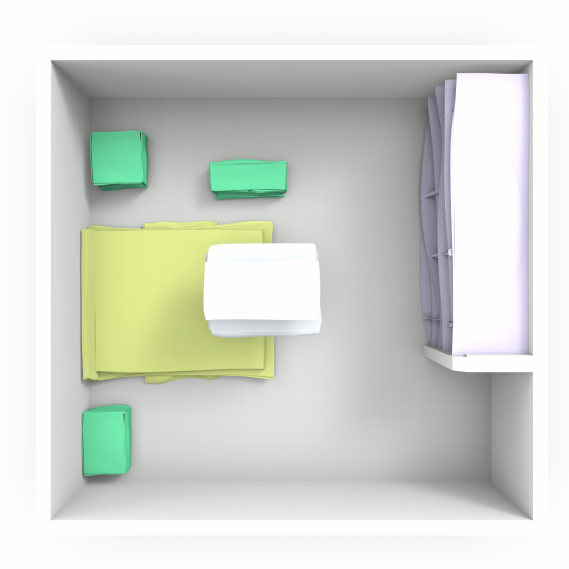}
    \includegraphics[width=0.158\linewidth]{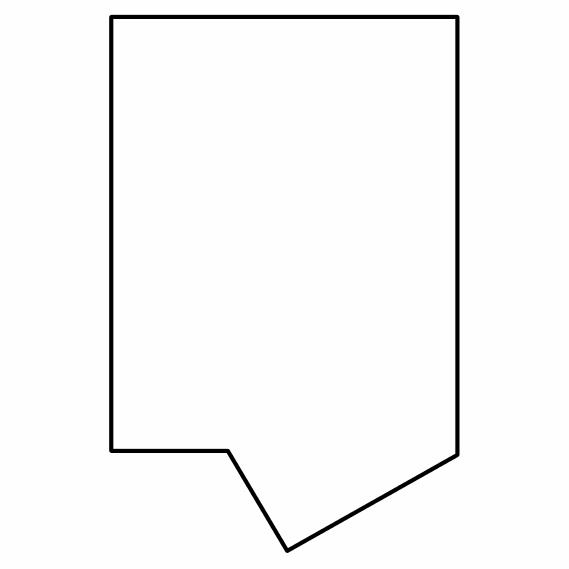}
    \includegraphics[width=0.158\linewidth]{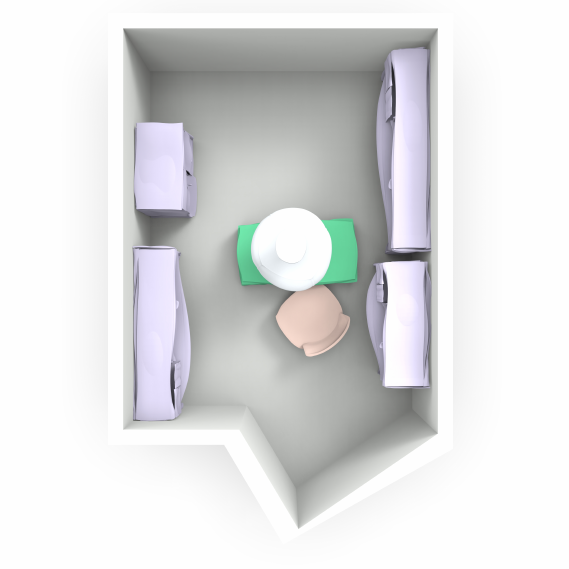}
    \\
    \includegraphics[width=0.158\linewidth]{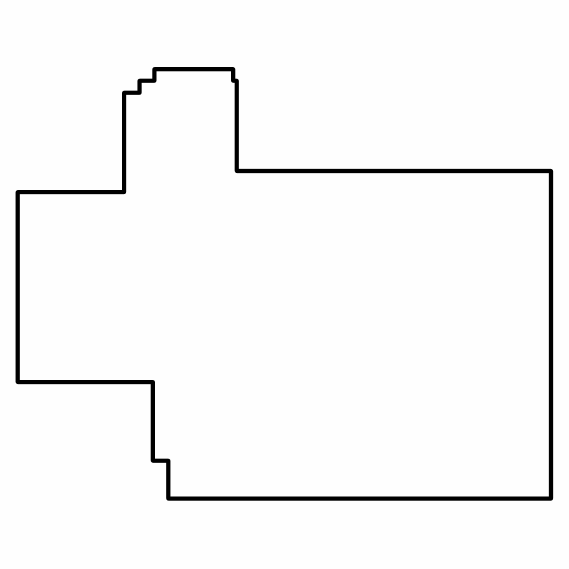}
    \includegraphics[width=0.158\linewidth]{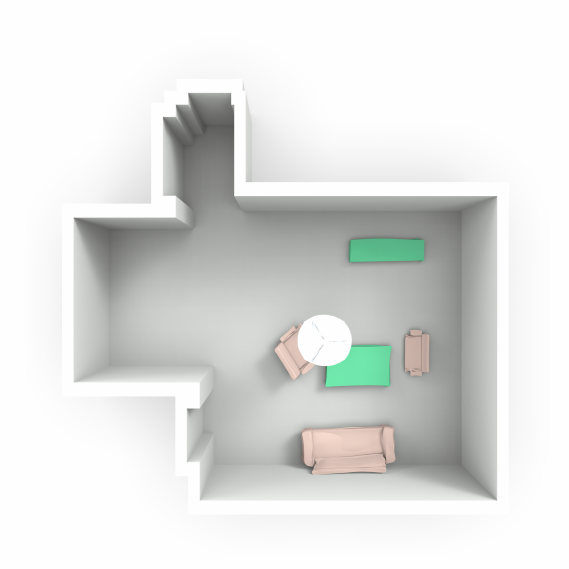}
    \includegraphics[width=0.158\linewidth]{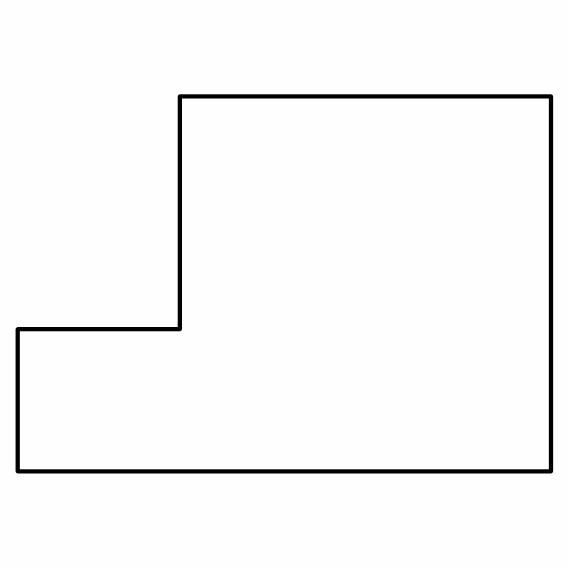}
    \includegraphics[width=0.158\linewidth]{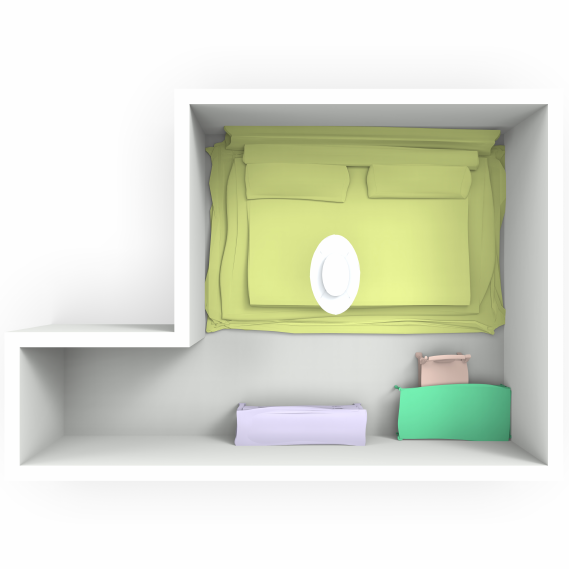}
    \includegraphics[width=0.158\linewidth]{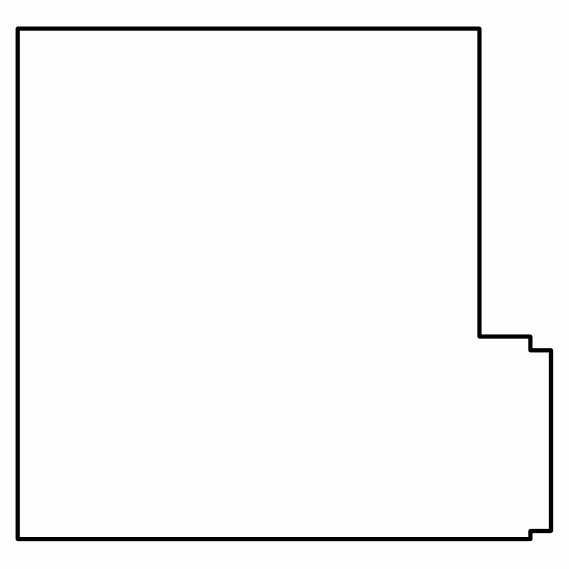}
    \includegraphics[width=0.158\linewidth]{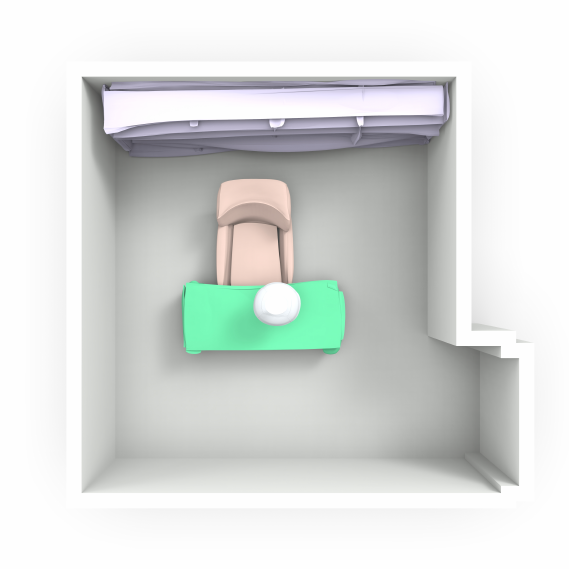}
    \\
    \includegraphics[width=0.158\linewidth]{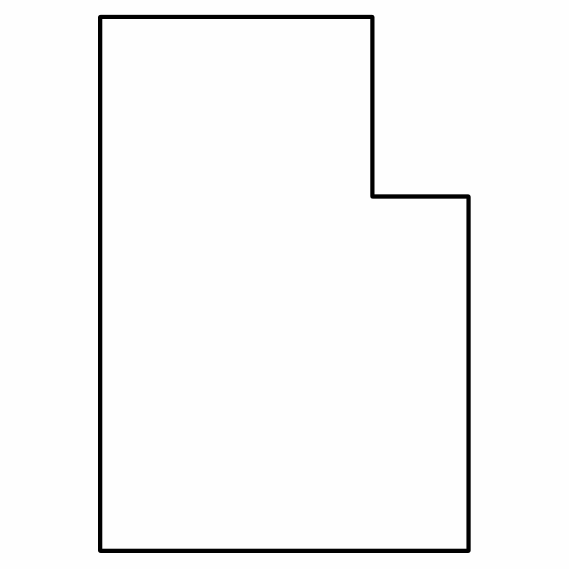}
    \includegraphics[width=0.158\linewidth]{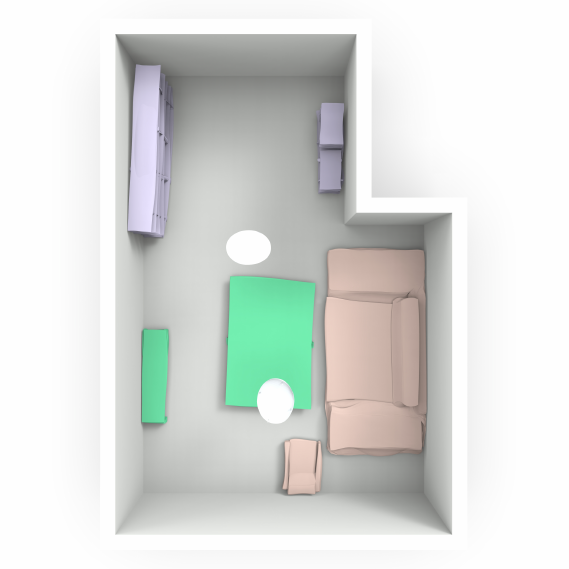}
    \includegraphics[width=0.158\linewidth]{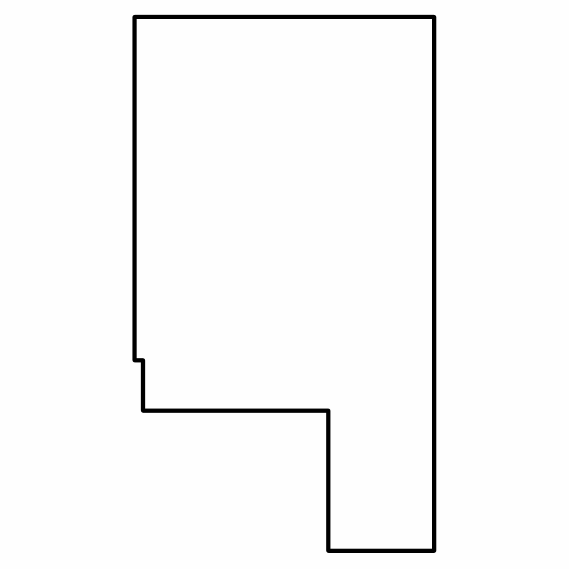}
    \includegraphics[width=0.158\linewidth]{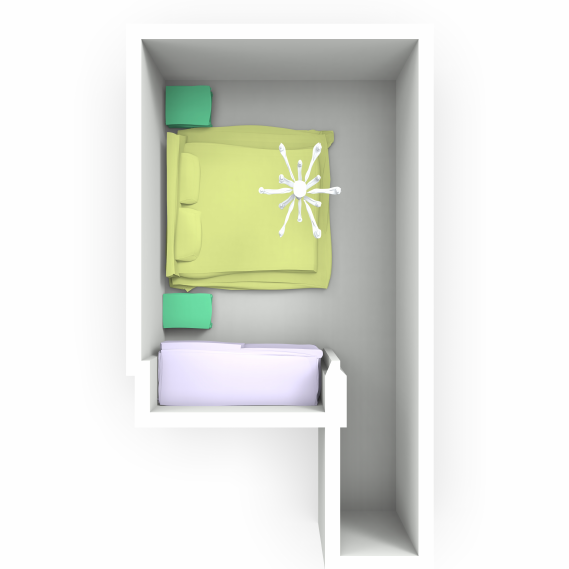}
    \includegraphics[width=0.158\linewidth]{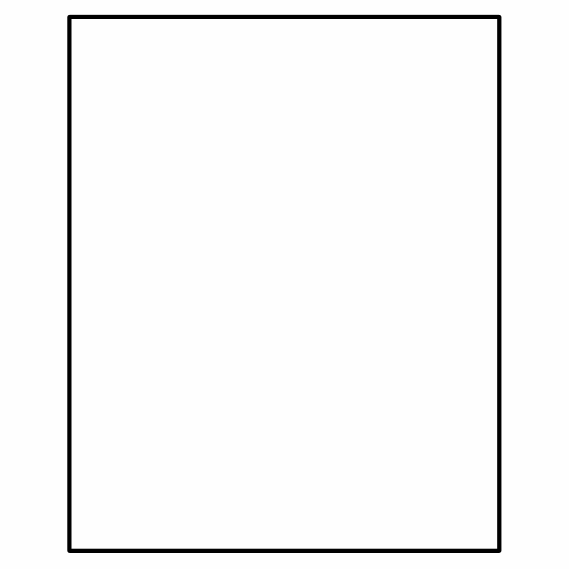}
    \includegraphics[width=0.158\linewidth]{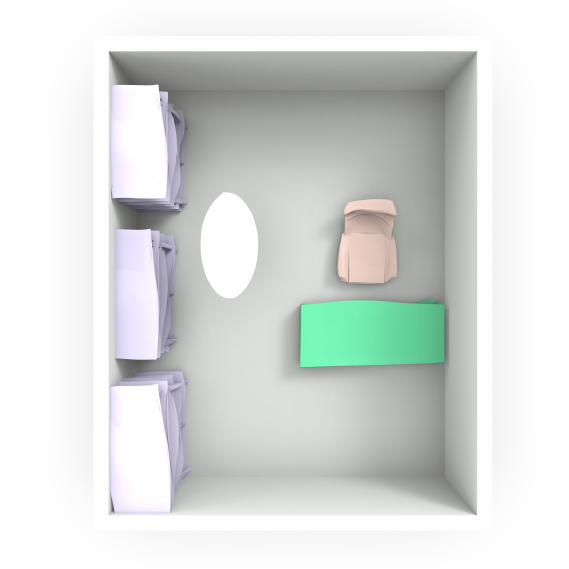}
    \\
    \includegraphics[width=0.158\linewidth]{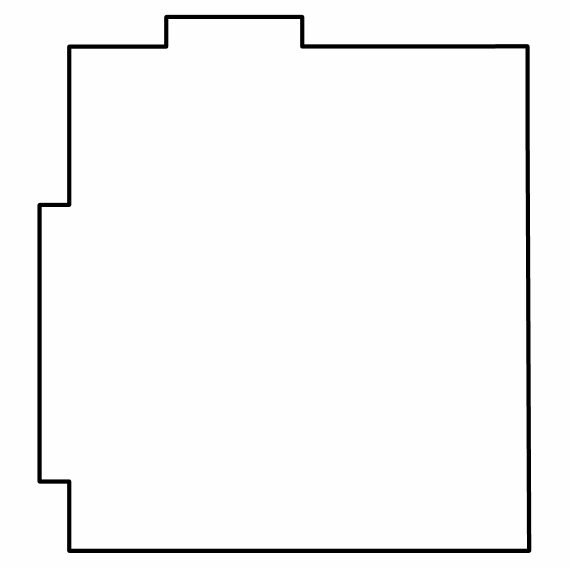}
    \includegraphics[width=0.158\linewidth]{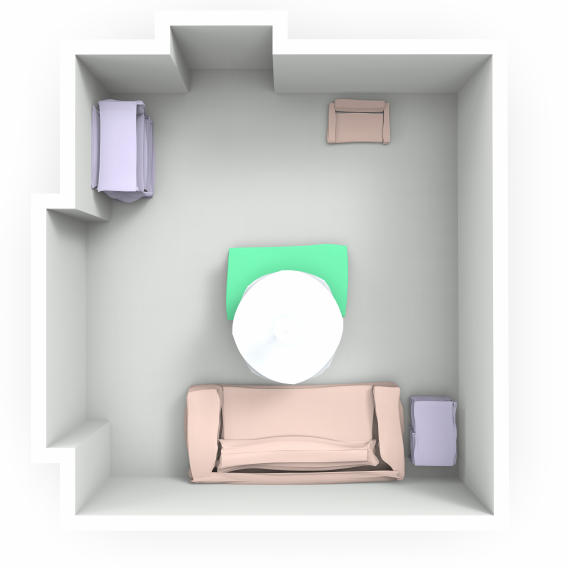}
    \includegraphics[width=0.158\linewidth]{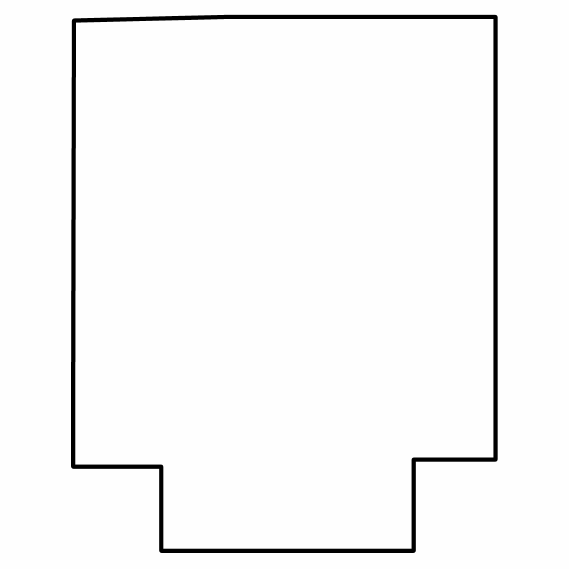}    
    \includegraphics[width=0.158\linewidth]{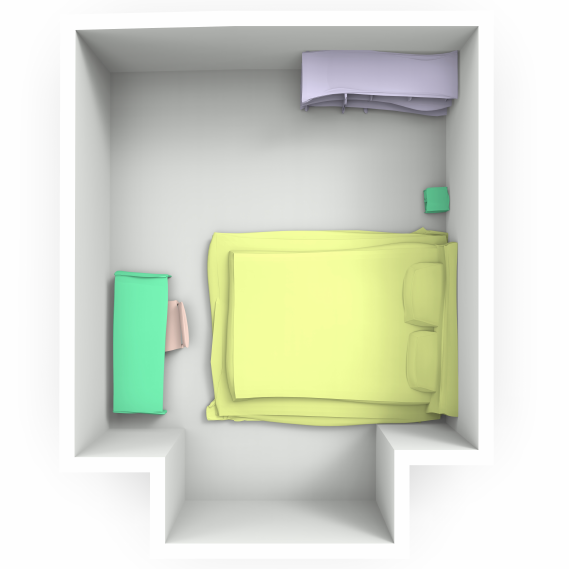}
    \includegraphics[width=0.158\linewidth]{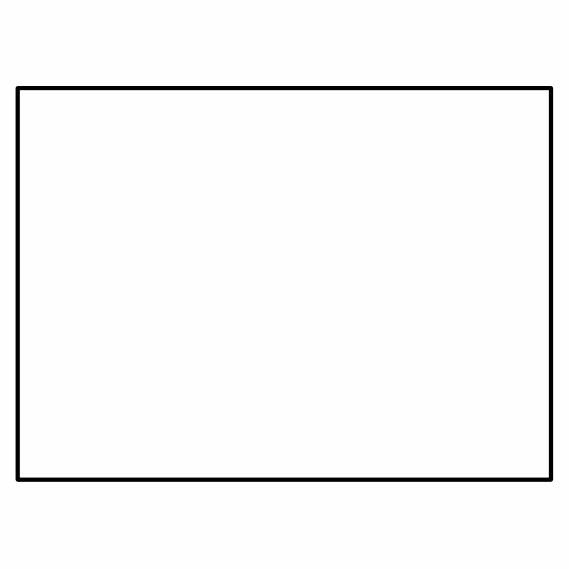}
    \includegraphics[width=0.158\linewidth]{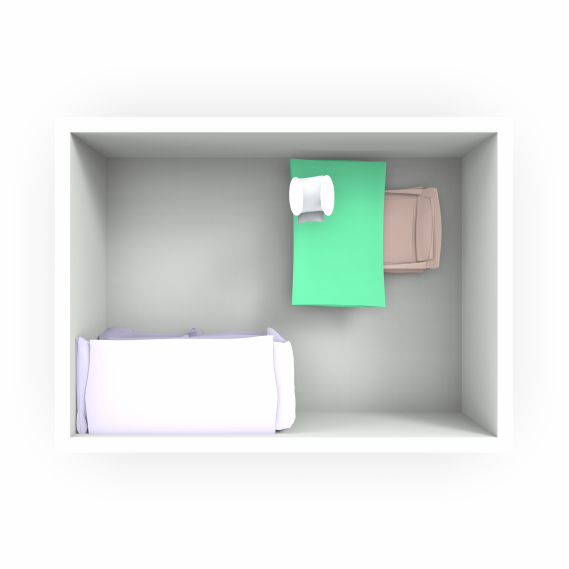}
    \caption{\titlecap{Room Generation results.}{Given the room boundary, we can utilize our trained decoder to generate new rooms. 
    Our network is able to take arbitrary room boundaries as input to generate object layouts and geometric details in a recursive manner. 
    The figure shows 12 generated rooms (4 living rooms, 4 bedrooms, and 4 libraries). 
    From the results, our network learns the continuous latent space successfully, which can capture the plausible part geometries and reasonable object layout that fits the room boundary simultaneously.}
    }
    \label{fig:generation1}
\end{figure}

From the results in Figure~\ref{fig:comgeneration}, our approach can be able to capture the functional regions' variation (or a local gathering of furniture) better. For example, our method can successfully predict four chairs with the same geometry surrounding a table, while the other baselines cannot. 
Besides, more generated indoor scenes are present in Figure~\ref{fig:generation1}. Given a room boundary, our network can generate the object layouts and the fine-grained geometry of furniture. The figure shows 12 generated rooms, including 4 living rooms, 4 bedrooms, and 4 libraries, which indicates the generated plausible part geometries and reasonable object layout can fit the given room boundaries.
Furthermore, to demonstrate the novelty of room generation, we show the top-5 nearest neighbors (Fig.~\ref{fig:generation}) in the training set by shape retrieval according to the CD on the sampled points (100,000). The presented results reveal that our generated rooms are different from the top-5 retrieved rooms on the object layouts and geometry details.

\textbf{Metrics for generation results.}
We have five metrics for generation results:
\begin{itemize}
    \item FID stands for Fréchet Inception Distance~\cite{Heusel2017GANsTB} between the generation results and the ground truth. The results and the ground truth are rendered into a top-down view similar to the input of~\cite{wang2018deep,ritchie2019fast}.
    \item $o_1$ is obtained by first calculating the \textbf{distribution of furniture categories} across the generation results and the ground truth, and calculating the \textit{Earth Mover's} Distance (EMD)~\cite{Rubner1998AMF} between them.
    \item $o_2$ is obtained by first calculating the \textbf{distribution of furniture categories for every room type} (\eg bedroom, living room, etc.), and taking the average EMD.
    \item $o_3$ is obtained by first calculating the distribution of the \textbf{co-occurrence of every two types of furniture for every room type}, and taking the average EMD. 
    \item \yjr{$o_4$ measures the distribution of the correlation of object pairs from generated rooms, which is obtained following two steps: firstly we calculate the offset of the $x$-$z$ positions of every possible object pair in each type of room, such as Table-Chair offset and Sofa-Table offset in living rooms. Secondly, we take out a square space around the origin of the 2-D plane, the edge length of which is 3.5m. This space is divided into a 1000$\times$1000 grid. Every offset that lies in this square is counted, and a gray-scale image can be drawn from this grid and offsets. 
    }
    \item \yjr{Orientation measures the radian distance between the rotation angle around the $y$-axis with the set of $\Theta = \{\theta | \theta = \frac{i\pi}{4}, i=-3,-2,-1,0,1,2,3,4\}$. In the 3D-Front dataset, we observe that the orientation of almost all objects are aligned to the $x$, $z$-axis or diagonal direction of them, meaning that the rotation angle around $y$-axis $\theta$ is in the set of $\Theta$. The function $\cos^2(x)$ is suitable since in our formulation $s = \cos^2(2\theta)$, it has peak value when $x$ is among these eight angles. For the orientation of an object, we take its rotation angle $\theta$ around $y$-axis as input, and calculate $s$, then report the average of $s$ for every object in our method and three baselines. }
\end{itemize}

\begin{figure}[h]
    \centering
    \rotatebox[origin=c]{90}{\qquad\qquad \small{Bed--Cabinet}}
    \includegraphics[width=0.18\linewidth]{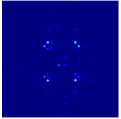}
    \includegraphics[width=0.18\linewidth]{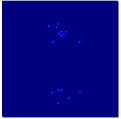}
    \includegraphics[width=0.18\linewidth]{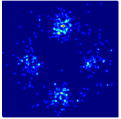}
    \includegraphics[width=0.18\linewidth]{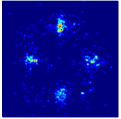}
    \includegraphics[width=0.18\linewidth]{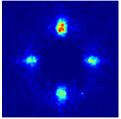}
    \vspace{-1.3cm}
    \\
    \rotatebox[origin=c]{90}{\qquad\qquad \small{Bed-NS}}
    \includegraphics[width=0.18\linewidth]{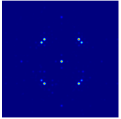}
    \includegraphics[width=0.18\linewidth]{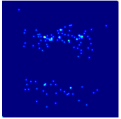}
    \includegraphics[width=0.18\linewidth]{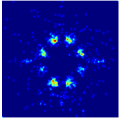}
    \includegraphics[width=0.18\linewidth]{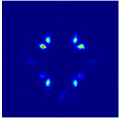}
    \includegraphics[width=0.18\linewidth]{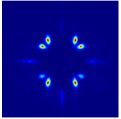}
    \vspace{-0.9cm}
    \\
    \rotatebox[origin=c]{90}{\qquad\qquad \small{Sofa-Table}}
    \includegraphics[width=0.18\linewidth]{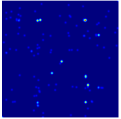}
    \includegraphics[width=0.18\linewidth]{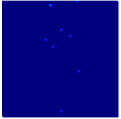}
    \includegraphics[width=0.18\linewidth]{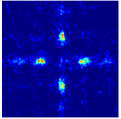}
    \includegraphics[width=0.18\linewidth]{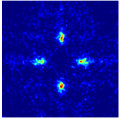}
    \includegraphics[width=0.18\linewidth]{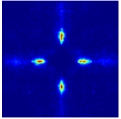}
    \vspace{-1.3cm}
    \\
    \rotatebox[origin=c]{90}{\qquad\qquad \small{Table-Chair}}
    \subfigure[GRAINS]{\includegraphics[width=0.18\linewidth]{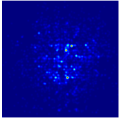}}
    \subfigure[DP]{\includegraphics[width=0.18\linewidth]{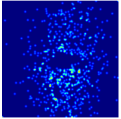}}
    \subfigure[ATISS]{\includegraphics[width=0.18\linewidth]{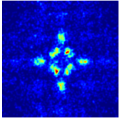}}
    \subfigure[Ours]{\includegraphics[width=0.18\linewidth]{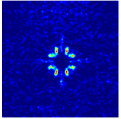}}
    \subfigure[GT]{\includegraphics[width=0.18\linewidth]{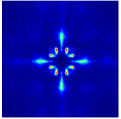}}
    \vspace{-0.8cm}
    \caption{\titlecap{Comparison on Room Generation for correlation metric $o_4$.} {We show the comparison of generation results on correlation metric $o_4$ of our method, Deep Priors (DP)~\cite{wang2018deep}, GRAINS~\cite{li2019grains}, ATISS~\cite{paschalidou2021atiss}, and GT (training data). We display some selected histograms on Bed-Cabinet, Bed-NS (Nightstands), Table-Chair and Sofa-Table. From the results we can see that our generated results can captures distribution of objects related to other objects. 
    }}
    \label{fig:eval-o4}
\end{figure}

We believe that FID is a global metric that measures the similarity between the ground truth and the generation results. On the other hand, $o_1,o_2,o_3,o_4$ are `structural' metrics that can prove whether the furniture arrangement patterns are learned. For example, in bedrooms, beds and nightstands often appear, and they often appear together. $o_2$ captures the appearance of beds and nightstands, while $o_3$ captures the co-occurrence of them.
Here, $o_1$ encourages the distribution of the overall furniture category to be close to the training set.
\yjr{$o_4$ measures the distribution of the correlation of object pairs from the generated rooms.
For this metric, we use the histogram to visualize the correlation of object pairs in Figure~\ref{fig:eval-o4}, \eg Bed-Cabinet, Bed-Nightstands, Sofa-Table, Table-Chair.
From the results in Figure~\ref{fig:eval-o4}, we can see that our generated results can successfully capture the distribution of objects related to other objects. 
All the results are reported in Table~\ref{tab:comgeneration}. From the numerical evaluations, we can see that our method outperform all baselines on FID, $o_1,o_2,o_3$ and has a similar performance on orientation metric with state-of-the-art work ATISS.
}

\begin{table}[!htbp] 
\centering
\caption{\titlecap{Generation comparison metrics between methods.}{\yjr{We compute Fréchet Inception Distance and four additional metrics $o_1,o_2,o_3$ and orientation that can measure the distribution of furniture in the generated rooms of our method, Deep Priors~\cite{wang2018deep}, GRAINS~\cite{li2019grains}, and  ATISS~\cite{paschalidou2021atiss}. We can see that our results outperforms the all baselines on FID, $o_1,o_2,o_3$ and has a similar performance on orientation metric with SoTA work ATISS.}}}
\label{tab:comgeneration}
\begin{adjustbox}{width={\linewidth},keepaspectratio}
\begin{tabular}{ccccccc}
\toprule[1pt]
Methods & FID & $o_1$ & $o_2$ & $o_3$ & orientation (degree)\\
\midrule
Ours & \textbf{139.4508}  &  \textbf{0.05}    &   \textbf{0.13}   &  \textbf{0.47}   &  0.9572 ($5.98^\circ$) \\ 
GRAINS~\cite{li2019grains} & 181.9106 & 0.30   &  0.52    &   0.79  & 0.9931 ($2.39^\circ$)\\
Deep Priors~\cite{wang2018deep} & 158.0010 & 0.40  &  0.41    &  1.04   & 0.9505 ($6.43^\circ$) \\
ATISS~\cite{paschalidou2021atiss} & 141.7889 & 0.47  &  0.19    &  0.87  & 0.9578 ($5.92^\circ$)\\
\bottomrule[1pt]
\end{tabular}
\end{adjustbox}
\end{table}

\textbf{\yjr{Perceptual study} on generation results.}
In addition to quantitative results, we conduct a \yjr{perceptual study} on the generation results of our method and two baseline methods(GRAINS and Deep Priors). To fairly compare the results, we randomly select 100-floor boundaries from the dataset and generate rooms for all three methods using these boundaries as conditions. For each participant, we prepare 20 questions. For every question, we ask the user to rank the results under three criteria: a) The layout of the furniture, a.k.a locations, and categories of the placed furniture. b)The coordination of the furniture (\eg a dining table and a beach chair do not coordinate with each other). c) The overall performance of the generated rooms. 
\yj{All the participants were local volunteers known to be reliable.}
The results of the \yjr{perceptual study} are shown in \autoref{tab:userstudy}. 

We can see that our method is the most preferred among the three methods. In the two baselines, Deep Priors performs better in layout, and GRAINS performs better in furniture coordination. We infer the reason for this result is Deep Priors tend to capture the whole furniture layout via top-down images and CNN, while GRAINS models structural relationships better with their hierarchical representation and RvNN-VAE.

\begin{table*}[t]
  \centering
  \caption{\titlecap{\yjr{Perceptual study} results on 3D scene generation.}{\yj{We show the average ranking scores (from 2 (the best) to 0 (the worst)) and the frequency of the method being ranked $1^{\rm st}$, $2^{\rm nd}$ and $3^{\rm rd}$: Deep Priors~\cite{wang2018deep}, GRAINS~\cite{li2019grains}, and ours. The results are calculated based on 433 trials. We see that our method achieves the best on all metrics.}}}
    \begin{tabular}{cccc}
    \toprule[1pt]
    Method & Deep Priors\cite{wang2018deep}/Rank($1^{\rm st},2^{\rm nd},3^{\rm rd}$) & GRAINS\cite{li2019grains}/Rank($1^{\rm st},2^{\rm nd},3^{\rm rd}$) & Ours/Rank($1^{\rm st},2^{\rm nd},3^{\rm rd}$)\\
    \midrule
    Layout                                  & 0.7611/(17.5\%, 41.2\%, 41.4\%) & 0.6482/(11.1\%, 42.7\%, 46.2\%) & \textbf{1.5907}/(71.4\%, 16.2\%, 12.4\%)\\
    \midrule
    \tabincell{c}{Furniture\\Coordination}  & 0.7212/(15.2\%, 41.6\%, 43.1\%) & 0.7389/(19.7\%, 34.5\%, 45.8\%) & \textbf{1.5398}/(65.0\%, 23.9\%, 11.0\%) \\
    \midrule
    Overall                                 & 0.6925/(14.8\%, 39.6\%, 45.6\%) & 0.6748/(12.2\%, 43.1\%, 44.7\%) & \textbf{1.6327}/(73.0\%, 17.2\%, 9.8\%) \\
    \bottomrule[1pt]
    \end{tabular}%
  \label{tab:userstudy}%
\end{table*}%

\subsection{Scene Interpolation}

\begin{figure}[h]
    \centering
    \includegraphics[width=0.155\linewidth]{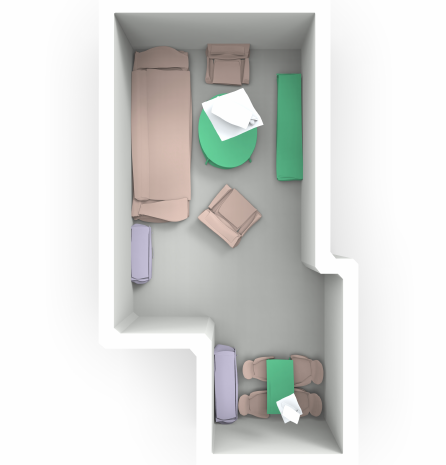}
    \includegraphics[width=0.155\linewidth]{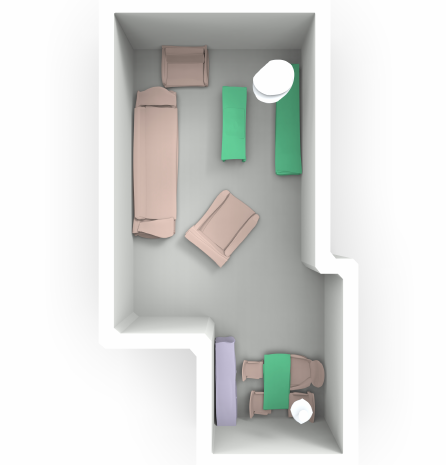}
    \includegraphics[width=0.155\linewidth]{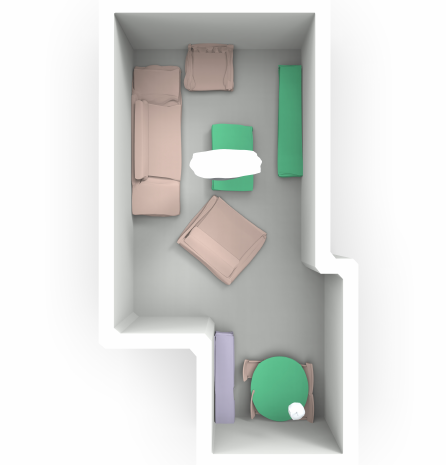}
    \includegraphics[width=0.155\linewidth]{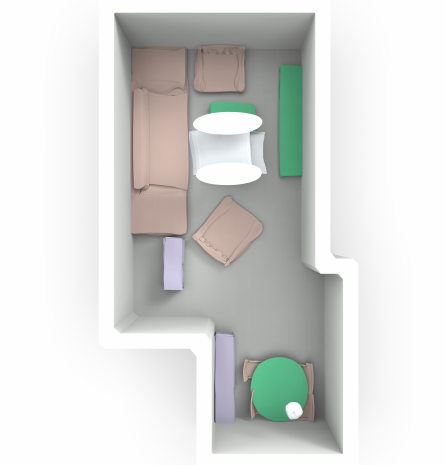}
    \includegraphics[width=0.155\linewidth]{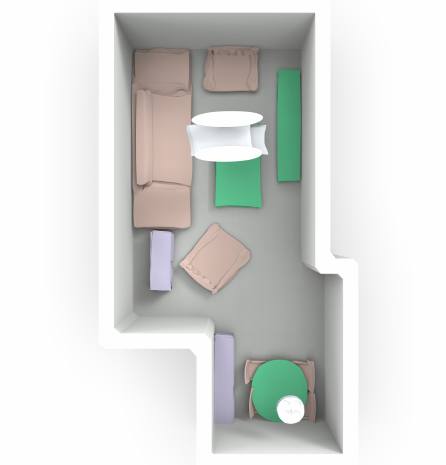}
    \includegraphics[width=0.155\linewidth]{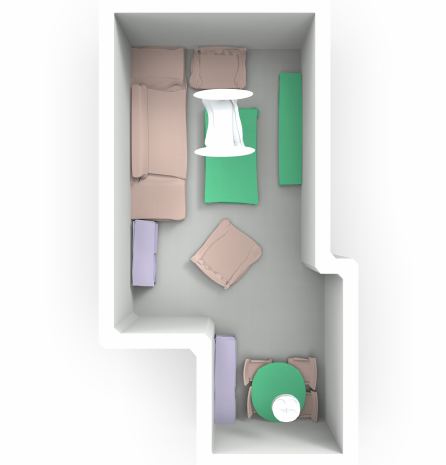}
    \\
    \includegraphics[width=0.155\linewidth]{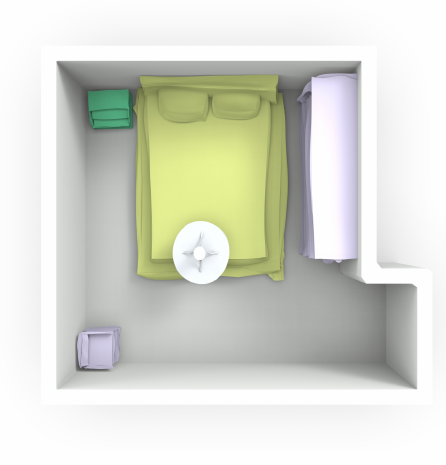}
    \includegraphics[width=0.155\linewidth]{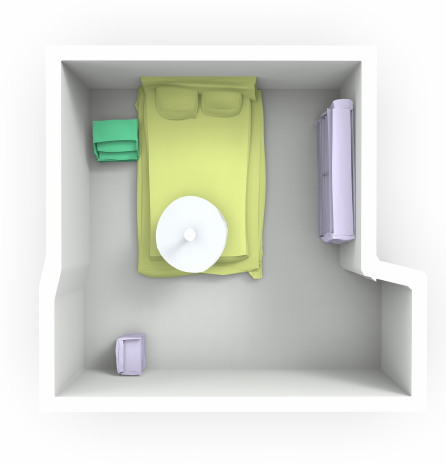}
    \includegraphics[width=0.155\linewidth]{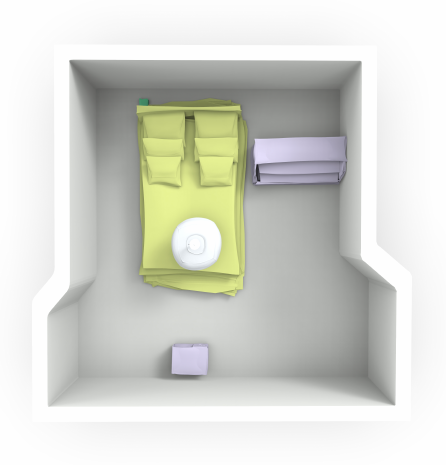}
    \includegraphics[width=0.155\linewidth]{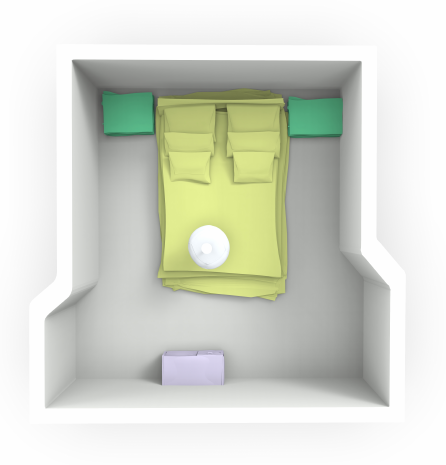}
    \includegraphics[width=0.155\linewidth]{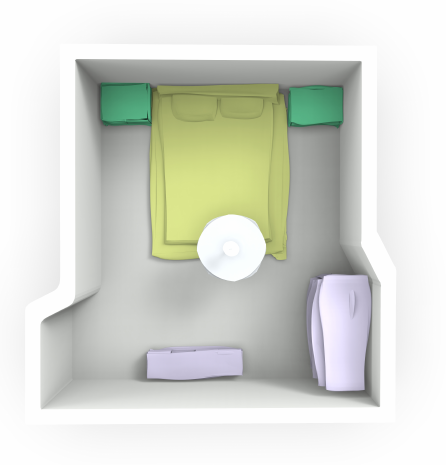}
    \includegraphics[width=0.155\linewidth]{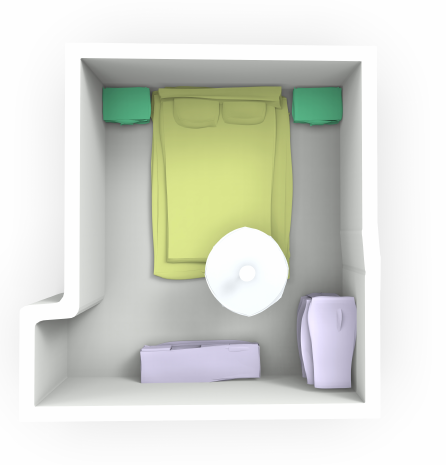}
    \\
    \includegraphics[width=0.155\linewidth]{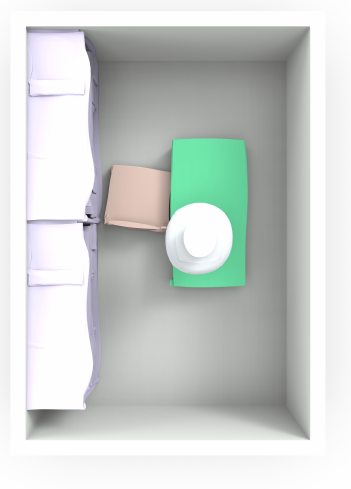}
    \includegraphics[width=0.155\linewidth]{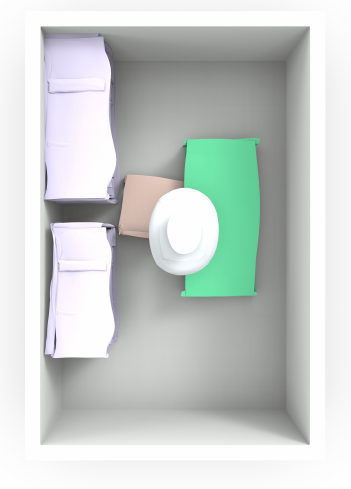}
    \includegraphics[width=0.155\linewidth]{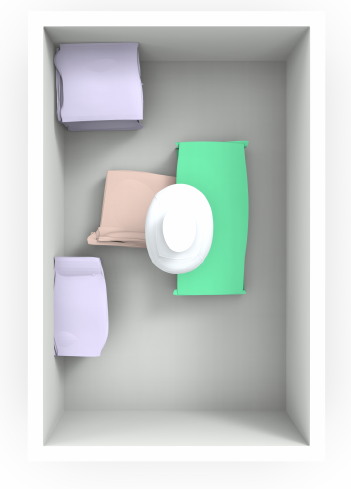}
    \includegraphics[width=0.155\linewidth]{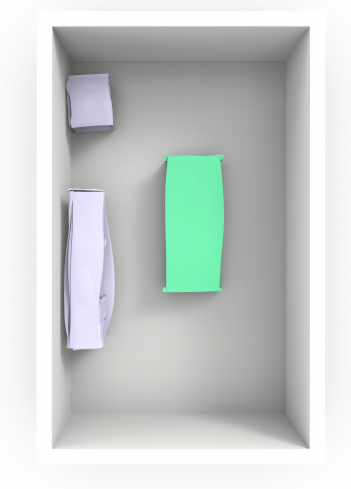}
    \includegraphics[width=0.155\linewidth]{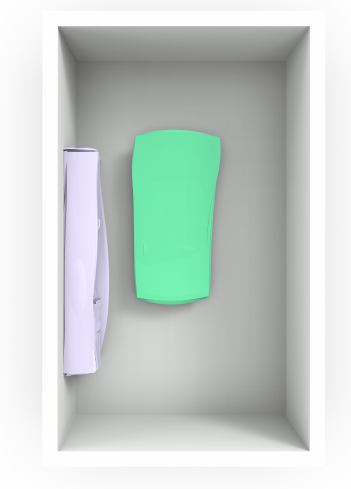}
    \includegraphics[width=0.155\linewidth]{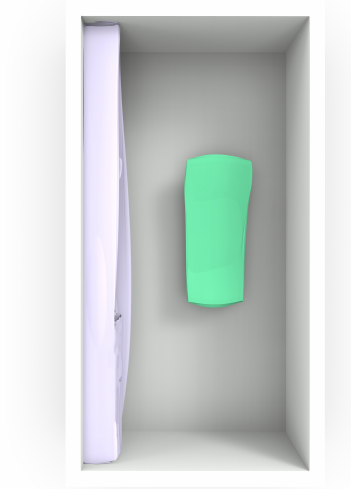}
    \\
    \subfigure[source]{\includegraphics[width=0.155\linewidth]{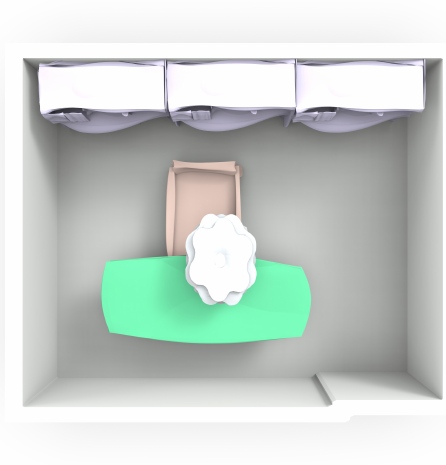}}
    \includegraphics[width=0.155\linewidth]{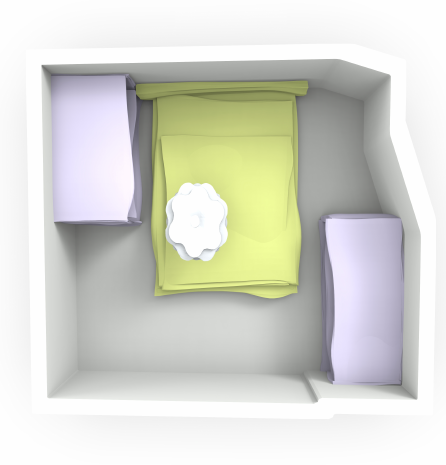}
    \includegraphics[width=0.155\linewidth]{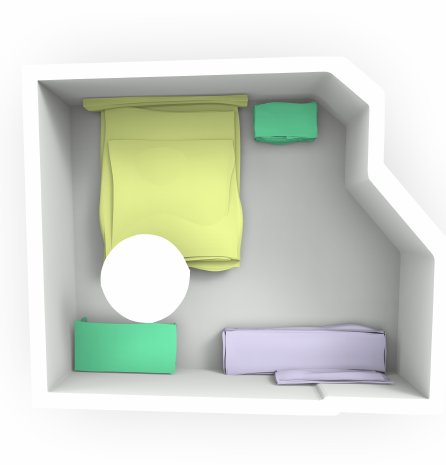}
    \includegraphics[width=0.155\linewidth]{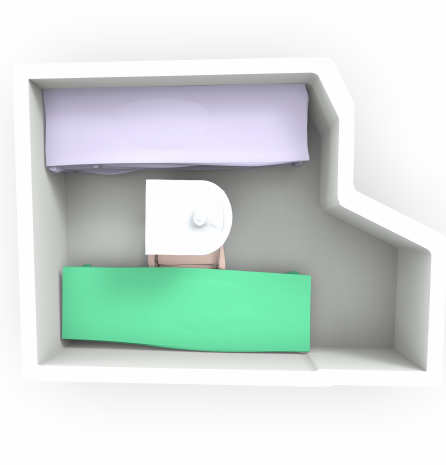}
    \includegraphics[width=0.155\linewidth]{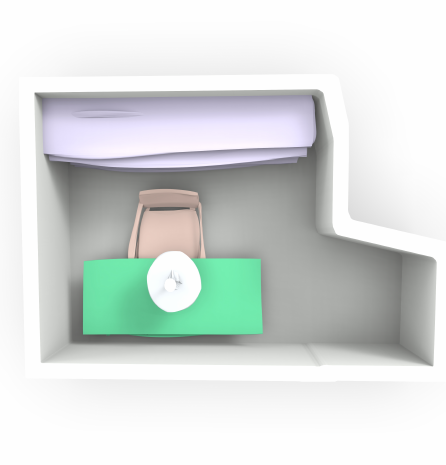}
    \subfigure[target]{\includegraphics[width=0.155\linewidth]{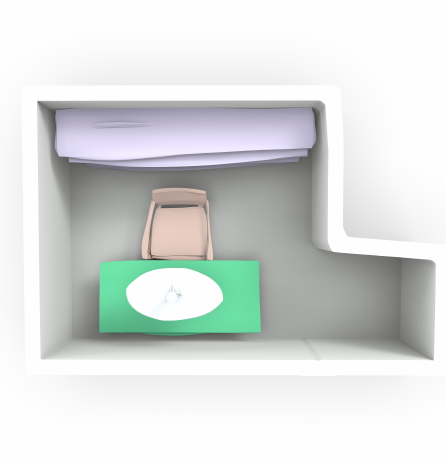}}
    \caption{\titlecap{Room Interpolation.}{We simultaneously interpolate on boundary and scene layout and feed them into the VAE decoder. Every interpolation step is a valid 3D scene layout, with the boundary deforming continuously and the layout changing according to the boundary. Note that the interpolation is only reasonable when the source and target are similar.}
    }
    \label{fig:interpolation}
\end{figure}

Interpolation is another direct application of our trained RvNN-VAE. To show the smoothness of the latent space our VAE has learned, we show some examples of interpolation in \autoref{fig:interpolation}. When interpolating, we first encode the floor boundary and the scene layout of source and target scenes into feature vectors and perform linear interpolation between source and target on the two features simultaneously before feeding them into our conditional VAE decoder. \yj{Thanks to our generative FloorNet to ensure the meaningful latent space, FloorNet can achieve a reasonable interpolated room boundary between source and target.}
Note that the interpolation between 3D indoor scenes is not as straightforward as 3D shapes, because our representation not only encodes the layout of objects, but also contains geometry information. We find that interpolation between two similar scenes has the best performance. For example, The source and target of the second row in \autoref{fig:interpolation} have similar floor boundaries and furniture layout.

As is shown in \autoref{fig:interpolation}, every step in the interpolation process is a valid 3D scene layout. The floor boundary gradually deforms from the source to the target, but in every step the layout changes according to the boundary rather than moving itself, which is what we expect. We also witness the deformation of object geometry in interpolation. For example, look at row 2 in \autoref{fig:interpolation}: \yjr{the small square-shaped shelf in the bottom of the source scene image stretches and becomes a longer one in the target scene.}

The interpolation results show that our latent space is smooth near any point representing a valid 3D indoor scene layout. This is a key feature for more applications like room editing or room completion.

\subsection{Applications}
The hierarchical graph representation of 3D indoor scenes and locally smooth latent space created by the RvNN-VAE enable us to perform multiple interesting applications.
In this section, we demonstrate three: a) room editing at multiple levels; b) room generation conditioned on 3D box layouts; and c) scene completion from a partial input room. 

\textbf{Room Editing.}
Editing 3D indoor scenes is not an easy task, because all the elements from functional regions to object parts are in relation to each other. Thus only editing one of them often makes the scene seem inharmonious (\eg Only editing one of four chairs surrounding a table makes the scene looks weird). But our RvNN-VAE learns the whole scene including the object part geometry and scene layout. With an edit to the scene and the latent space, we can find a latent code in the space that both satisfies the edit and decodes to a valid scene layout.

\begin{figure*}[t!]
    \centering
    \subfigure[Origin]{\includegraphics[width=0.105\linewidth]{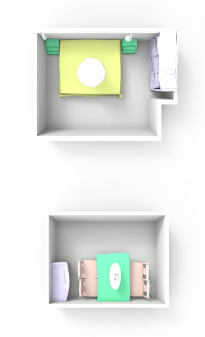}}
    \subfigure[Op1.]{\includegraphics[width=0.105\linewidth]{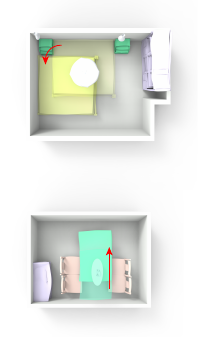}}
    \subfigure[Edited]{\includegraphics[width=0.105\linewidth]{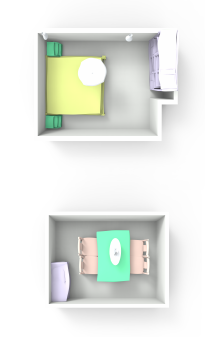}}
    \subfigure[Op2.]{\includegraphics[width=0.105\linewidth]{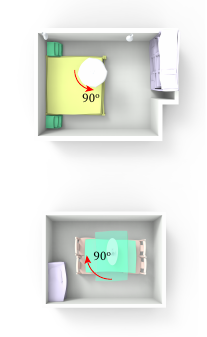}}
    \subfigure[Edited]{\includegraphics[width=0.105\linewidth]{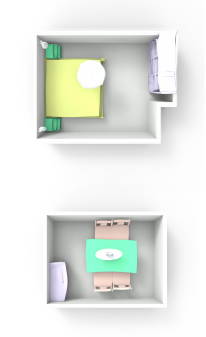}}
    \subfigure[Op3.]{\includegraphics[width=0.105\linewidth]{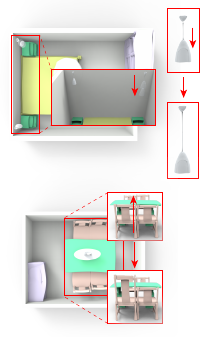}}
    \subfigure[Edited]{\includegraphics[width=0.105\linewidth]{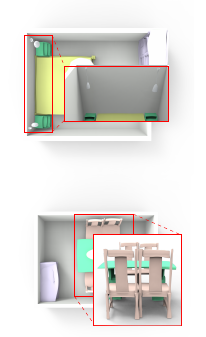}}
    \subfigure[Op4.]{\includegraphics[width=0.105\linewidth]{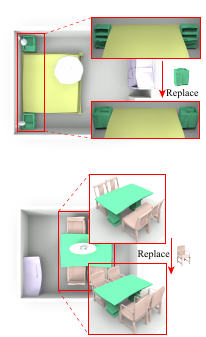}}
    \subfigure[Edited]{\includegraphics[width=0.105\linewidth]{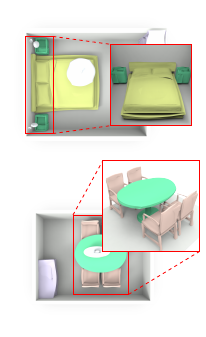}}
    \caption{\titlecap{Room Editing.}{\yjr{The first column shows the original scene, followed by pairs of columns demonstrating the edits and their results. There are four edits to each of the two scenes. The first and second edits only alter the locations and orientations of the objects, the third edit deforms object parts, and fourth edit replaces the geometry of objects. From the results, we can observe that every object related to the edited object moves or deforms according to the edit. }}}
    \label{fig:edit}
\end{figure*}

Because of the local smoothness of our latent space, similar scenes are close to each other in the space. After editing a scene, we can search for a latent code in the space which is close to the original scene and satisfies the editing. So we \yj{apply the gradient descent (using the Adam~\cite{kingma2014adam} optimizer)} to minimize the objective function:
$\rVert z - z_*\rVert^2_2 + q_{chs}(T(B_e^z)\mathbf{U},T(B_e^t)\mathbf{U}) + \mathcal{L}_{struc}(d(z))$, where $z$ is the latent vector we need to optimize on, $z_*$ is latent vector of the unedited original scene, $B_e^t$ is the edited box, and $B_e^z$ is the corresponding box in the decoded scene of $z$. (Note that both the edit of the object and the edit of a part are actually edited on boxes but at different levels.) $\mathbf{U}$ is a pre-computed set of samples on the unit cube. This loss function encourages our edited scene to be as close as the original scene while preserving the edited features.
\autoref{fig:edit} shows some results of our room editing application.

\yjr{In \autoref{fig:edit}, each row shows a scene that has been edited four times. The first column shows the original scene, the second column shows the first edit on a certain object, and the third column shows the result of the editing. There are three groups of edits and results. The first two edits are rigid, meaning we only edit the locations or orientations of the objects. In the third edit, we deform object parts to demonstrate our network can process fine-grained features down to object parts. For the final edit (replacing the geometry), it can lead to the other object changes happening in the scene, which demonstrates that our method is able to learn the correlations of objects from the data.
}

For example, The second row shows three edits on a living room. Firstly, we move the table to the left. The result is that the chairs surrounding it move with it, and all other objects remain the same. Secondly, we rotate the table counterclockwise by 90 degrees, and the chairs also rotate around the center of the table. Thirdly, we stretch the back of one chair to make it taller, and all the other chairs become taller too. 

\begin{figure}[h]
    \centering
    \includegraphics[width=0.19\linewidth]{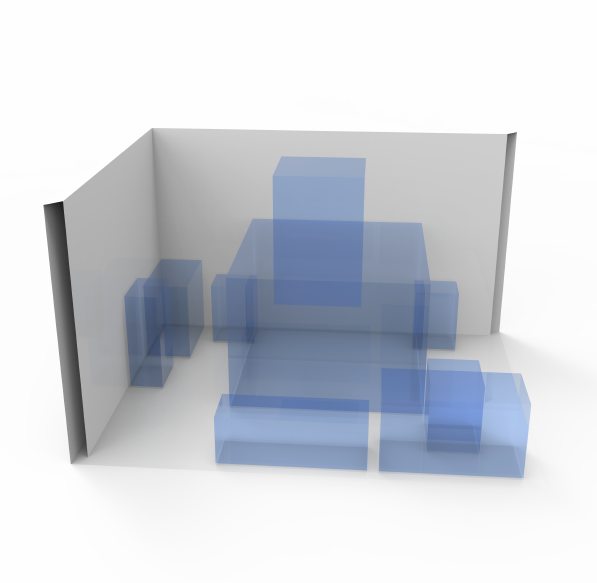}
    \includegraphics[width=0.19\linewidth]{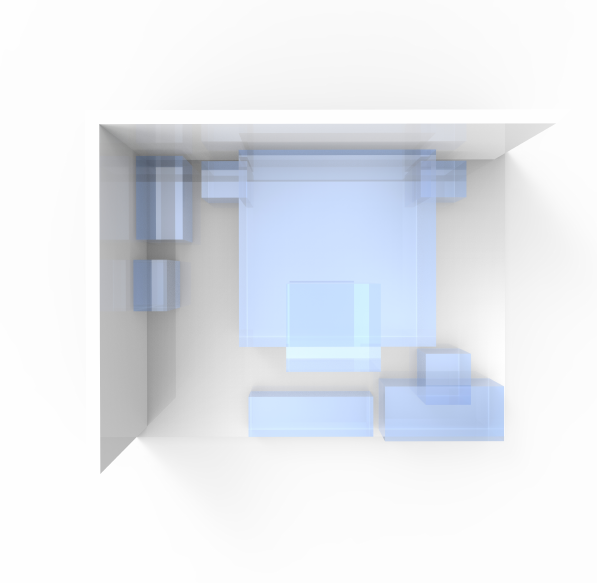}
    \includegraphics[width=0.19\linewidth]{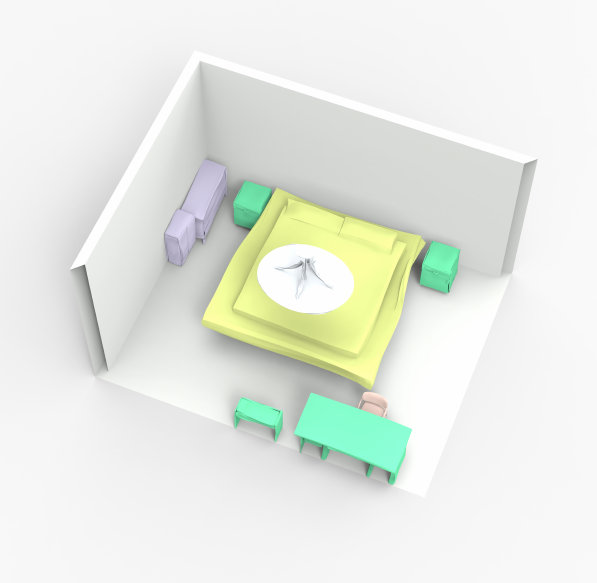}
    \includegraphics[width=0.19\linewidth]{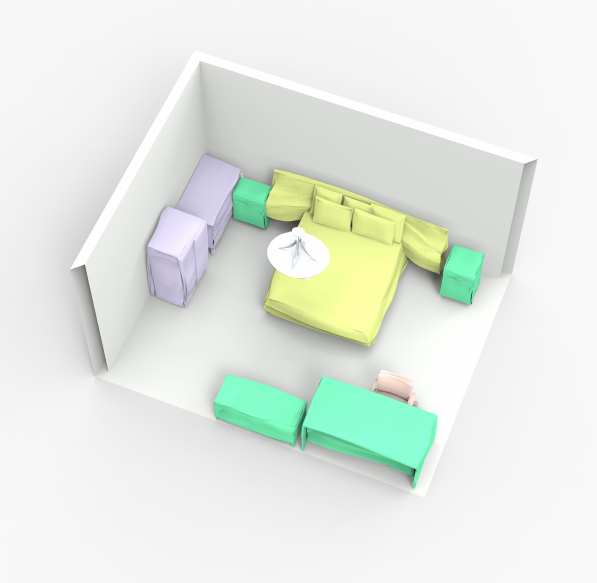}
    \includegraphics[width=0.19\linewidth]{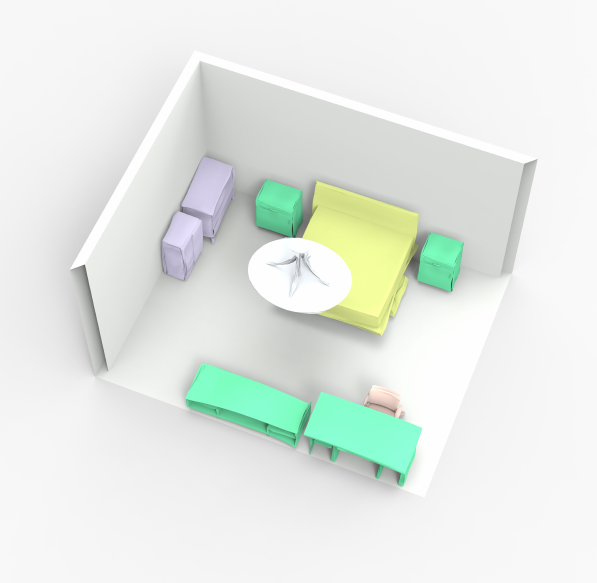}\\
    \includegraphics[width=0.19\linewidth]{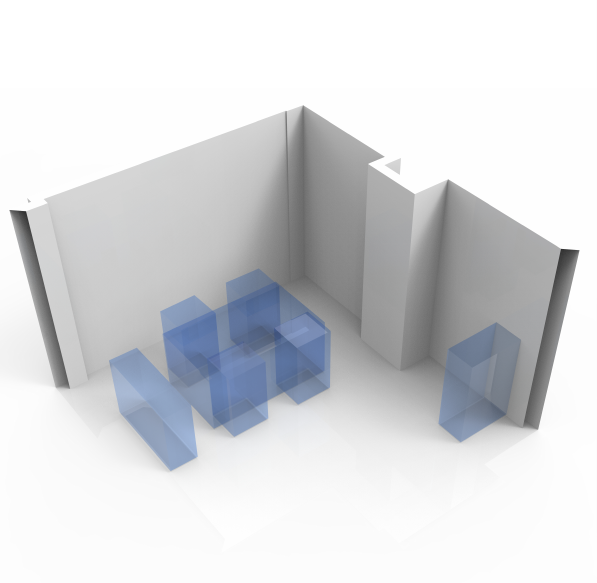}
    \includegraphics[width=0.19\linewidth]{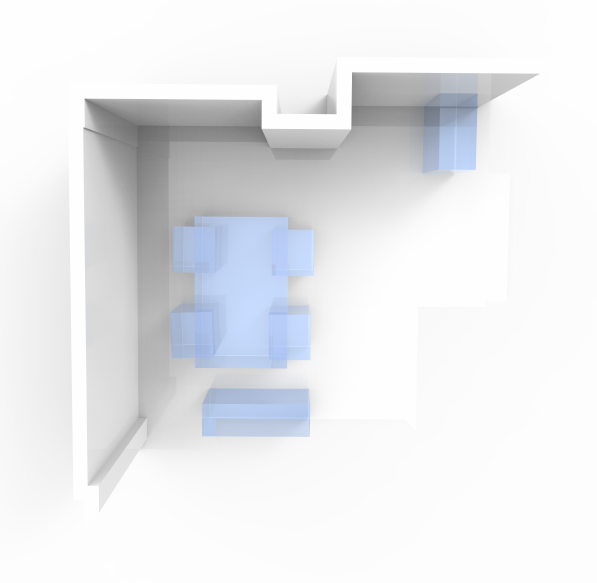}
    \includegraphics[width=0.19\linewidth]{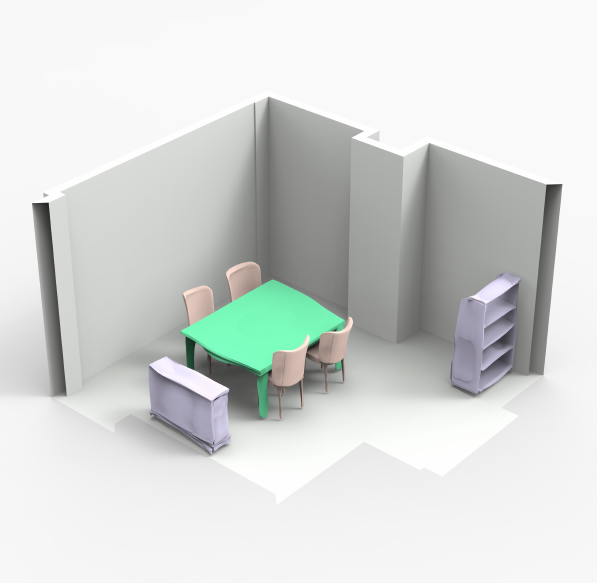}
    \includegraphics[width=0.19\linewidth]{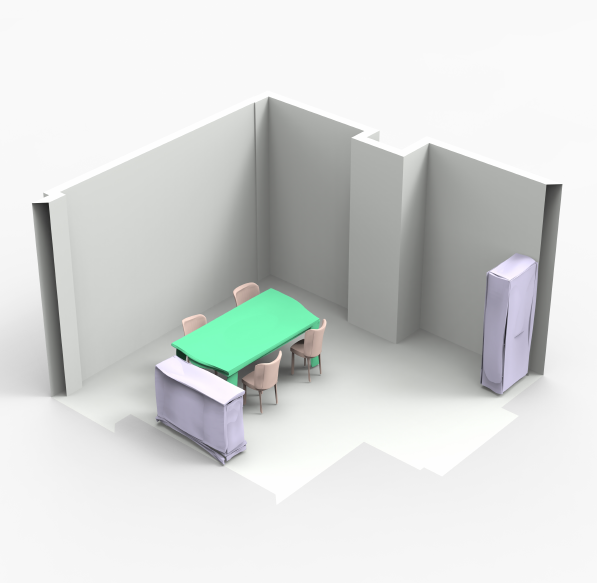}
    \includegraphics[width=0.19\linewidth]{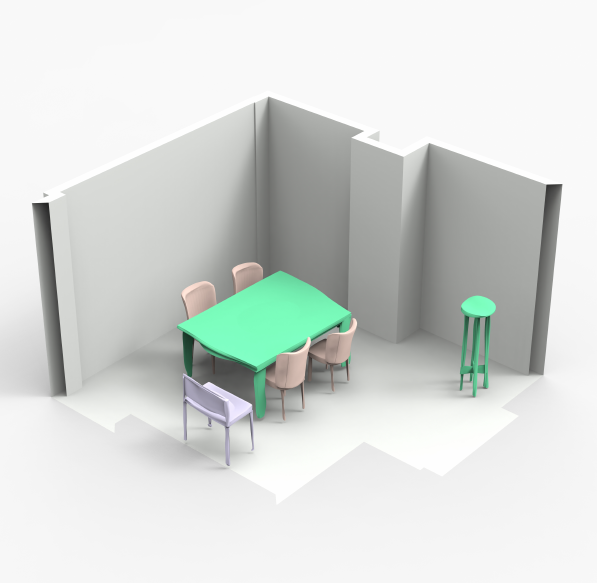}\\
    \includegraphics[width=0.19\linewidth]{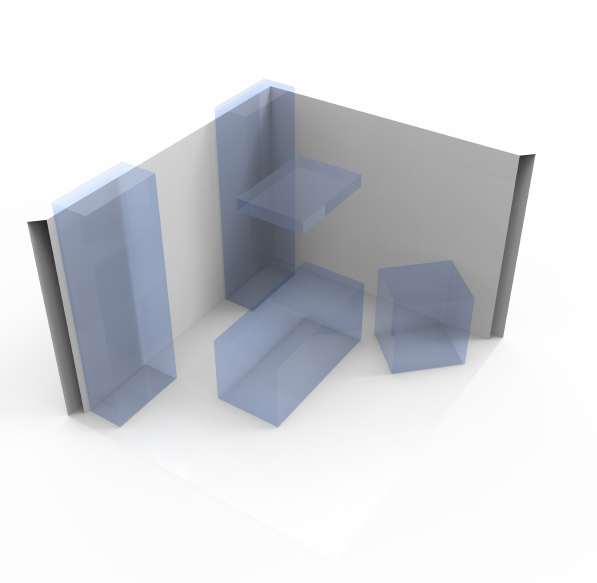}
    \includegraphics[width=0.19\linewidth]{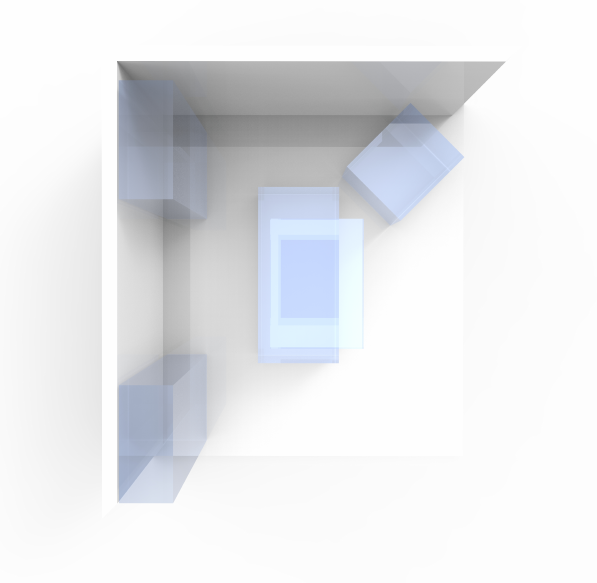}
    \includegraphics[width=0.19\linewidth]{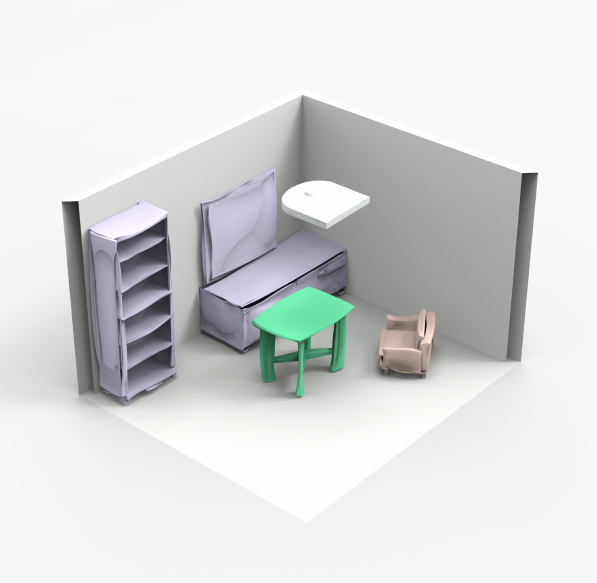}
    \includegraphics[width=0.19\linewidth]{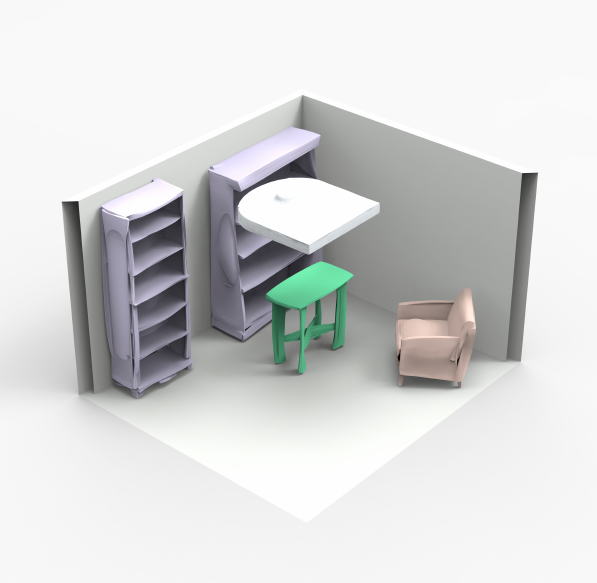}
    \includegraphics[width=0.19\linewidth]{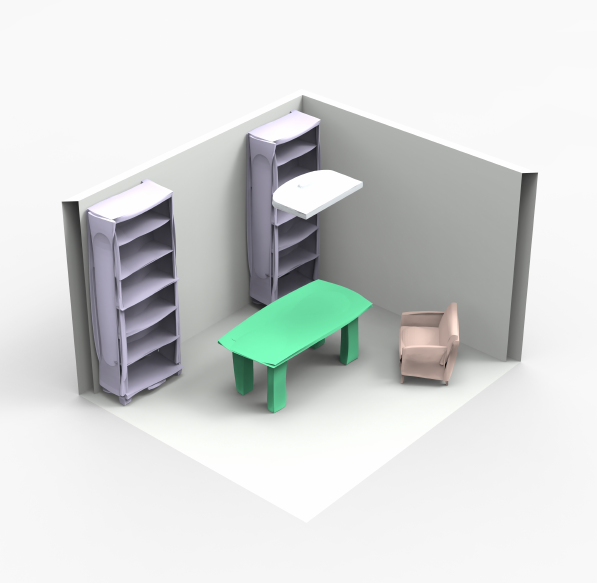}\\
    \subfigure[\hspace{-1mm}3D box-layout(2 Views)]{\includegraphics[width=0.185\linewidth]{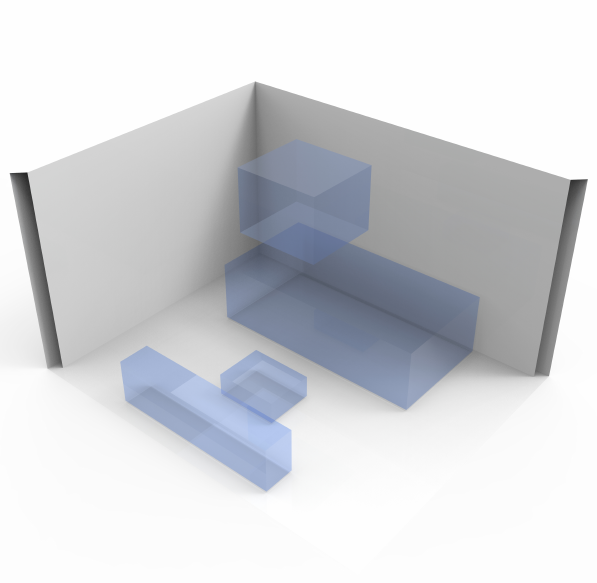}
    \includegraphics[width=0.185\linewidth]{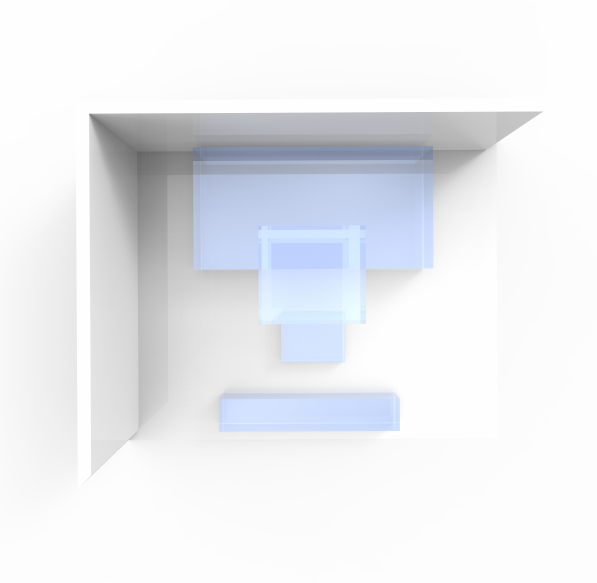}}
    \subfigure[Generated Room]{
    \includegraphics[width=0.185\linewidth]{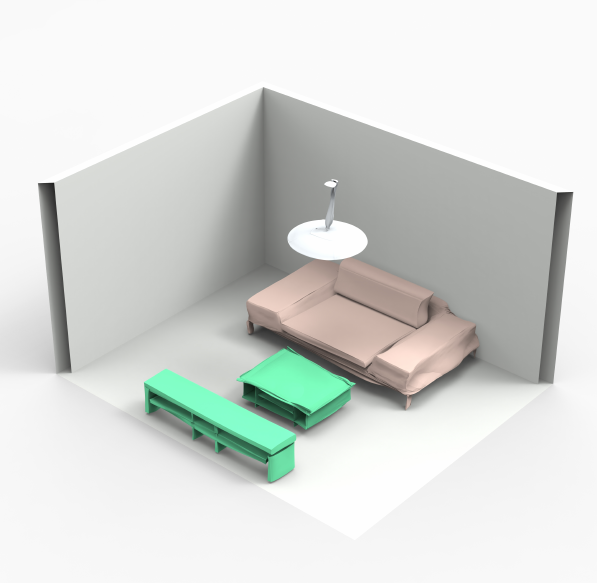}
    \includegraphics[width=0.185\linewidth]{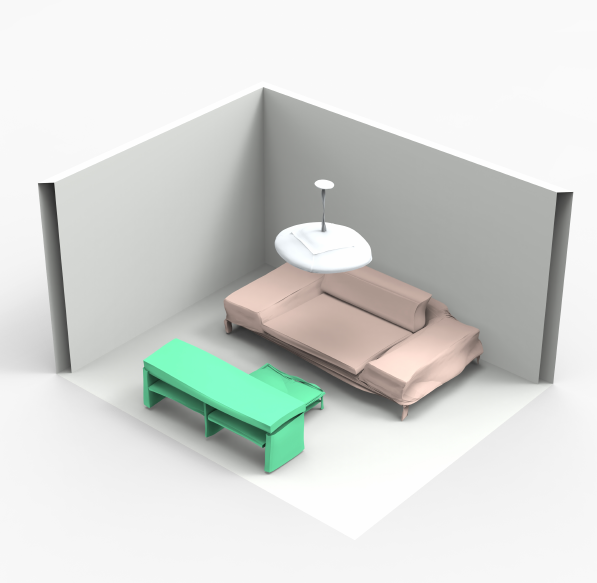}
    \includegraphics[width=0.185\linewidth]{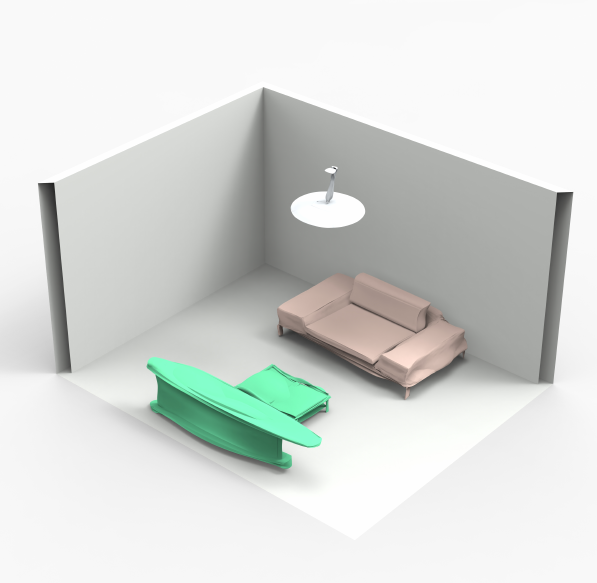}}
    \caption{\titlecap{3D scene generation from 3D box layout.}{We input a hierarchy consisting of 3D boxes into our RvNN-VAE with geometry information sampled from random distribution, then we encode the hierarchy, sample a latent vector and decode it into a complete 3D scene. We can see the positions of objects in the generated results are similar to the box layout, and the detailed geometry of the scenes looks harmonious.}}
    \label{fig:3dbox2scene}
\end{figure}

\textbf{Room Generation from 3D Box Layout.}
Sometimes more conditions than floor boundaries are given when generating 3D indoor scenes. A widely used condition in interior design is the 2D floor plan, which basically contains the information on floor boundaries and the 2D bounding boxes of the furniture. We can extend 2D floor plans to 3D box layouts. In 3D box layouts, we provide room boundary and 3D bounding boxes of objects, but the semantic types and part geometry of objects are unknown. We are required to generate a room, where the furniture in it needs to coordinate with the input boxes.

To complete the task, we can build a hierarchy with no structural edges and part-level nodes. In the object-level nodes, we only include the box features, and fill the geometry features with random numbers. We then construct a hierarchy with these boxes (note there are no structural relations in this hierarchy) and feed this hierarchy into the encoder of our VAE, then we sample from the Gaussian distribution our encoder outputs. The sampled latent vector is then mapped back into a hierarchy-graph representation of our generated 3D scene. Some results of this procedure are displayed in \autoref{fig:3dbox2scene}. 

In \autoref{fig:3dbox2scene} we include two views of input box layouts for clarity. For each input box layout, we generate three results. We can see clearly that all the objects in the results are placed roughly in the position of the input boxes, but the geometry of the objects is different. The part geometry is completely generated by our network, and the generated geometry looks harmonious (\eg in row 2 the size of the chairs corresponds to the size of the table).

\begin{figure}[h]
\centering
    \includegraphics[valign=m,width=0.24\linewidth]{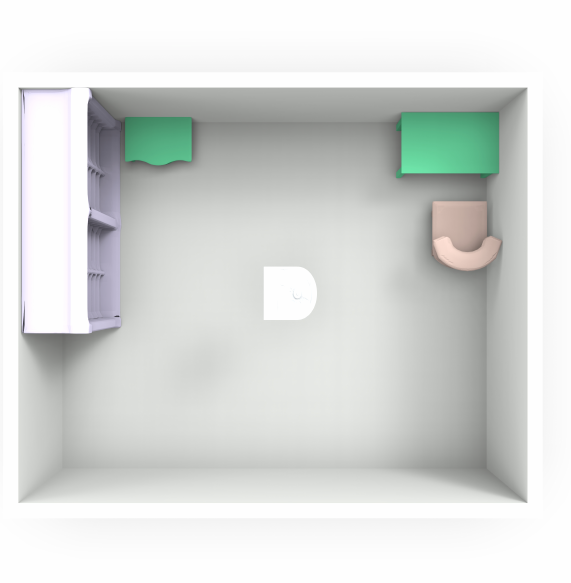}
    \includegraphics[valign=m,width=0.24\linewidth]{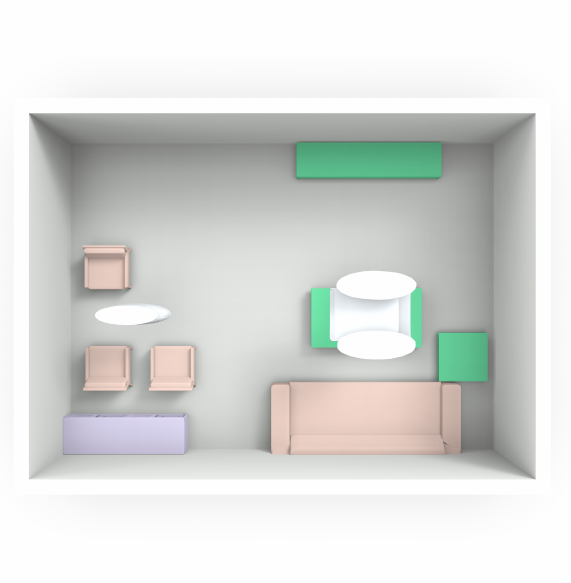}
    \includegraphics[valign=m,width=0.24\linewidth]{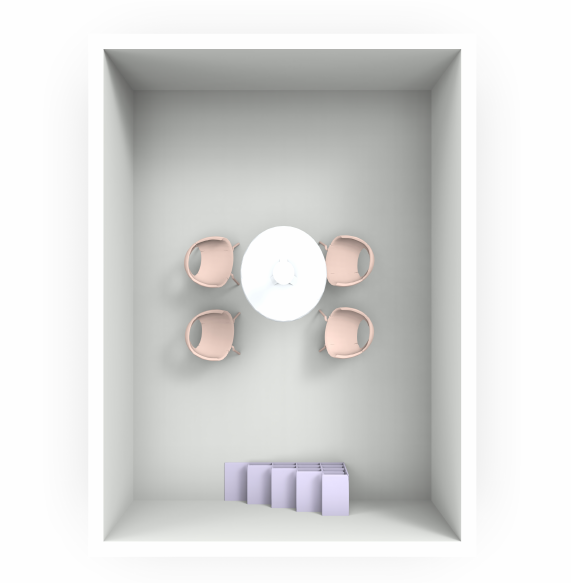}
    \includegraphics[valign=m,width=0.24\linewidth]{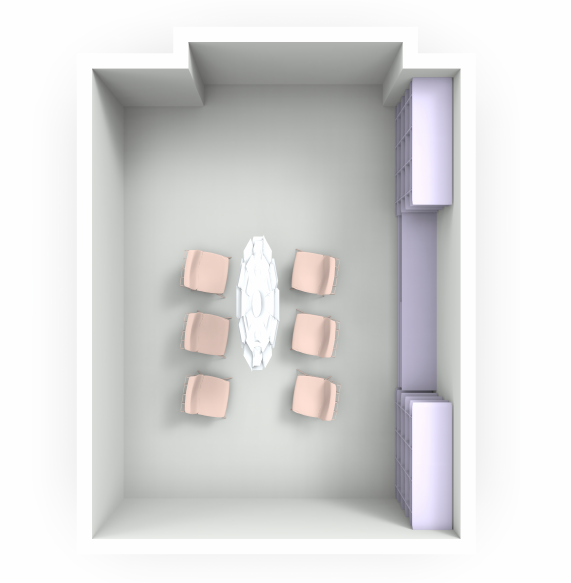}\\
    \includegraphics[valign=m,width=0.24\linewidth]{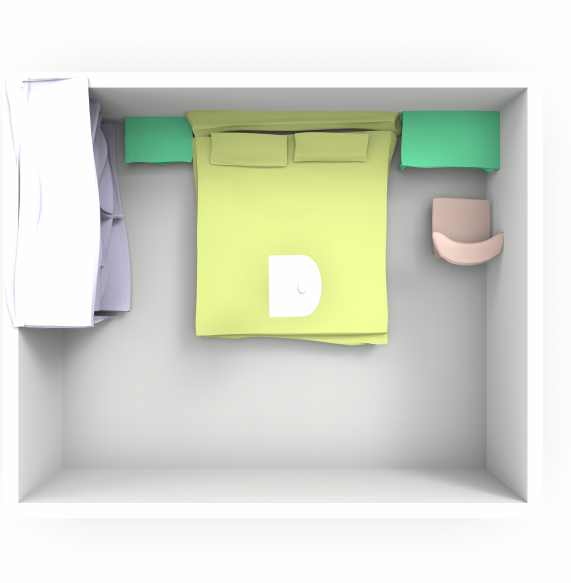}
    \includegraphics[valign=m,width=0.24\linewidth]{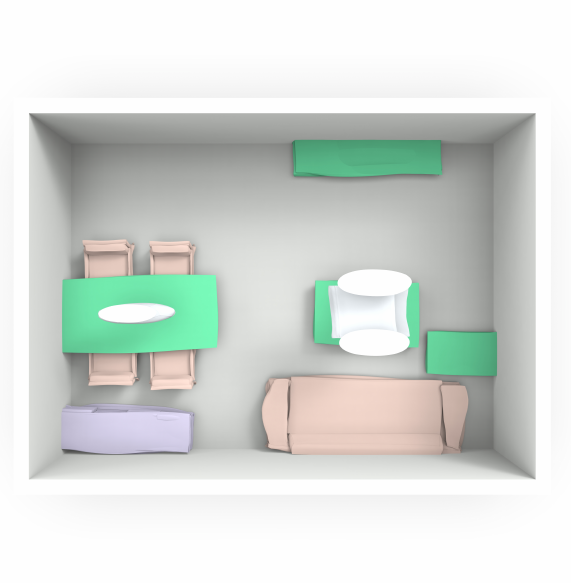}
    \includegraphics[valign=m,width=0.24\linewidth]{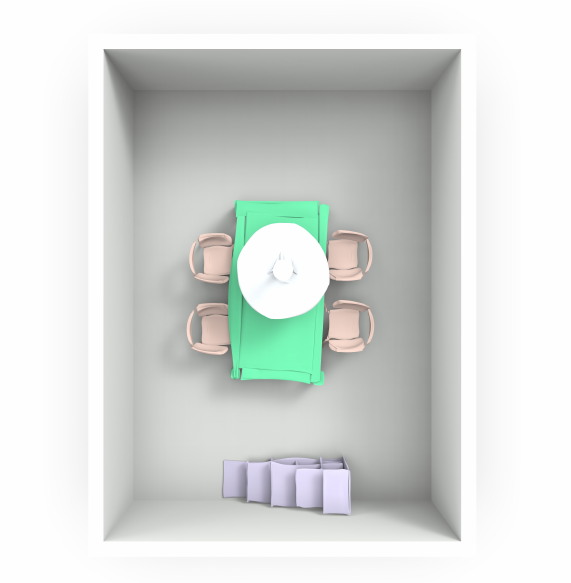}
    \includegraphics[valign=m,width=0.24\linewidth]{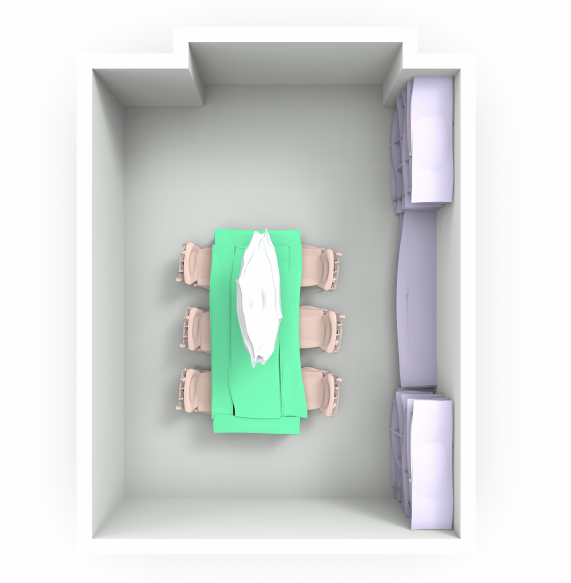}
    \caption{\titlecap{Room Completion.}{Given a partial room, our method can complete the room layout. Each of the top four images shows one partial scene, and the bottom four images show the results of our room completion. We can see not only the deleted object is added, its children are added too.}}
    \label{fig:scenecompletion}
\end{figure}

\textbf{Room Completion.}
Room completion is a somehow more challenging task than room generation conditioned on 3D box layout. With a partial scene, our method needs to predict all of the missing objects in the room. \jm{To solve this problem, we make use of the latent space learned by our VAE. As the reconstruction loss is used to train the VAE, in theory every point in the latent space can be decoded into a complete scene. So the room completion pipeline consists of two parts: using the encoder to map a partial scene to a point in the latent space, and using the decoder on the point to get the complete scene. To test the quality of room completion,} we can start by deleting some of the nodes in our hierarchical graph representation and its children, and feed this partial tree into our VAE, then decode the sampled latent vector into a complete scene. The results are shown in \autoref{fig:scenecompletion}.

The top four images and the bottom four images of \autoref{fig:scenecompletion} show the partial and completed scenes. The first, third, and fourth columns show that our method can complete scenes missing one key object. The second column shows that the children of the missing object can also be added back, while the other parts of the room remain unchanged.

\begin{figure*}[h]
    \centering
    \includegraphics[width=0.15\linewidth]{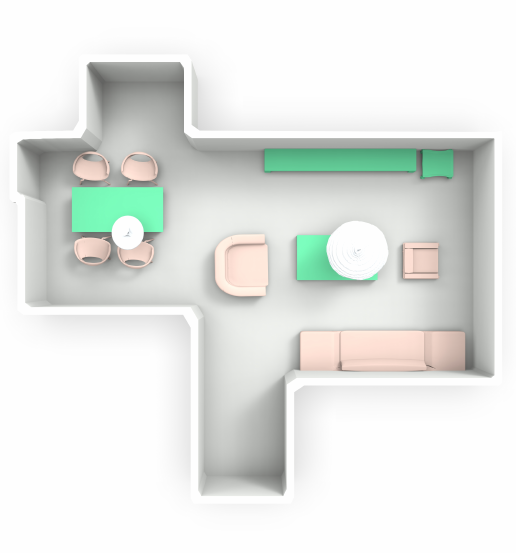}
    \includegraphics[width=0.15\linewidth]{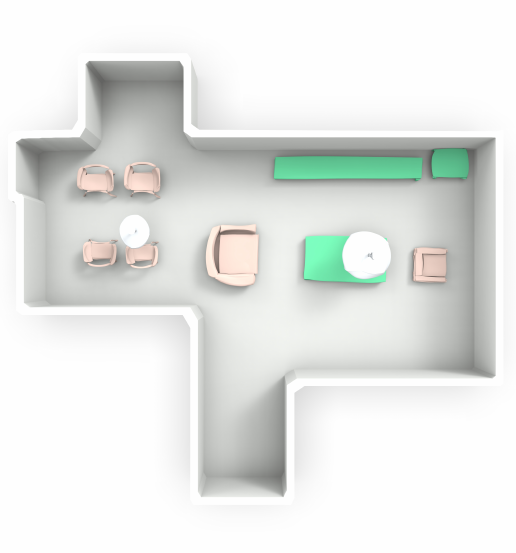}
    \includegraphics[width=0.15\linewidth]{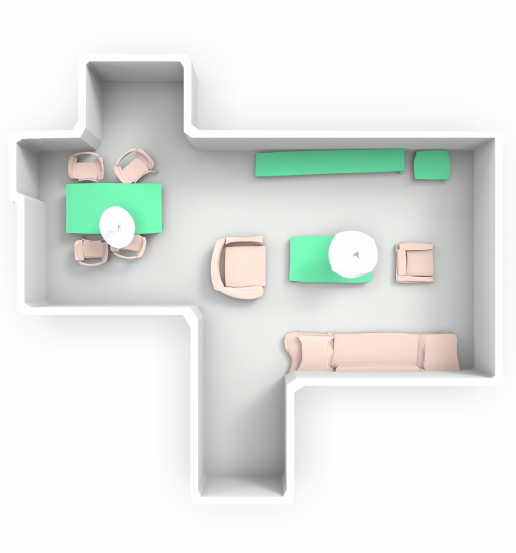}
    \includegraphics[width=0.15\linewidth]{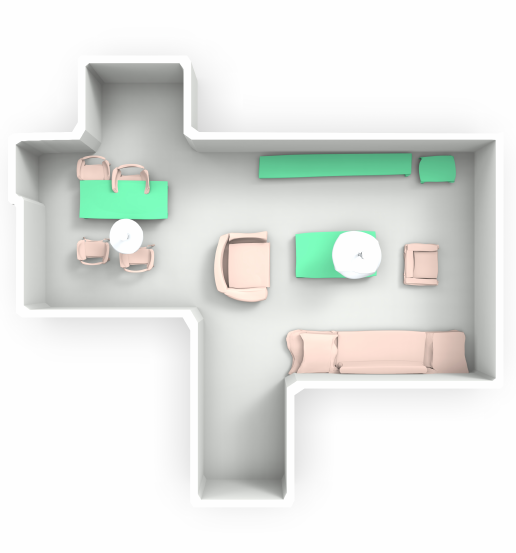}
    \includegraphics[width=0.15\linewidth]{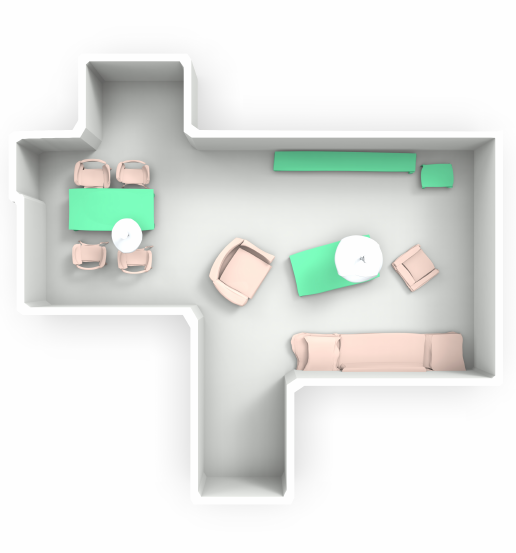}
    \includegraphics[width=0.15\linewidth]{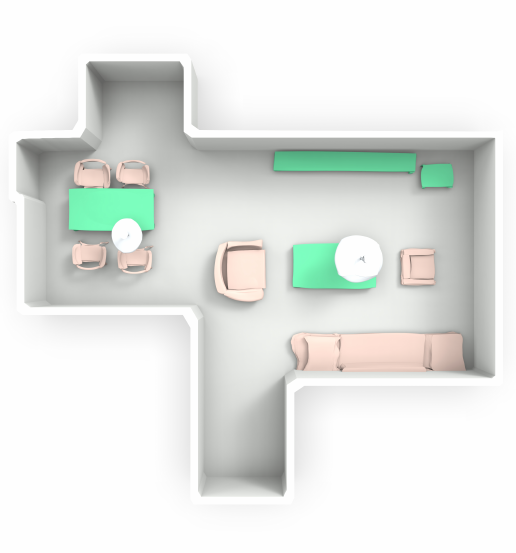}\\
    \subfigure[Input]{\includegraphics[width=0.14\linewidth,angle=90]{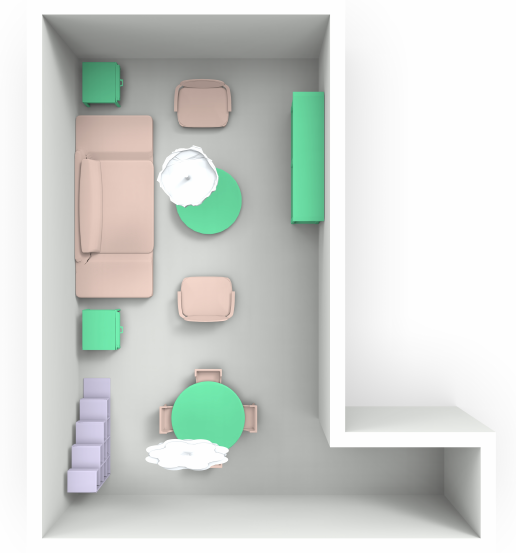}}
    \subfigure[w/o FR]{\includegraphics[width=0.14\linewidth,angle=90]{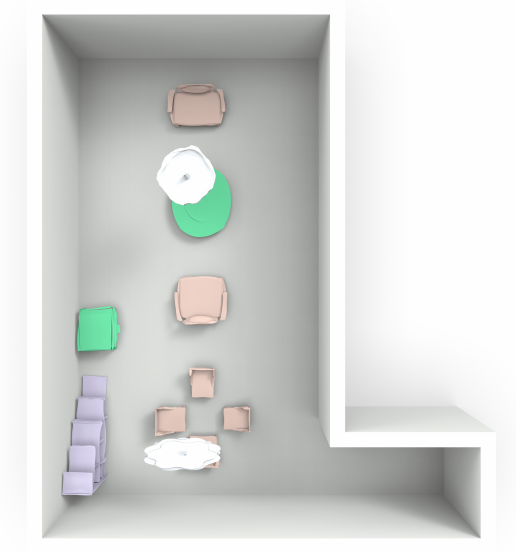}}
    \subfigure[w/o ROE]{\includegraphics[width=0.14\linewidth,angle=90]{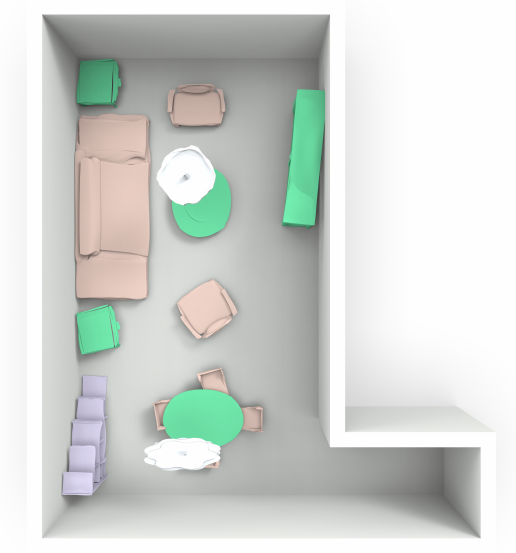}}
    \subfigure[w/o OOE]{\includegraphics[width=0.14\linewidth,angle=90]{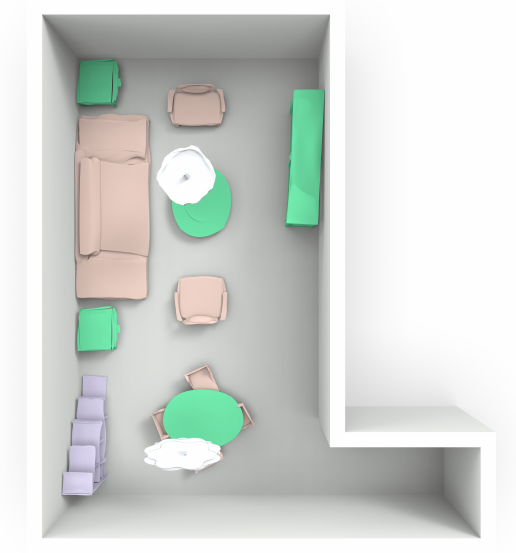}}
    \subfigure[w/o HE]{\includegraphics[width=0.14\linewidth,angle=90]{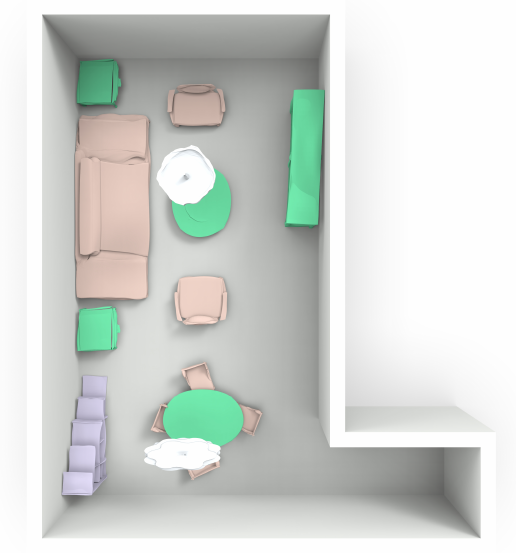}}
    \subfigure[Ours]{\includegraphics[width=0.14\linewidth,angle=90]{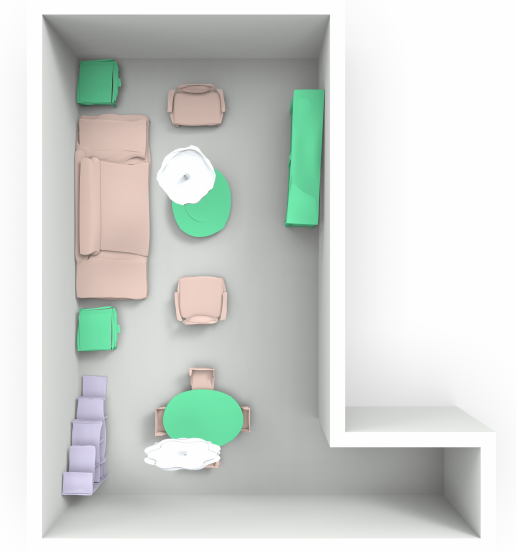}}
    \caption{\yjr{\titlecap{Ablation Study on Key Structural Designs of the Scene Hierarchy.}{We perform ablation studies that respectively remove the introduced functional region level and each type of relation edges. It is clear that: in (b), the version without functional regions (FR) drops some objects; in (c), the version without room-object edges (ROE) predicts misaligned objects; in (d), the version without object-object edges (OOE) struggles when objects need to be placed symmetrically; and (e), the version without hyper-edges (HE) fails to place parallel objects. In contrast, our full method does not have any of these problems.}}}\vspace{-2mm}
    \label{fig:abla-edge}
\end{figure*}

\begin{table}[t!]
  \centering
  \caption{\yj{\titlecap{Quantitative Scene Reconstruction Performance of the Ablation Studies.}{We show the CD($\times 10^{-5}$) and EMD($\times 10^{-4}$) metrics of the reconstruction results and FID, $o_{1}, o_{2}, o_{3}$ metrics of the generation results from our full method and the ablated versions, each of which is trained without a certain element of our key scene hierarchy designs. We can see from the table that the performance of our full method is the best.}}}
  \begin{adjustbox}{width={0.47\textwidth}, keepaspectratio}
    \begin{tabular}{ccccccc}
    \toprule[1pt]
    \multirow{2}[2]{*}{Methods} & \multicolumn{2}{c}{Reconstruction} & \multicolumn{4}{c}{Generation} \\
          & CD & EMD & FID   & $o_1$   & $o_2$    & $o_3$ \\
    \midrule
    separate training      & 315.2 & 423.5 & 139.450 &    0.065   &      0.126     &  1.348 \\
    end-to-end training    & 473.9 & 676.8 & 170.079 &    0.119   &      0.140     &  1.344 \\
    w/o object-object edge & 321.3 & 440.8 & 145.398 &    0.119   &      0.137     &  1.625 \\
    w/o room-object edge   & 344.9 & 470.9 & 150.478 &    0.069   &      0.131     &  1.357 \\
    w/o hyper-edge         & 315.8 & 443.5 & 130.276 &    0.080   &      0.126     &  1.553 \\
    w/o functional region  & 398.1 & 563.3 & 135.940 &    0.096   &      0.140     &  1.595 \\
    Ours (Full)            & 310.7 & 389.0 & 106.005 &    0.050   &      0.130     &  0.470 \\
    \bottomrule[1pt]
    \end{tabular}%
  \end{adjustbox}
  \label{tab:abla-num}%
\end{table}%

\subsection{Ablation Study}\label{sec:abla}
We introduce functional regions and many node edges, including the proposed hyper-edges, in our method. To show that these elements are indeed beneficial, we perform a set of ablation studies. For each ablation study, we remove a certain element from our method, use this version to reconstruct 3D indoor scenes, and compare its performance to our full method. We also validate our training strategy. Again, we use Chamfer Distance and Earth Mover's Distance to measure the quality of reconstruction, \yj{and use FID, $o_{1}, o_{2}, o_{3}$ to measure the ability of generation}. We show the quantitative evaluations in \autoref{tab:abla-num}, where we can see clearly that all of the ablated versions perform worse than our full method, especially for the distribution $o_{3}$ of co-occurrence of two furniture for each room, there is a very large margin compared to other ablated versions.

\textbf{Structures in Hierarchical Graph Representation.}
We show the reconstruction results of our method without a certain structure in the hierarchical graph representation in \autoref{fig:abla-edge}. 
We consider some key structures in the hierarchy: functional regions, room-object edges, object-object edges, and hyper-edges.
We can see that removing each of them introduces some specific flaws in the reconstruction results. 

As we can observe from \autoref{fig:abla-edge}, without functional regions, the network fails to predict some objects because the hierarchy is missing a whole level of nodes, and is not organized tightly. Without room-object edges to keep the objects aligned, we see that the chairs and the tables are often misaligned. Without object-object edges to preserve the symmetrical relation between objects, the chairs around the table (which are in rotational or reflective symmetry) fail to preserve the symmetrical relation. Without hyper-edges, we also find that the rotational symmetry fails to preserve, and in addition we observe that the parallel relationship breaks in the top image of the fifth column of \autoref{fig:abla-edge}. Our full method helps address the above issues.

\begin{figure}[t!]
    \centering
    \includegraphics[width=0.24\linewidth]{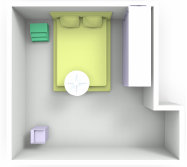}
    \includegraphics[width=0.24\linewidth]{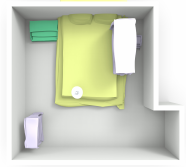}
    \includegraphics[width=0.24\linewidth]{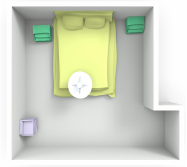}
    \includegraphics[width=0.24\linewidth]{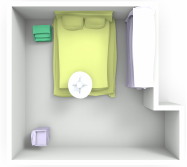}
    \\
    \subfigure[]{\includegraphics[width=0.24\linewidth]{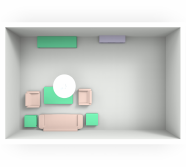}}
    \subfigure[]{\includegraphics[width=0.24\linewidth]{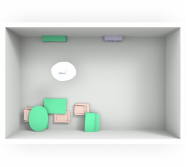}}
    \subfigure[]{\includegraphics[width=0.24\linewidth]{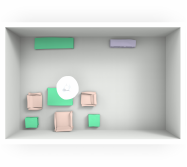}}
    \subfigure[]{\includegraphics[width=0.24\linewidth]{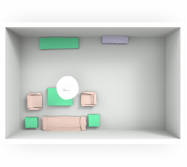}}
    \vspace{-2mm}
    \caption{\yj{\titlecap{Training Strategy.}{We compare 3 training methods: end-to-end, separate and two-stage training on reconstruction results. (a) is the input scene, (b) is reconstruction via end-to-end training of total networks, (c) is the reconstruction via separate training (i.e., first train object-part network, and then train room-layout network), (d) is reconstruction via two-stage training with fine-tuning the object-to-part networks. All networks have converged, and the results of two-stage training are much better than the others. Compared to separate training (c), our full model can achieve reasonable and realistic results due to fine-tuning the object-part networks, which optimizes the relation and geometry between objects.}}}\vspace{-2mm}
    \label{fig:ablae2e}
\end{figure}

\textbf{Training Strategy.}
\yj{When training our network, we use two-stage training. More precisely, we first train the networks that encode and decode the part-to-object $Enc_{p2o}, Dec_{o2p}$. 
Then, we start training the two other networks $Enc_{o2r}, Dec_{r2o}$ while fine-tuning the $Enc_{p2o}, Dec_{o2p}$ networks. 
We find this training strategy gives better results than training the four networks from scratch jointly or separately training the room-object and object-part networks. 
This is also intuitively reasonable as the object-part networks and the room-object networks are relatively entangled to learn the consistency between the objects within a functional region and performing a joint training from scratch makes the network too deep to be effectively trained.
\autoref{fig:ablae2e} shows some qualitative result comparisons, the end-to-end training means that we train the whole networks without fine-tuning object-to-part networks, the separate training means that we first train the object-part network and then train the room-object network. The losses of both networks in \autoref{fig:ablae2e} have converged, and we can conclude from the images that the layouts and the geometry of the two-stage training reconstruction results are considerably better and more realistic than the others.}

%% file: sec/7Conclusion.tex
\section{Conclusion, Limitations and Future Work}
\label{sec:conclusion}

\begin{figure}[t!]
    \centering
    \subfigure[]{\includegraphics[width=0.15\linewidth]{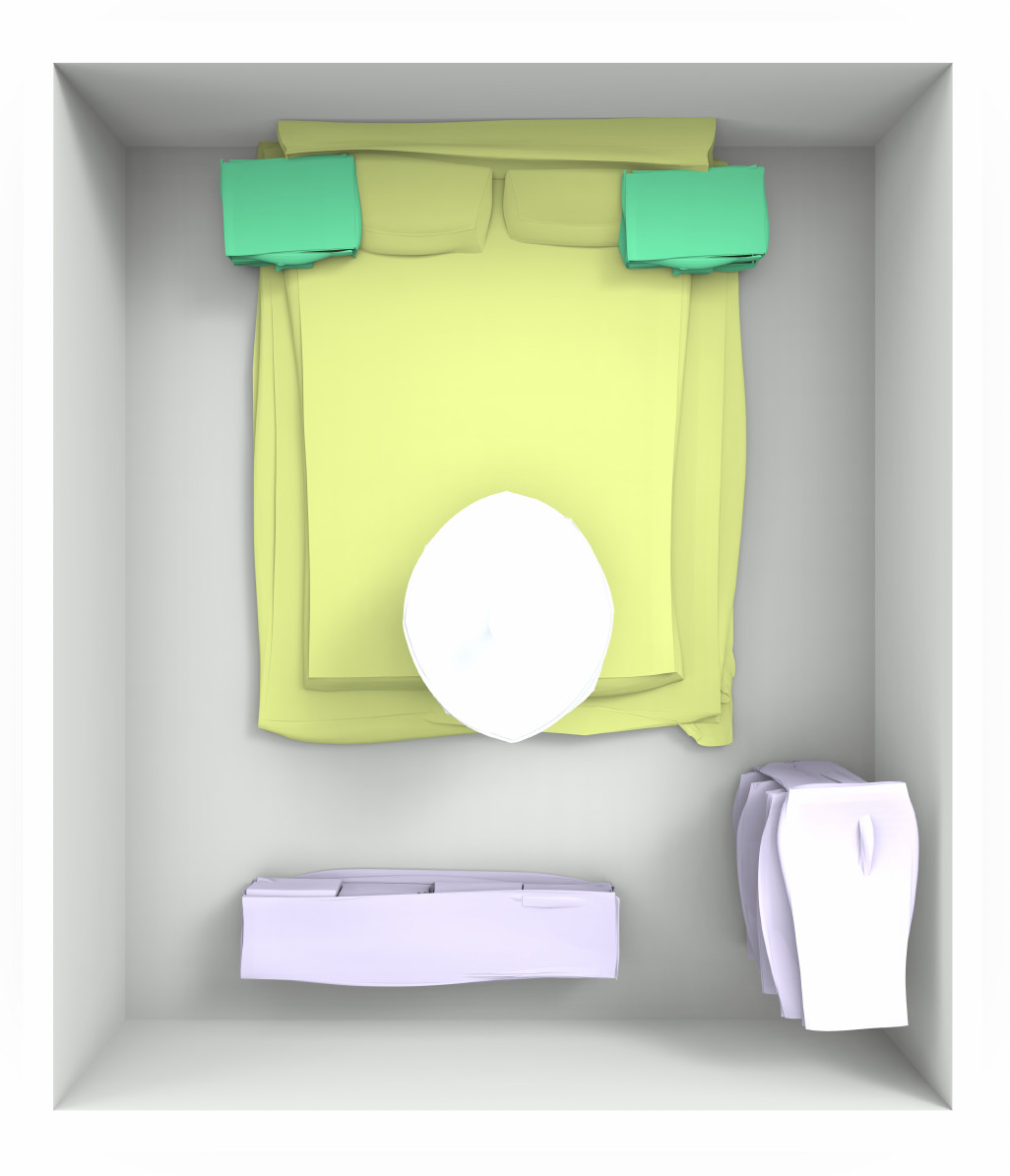}}
    \subfigure[]{\includegraphics[width=0.2\linewidth, angle=90]{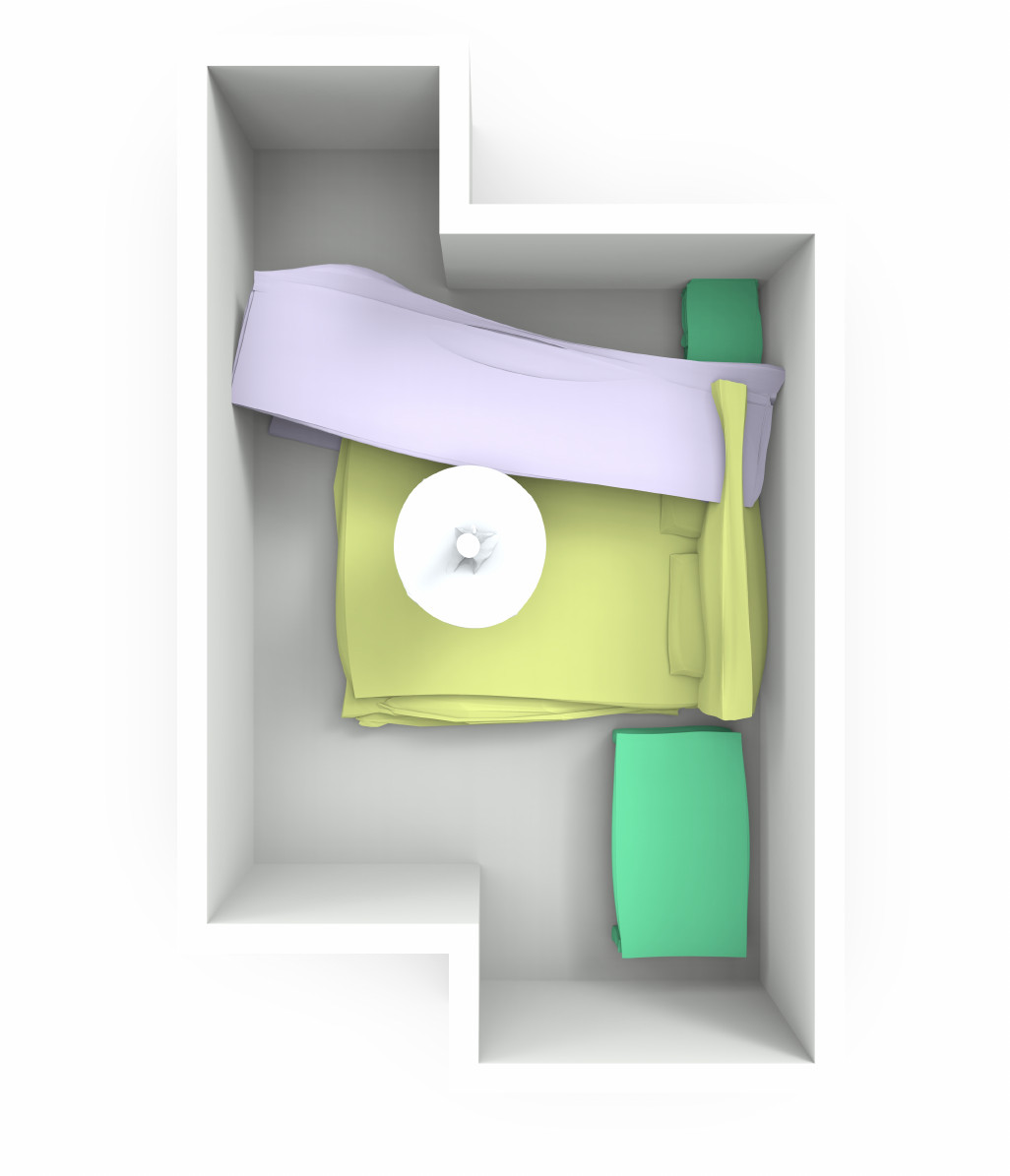}}
    \subfigure[]{\includegraphics[width=0.15\linewidth]{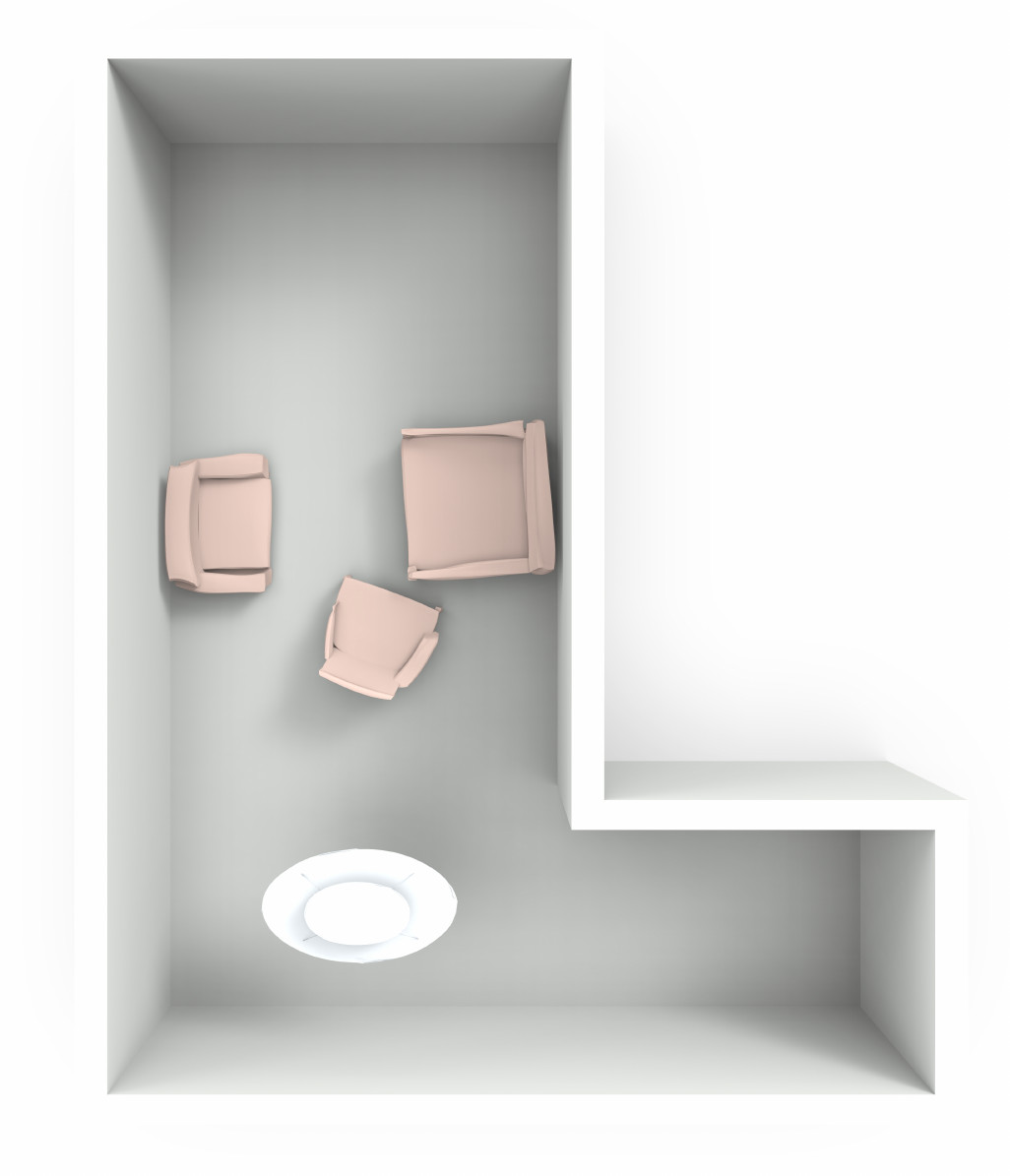}}
    \subfigure[]{\includegraphics[width=0.185\linewidth,angle=90]{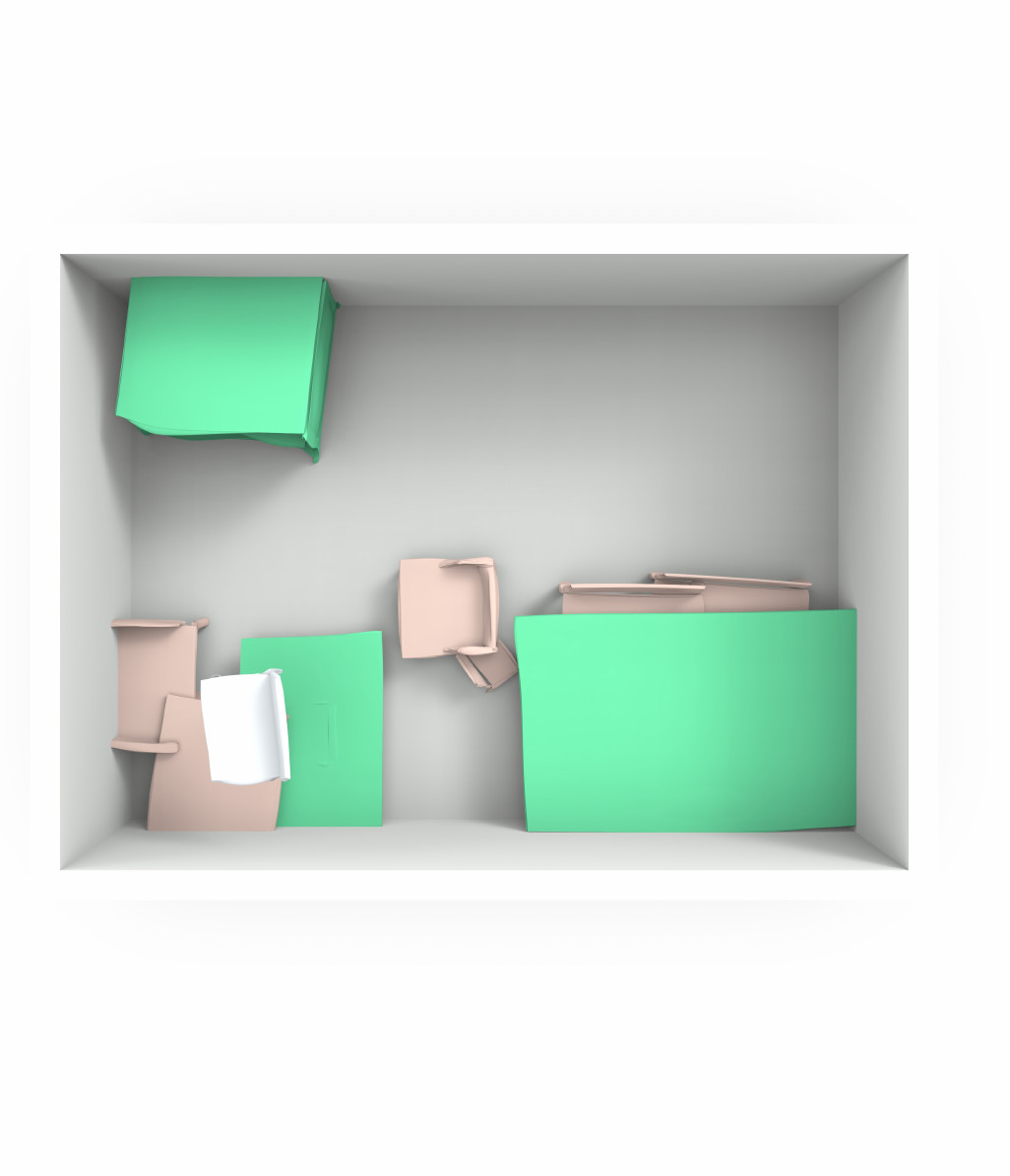}}
    \subfigure[]{\includegraphics[width=0.165\linewidth]{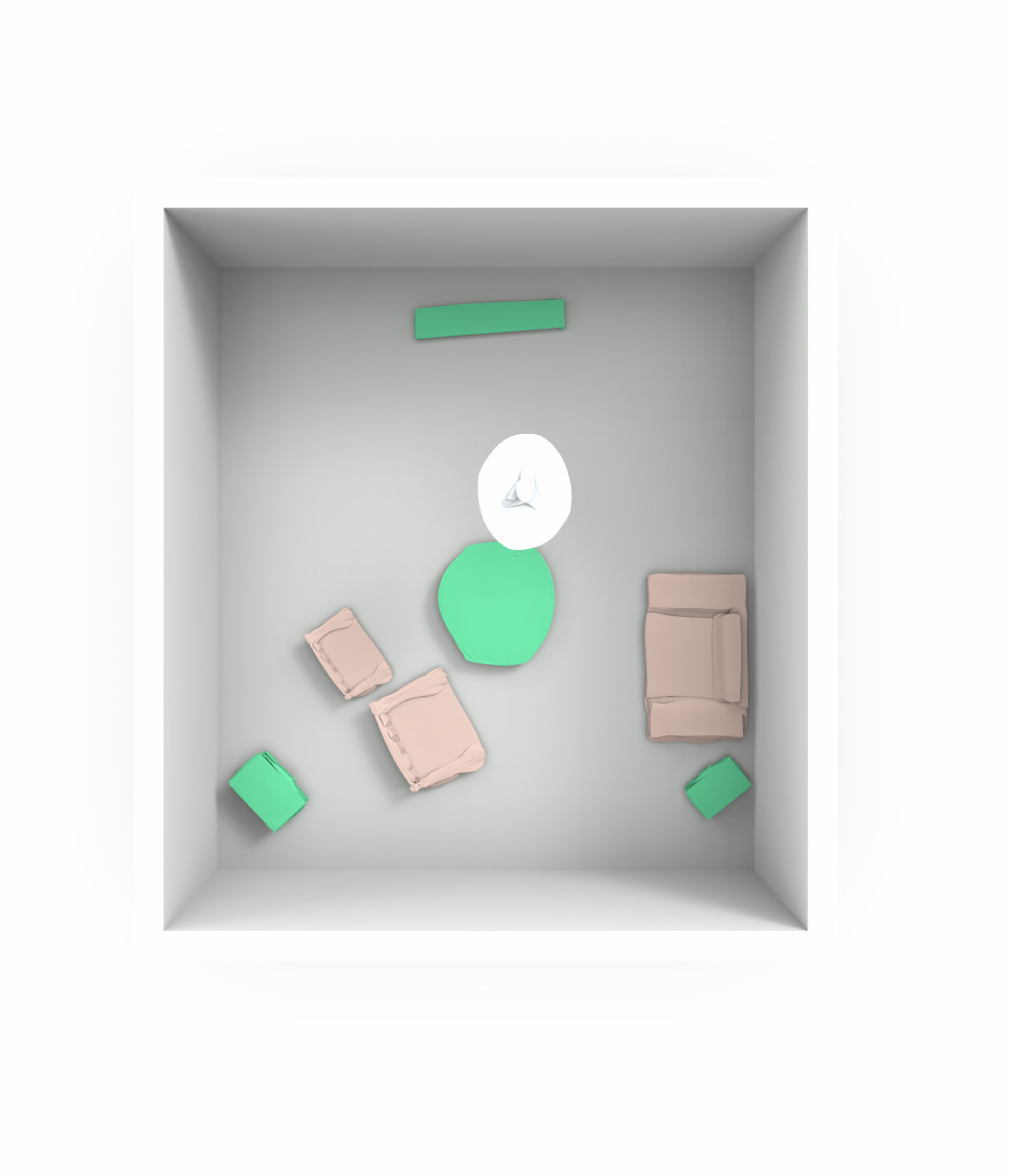}}
    \vspace{-2mm}
    \caption{\titlecap{Failure Cases.}{We present exemplar failure cases of our scene generation results.
    We may see some problematic prediction results, such as the invalid intersection between objects (a), some duplicate or missing objects in the scene ((b), (c)), prediction errors for the semantic categories of objects (\eg a chair is predicted to be a lamp and attached to the ceiling in (d)), or unaligned object orientations with the room boundary walls (e).}}\vspace{-3mm}
    \label{fig:failure}
\end{figure}

We propose a hierarchical graph network \name on 3D indoor scene generation.
Our method conducts learning over a structural scene hierarchy containing multiple conceptual levels of entities: the room, functional regions, objects, and object parts.
We train a recursive variational autoencoder that learns to map 3D scene hierarchies to a latent manifold, on which we can generate diverse 3D scene meshes by randomly sampling over the learned distribution, interpolate smoothly between input 3D scene data, and perform many downstream applications, such as scene editing, conditional generation, and completion tasks.
At the core of our innovations, we propose a level of functional regions between the room root node and its constituent objects for more effective and efficient learning and devise a rich set of binary edges and $n$-ary hyper-edges among the nodes in the hierarchy for better modeling the structural and relational constraints for a valid 3D scene generation.
We conducted extensive evaluations and comparisons to strong state-of-the-art methods, demonstrating our superior performance.

\textbf{Limitations and Future Works.}
There are some limitations of our \name networks: a) our networks are basically VAEs, and naturally need a large amount of data of high consistency and quality for training; \yjr{b) the construction of our hierarchical graph representation heavily relies on annotated data such as object part hierarchy, especially for some small objects (\eg cups, vases, books), more labeling is required to allow our approach to be easily extended to more object types. Also, our framework is not designed for the vertical edges within object-level, \eg supporting, etc., which is limited for representing complex structures on the $z$-axis. But our model is easily generalizable to this design via adding an extra edge, which can be our future feature as deep exploration along the direction}; 
c) our introduced edges and functional regions can successfully model most of the layouts in the 3D-FRONT dataset, but we cannot guarantee the structures are applicable to all scene data; d) we use non-rigid registration and ACAP feature to encode the detailed geometry for the room boundary layout and the object part geometry, which do not always provide realistic model generation in our results;
e) doors and windows are often an important part of a complete room. We use the deformation to represent the room boundary. Since doors and windows are located on the wall, they are not compatible with our framework. Our proposed framework needs to be extended to handle such relationships for windows and doors, which should be achievable from a technical point of view.
\yjr{f) our network is not able to generate photorealistic rooms since the texture is not considered. So it would be interesting as future work to align the PartNet hierarchy with 3D-FRONT models, using the former as structure supervision and the latter for detailed geometry and texture.}

We show the several failure cases of our method in \autoref{fig:failure}. Since we do not explicitly discourage collision between objects, the current approach may generate objects with intersection among them in some cases, \eg \autoref{fig:failure} (a). Some points in our latent space may be mapped to generate imperfect scenes with missing/duplicate objects or incorrect semantic labels.
See \autoref{fig:failure} (b, c, d) and the figure caption for more detailed explanations. 
Lastly, although we introduce room-object edges to align objects with rooms, there would still be some occasional failure cases, as presented in \autoref{fig:failure} (e). 
Future works may address these issues and thus further improve the 3D scene generation performance of our framework.

%% file: sec/supp_1data.tex
\section{Dataset Preparation}
\label{sec:dataset}

In 3D-Front~\cite{fu20203dfront}, there are 6,815 houses and 51,708 rooms, a newly released dataset of 3D indoor scenes. 
They were created by professional designers directly. 
The dataset contains several rooms per house with a room type associated with each one. 
However, there are 18,797 rooms furnished with objects from 3D-FUTURE~\cite{fu20203dfuture}, a dataset of textured 3D furniture models, and each model is labeled with its furniture category.
\yj{Hence, each room is only segmented at the object level.} 
\yjr{For the hierarchical structure of indoor scenes, we limit each parent node to a maximum of 10 child nodes, which is enough to describe many complex scenes. In the 3D-Front dataset, there are more than 100 furniture in very few scenes (12). Therefore, we have cleaned the data, which finally contains 14654 indoor scenes.}

\begin{figure}[!htbp]
\centering
    \includegraphics[width=0.99\linewidth]{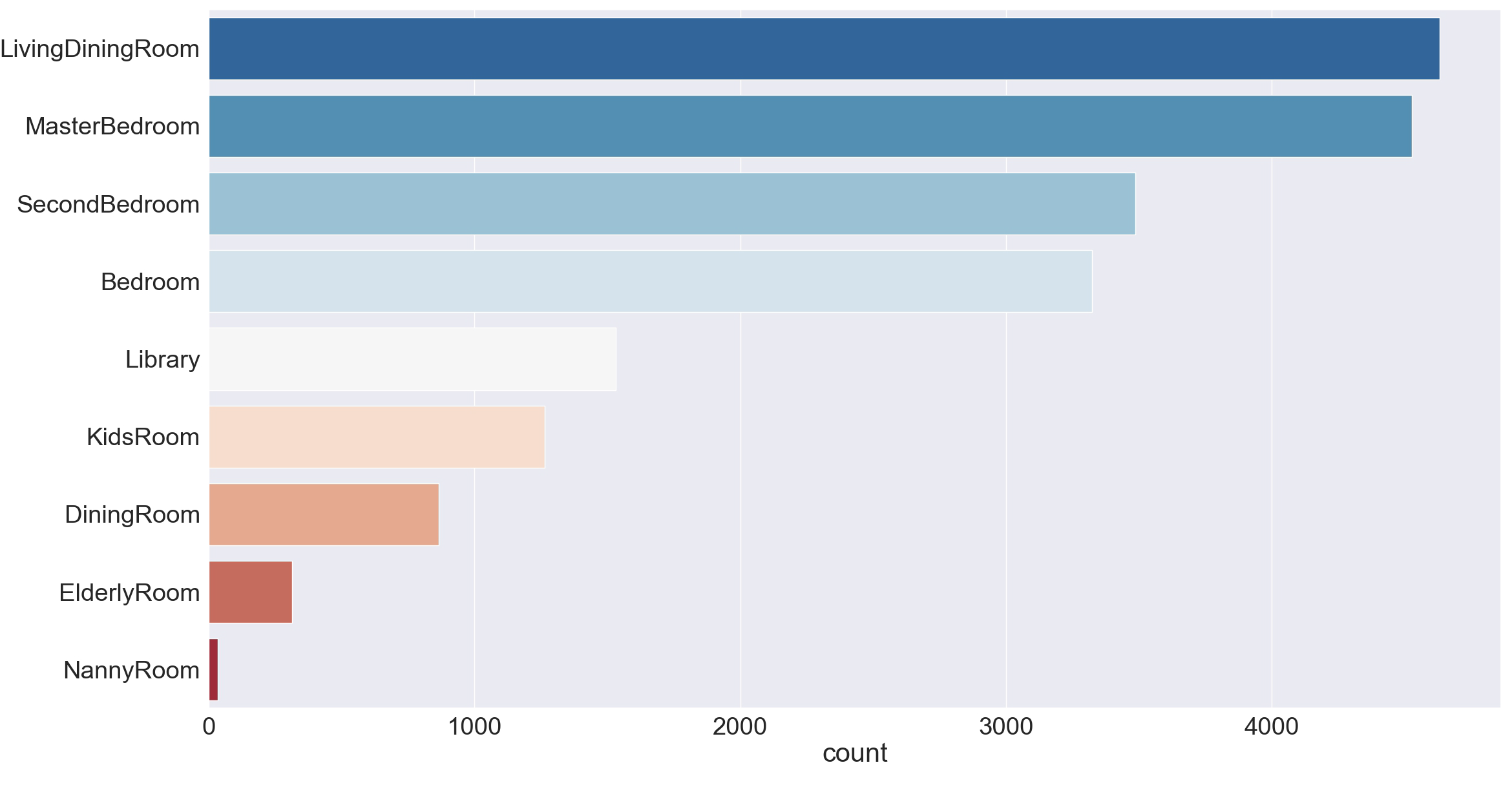}
    \caption{\titlecap{3D-FRONT room type distribution.}{We plot the distribution of room types in the original 3D-FRONT dataset. From the histogram, we see that most of the rooms are living rooms and bedrooms. Note that there are some room types excluded because there is no furniture in these types of rooms.}}
    \label{fig:frontroomtype}
\end{figure}

The distribution of room types in 3D-FRONT is shown in \autoref{fig:frontroomtype}. Note that other room types (\eg balcony, toilet) are excluded because they do not have any furniture.
\begin{figure}[!htbp]
    \subfigure[Original 3D-FRONT]{\includegraphics[width=0.4\linewidth, angle=90]{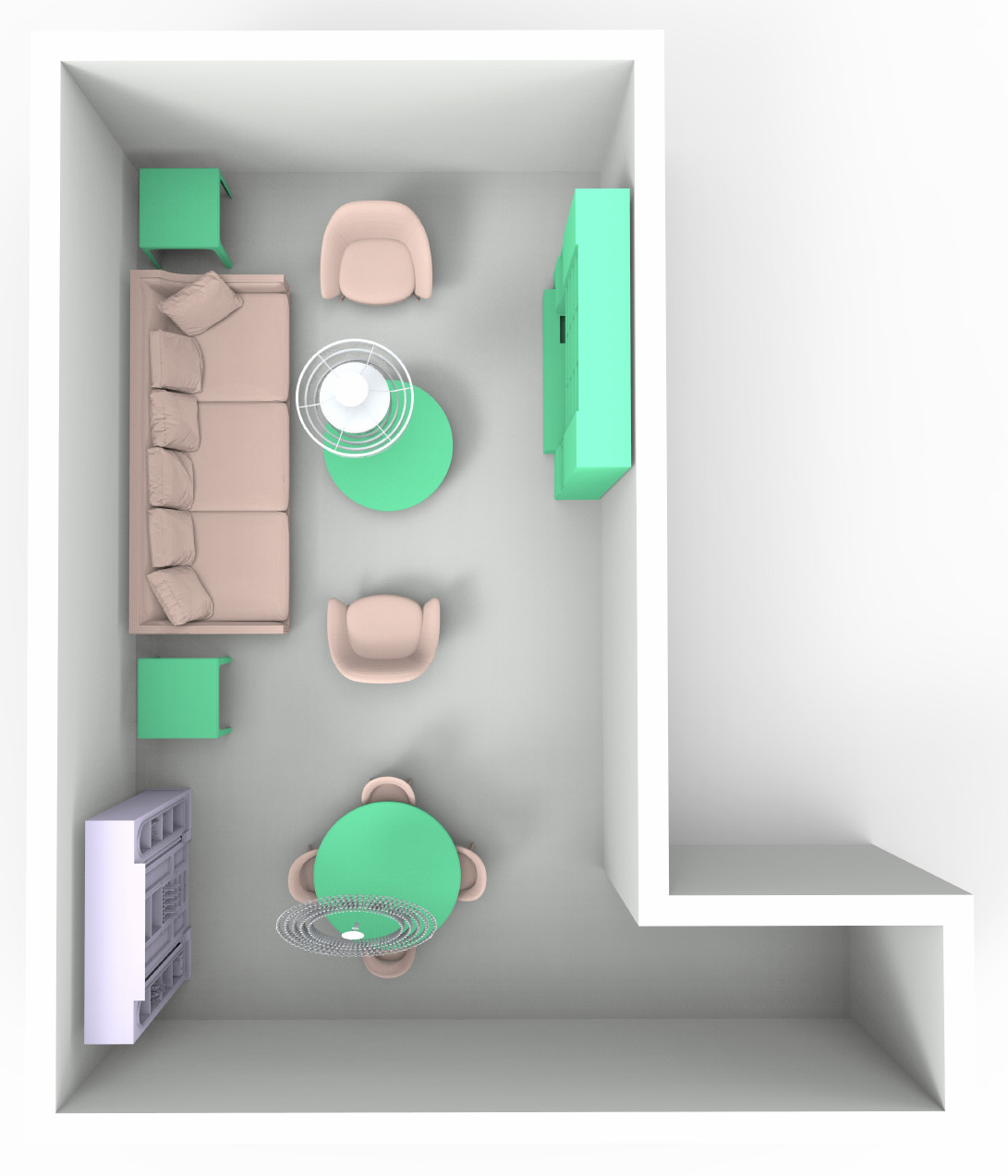}}
    \subfigure[Replaced 3D-FRONT]{\includegraphics[width=0.4\linewidth, angle=90]{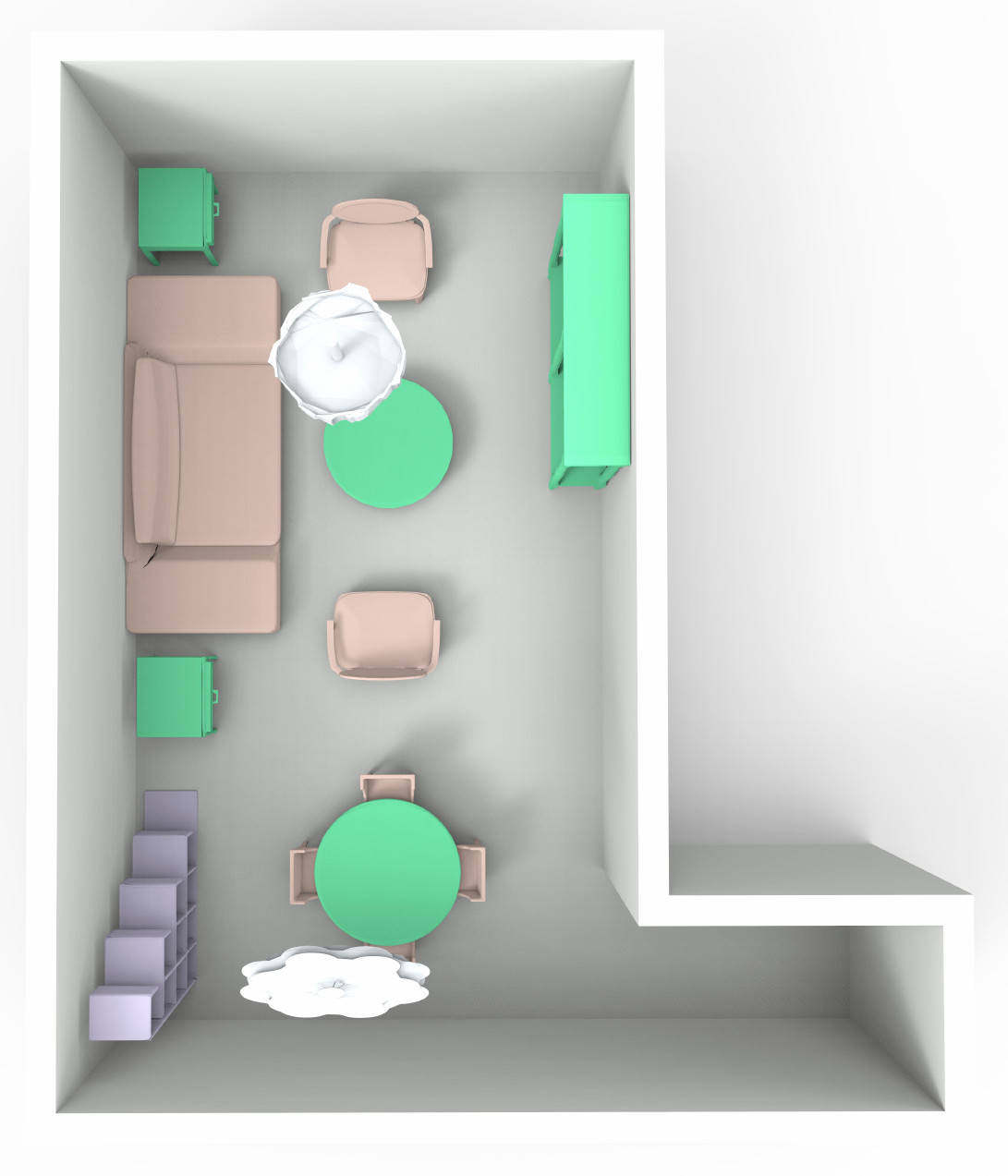}}
    \caption{\titlecap{Furniture Replacement for 3D-FUTURE.}{We decide to replace 3D-FUTURE~\cite{fu20203dfuture} objects with PartNet~\cite{mo2019partnet} objects to ensure objects contain consistent part-level annotations. The criteria for selecting substitute models are based on semantic category and Chamfer Distance. The figure on the left shows the original scene and the one on the right shows the scene after the mesh replacement. We can see the scene still looks good after replacement.}}
\end{figure}

\begin{table*}[!t]
  \centering
  \caption{Data Statistics. All the training and testing samples are split with a ratio of 4:1.}
  \begin{adjustbox}{width={\textwidth},keepaspectratio}
    \begin{tabular}{c|ccc|cccccc|c|c}
    \toprule[1pt]
    \multirow{2}[4]{*}{Category} & \multicolumn{3}{c|}{LivingDingingRoom} & \multicolumn{6}{c|}{BedRoom}                  & Library & \multirow{2}[4]{*}{Total} \\
\cmidrule{2-11}          & LivingDingingRoom & LivingRoom & DiningRoom & ElderlyRoom & Bedroom & KidsRoom & MasterBedroom & SecondBedroom & NannyRoom & Library &  \\
    \midrule
    \#Training & 1829  & 892   & 463   & 199   & 1533  & 809   & 3012  & 2094  & 13    & 900   & 11744 \\
    \#Testing & 446   & 217   & 114   & 55    & 382   & 199   & 755   & 522   & 3     & 217   & 2910 \\
    \bottomrule[1pt]
    \end{tabular}%
    \end{adjustbox}
  \label{tab:datastats}%
\end{table*}%

\subsection{Mesh replacement of 3D-FUTURE.}
Due to the hierarchical data representation we use, we need all the furniture models to be segmented into parts with consistent semantic labels. But many of the 3D-FUTURE models cannot satisfy this condition (\eg a table and a vase on it are modeled together and labeled as `table', the back surface of a pillow on a bed is missing because it can not be seen, etc.) So we decide to replace the 3D-FUTURE models with PartNet models~\cite{mo2019partnet}. PartNet has a total of 26,671 3D objects from 24 object categories, while 3D-FUTURE has 9,992 objects from 34 categories. All the categories in the 3D-FUTURE dataset can be mapped into one in PartNet. Besides, PartNet provides detailed part segmentation and hierarchical decomposition of objects, which is exactly what we need.

The replacement is performed when two conditions are met:
\begin{itemize}
    \item The substitute model and the original model are of the same category.
    \item The substitute model has a minimum Chamfer distance from the original model. 
\end{itemize}

\yj{Note that all shapes from the two datasets are aligned for each category and scaled into a unit sphere by its scaling factor respectively, making Chamfer distance measure meaningful.} We find this simple strategy can produce fairly reasonable replacement results.

\subsection{Extraction of Functional Regions.}
Functional regions are key components in our hierarchical graph representation, but it does not exist in the dataset. To extract them, we have two key observations:
\begin{itemize}
    \item All the objects in a functional region must gather in a small area.
    \item A functional region must have a key object that decides the function of this region, and the key object is often the largest (\eg A region for dining always has a big table and some chairs surrounding it).
\end{itemize}

Based on these observations, we sample 10,000 points on the surface of each object forming a point cloud and use the DB-Scan~\cite{DBScan} algorithm on it to find clusters of objects. For each cluster, we assign a semantic label according to its largest object. After processing the dataset, we find only 5 types of the functional regions: Living\_region, Dining\_region, Office\_region, Ceil\_region and Cabinet\_region. 
\yj{Note that we have manually checked all functional regions and corrected some clustering results for reasonable functional clustering.}

\subsection{Extraction of Hyper-edges.}
Like functional regions, we also need to extract hyper-edges from the dataset. 
To extract parallel hyper-edges, we start with a set of sets $\mathbf{O_{He}} = \{\{O_1\},\cdots,\{O_N\}\}$. For each pair of sets in $\mathbf{O_{He}}$, we merge them if and only if the loss $\mathcal{L}_{\mathbf{e}^{hyper}_2}$ of the combined set is not greater than a constant threshold. We perform the merging process until there are no new pairs to merge.

To extract N-fold rotational symmetry hyper-edges, we start with existing regular rotational symmetry pairs of objects. we merge rotational symmetry pairs if the distance between their rotation centers is not greater than a constant threshold. Here we do not include the condition of rotation angles (they exist in the definition of hyper-edges) because we observe that if multiple objects in the dataset share a rotation center, they are very likely to be in an N-fold rotational hyper-edge.

\section{Implementation Details}
\label{sec:implementation}

Our network is implemented by PyTorch~\cite{paszke2017automatic}. The whole framework is borrowed from DSG-Net~\cite{yang2020dsm}. Our network is trained in a two-stage manner, where the network trained in the first stage is also fine-tuned during the second stage. 
The training process is optimized by Adam solver~\cite{kingma2014adam}, and the learnable parameters are initialized randomly with Gaussian Distribution. 
We adopt the batch size of 64 and use a learning rate starting from 1e-3 and decaying every 2000 steps with a decay rate of 0.9 until loss converges with 500 epochs. The weights of energy terms are following DSG-Net~\cite{yang2020dsm}.
Besides that, our network does not apply the BatchNorm~\cite{ioffe2015batch} and consists of linear layers and graph convolutions. 
Due to the hierarchy of indoor scenes can be very different, we use the same training strategy as StructureNet/DSG-Net that forwards each scene sequentially and backward with a batch.
All the processed data is randomly split into the training set and test set with a ratio of 4:1 for each room type, the detailed data statistics are listed in Table~\ref{tab:datastats}.
All our experiments are carried out on a PC with a single GeForce RTX 2080Ti GPU and an Intel i9-9900K CPU.
\yjr{With the pretrained DSG-Net for each object category, it takes about 7 days for our network to converge. Once trained, the generation process would take some time from a single latent code since the whole indoor scene includes more geometric details and complex structures, and it takes about 5025ms on average for generating a room. We split the time into two parts: runtime of latent code to all node features and runtime of leaf node features to concrete geometries.
We report the average runtimes on 100 scenes as 203ms and 4822ms for these two steps respectively.
}
Our code and dataset where each scene is represented as a hierarchical tree structure will be released.

%% file: sec/supp_2recon.tex
\section{Scene Reconstruction}\label{sec:recon}
\begin{figure*}[h]
    \centering
    \subfigure[]{
    \begin{minipage}[b]{0.24\linewidth}
    {
    \includegraphics[width=0.48\linewidth]{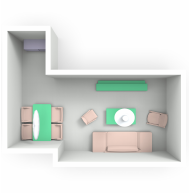}
    \includegraphics[width=0.48\linewidth]{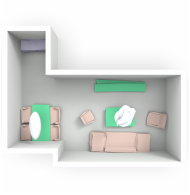}
    \includegraphics[width=0.48\linewidth]{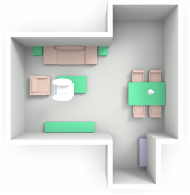}
    \includegraphics[width=0.48\linewidth]{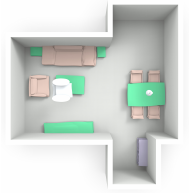}
    }
    \end{minipage} }
    \subfigure[]{
    \begin{minipage}[b]{0.24\linewidth}
    {
    \includegraphics[width=0.48\linewidth]{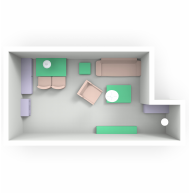}
    \includegraphics[width=0.48\linewidth]{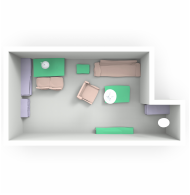}
    \includegraphics[width=0.48\linewidth]{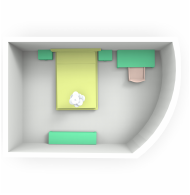}
    \includegraphics[width=0.48\linewidth]{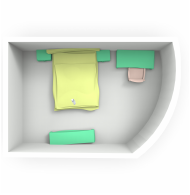}
    }
    \end{minipage}}
    \subfigure[]{
    \begin{minipage}[b]{0.24\linewidth}
    {
    \includegraphics[width=0.48\linewidth]{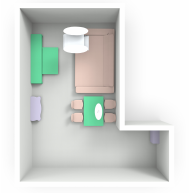}
    \includegraphics[width=0.48\linewidth]{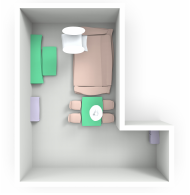}
    \includegraphics[width=0.48\linewidth]{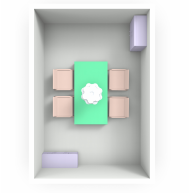}
    \includegraphics[width=0.48\linewidth]{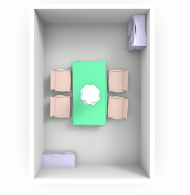}
    }
    \end{minipage}}
    \subfigure[]{
    \begin{minipage}[b]{0.24\linewidth}
    {
    \includegraphics[width=0.475\linewidth,angle=90]{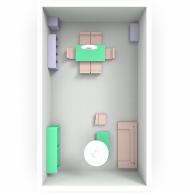}
    \includegraphics[width=0.475\linewidth,angle=90]{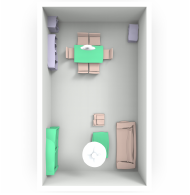}
    \includegraphics[width=0.48\linewidth]{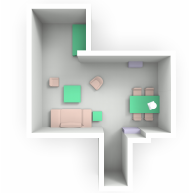}
    \includegraphics[width=0.48\linewidth]{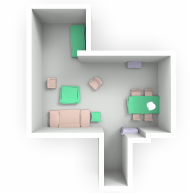}
    }
    \end{minipage}}
    \caption{\titlecap{The gallery of shape reconstruction results.}{For each sub-figure of results, the left column shows the ground-truth target, and the right column presents our reconstruction result. We observe that our method can capture complex shape structures and detailed part geometry at the same time.}
    }
    \label{fig:reconstruction}
\end{figure*}

Firstly, we perform an experiment of reconstructing 3D indoor scenes with our network. This experiment is capable of proving the RvNN-VAE network can efficiently learn the features of our hierarchical graph representation and create a correct latent space. In this experiment, we modify our network to be an autoencoder rather than a variational autoencoder.
Some of the reconstruction results are shown in \autoref{fig:reconstruction}.

We use Chamfer Distance and \textit{Earth Mover's} Distance to measure the error of the reconstruction results. We first normalize the reconstruction and the ground truth models, sample a certain number of points on their surfaces, and compute the distances between these two point clouds. For Chamfer Distance, we sample 100,000 points, and for Earth Mover's Distance, we sample 10,000 points.

\subsection{Comparison with implicit-based methods and GRAINS.}
Occupancy Network~\cite{mescheder2019occupancy} and ConvONet~\cite{peng2020convolutional} are two representative methods for reconstruction. They learn a function that maps an arbitrary point in 3D space to an occupancy probability between 0 and 1, and can be used for 3D reconstruction of an object based on observations of that object (\eg, image, point cloud, etc.).
For the capability of representation, ConvONet uses a 3D feature volume ($64^3\times$32-d) to store the 3D scene, but ours only uses a single latent code (256-d) to represent the 3D scene, which has a huge difference in the capability of representation. Hence, it is not possible to make a fair comparison, and we choose Occ-Net (which also represents a shape as a single code) as the baseline to compare with the implicit-based method.
Since the original Occ-Net is trained on the ShapeNet, we retrain the Occ-Net on 3D-FRONT until convergence for a fair comparison.

GRAINS~\cite{li2019grains} provides an RvNN-VAE like our method. Theoretically, it can be used for reconstruction, but its representation is based on bounding boxes, and can only retrieve 3D models from a database using the predicted bounding boxes. Thus we cannot expect it to produce high-quality results. We include GRAINS here only for reference.

\begin{table}[!htbp] 
\centering
\caption{\titlecap{Reconstruction Error Comparison.}{We compute Chamfer Distance and Earth Mover's Distance between reconstruction results of each method and the ground truth. From the results, we can conclude that our method has better reconstruction performance than the two baselines, proving that the latent space our network learns can represent the 3D indoor scenes in the dataset.}}
\label{tab:comreconstruction}
\begin{tabular}{cccc}
\toprule[1pt]
Methods & OCC-Net\cite{mescheder2019occupancy} & GRAINS~\cite{li2019grains} & Ours \\
\midrule
CD ($\times 10^{-5}$) & 594.8 & 2490.1 & 310.7 \\
EMD ($\times 10^{-4}$) & 1747.3 & 2252.0 &  389.0 \\
\bottomrule[1pt]
\end{tabular}
\end{table}

\begin{figure}[t]
    \centering
    \includegraphics[width=0.24\linewidth]{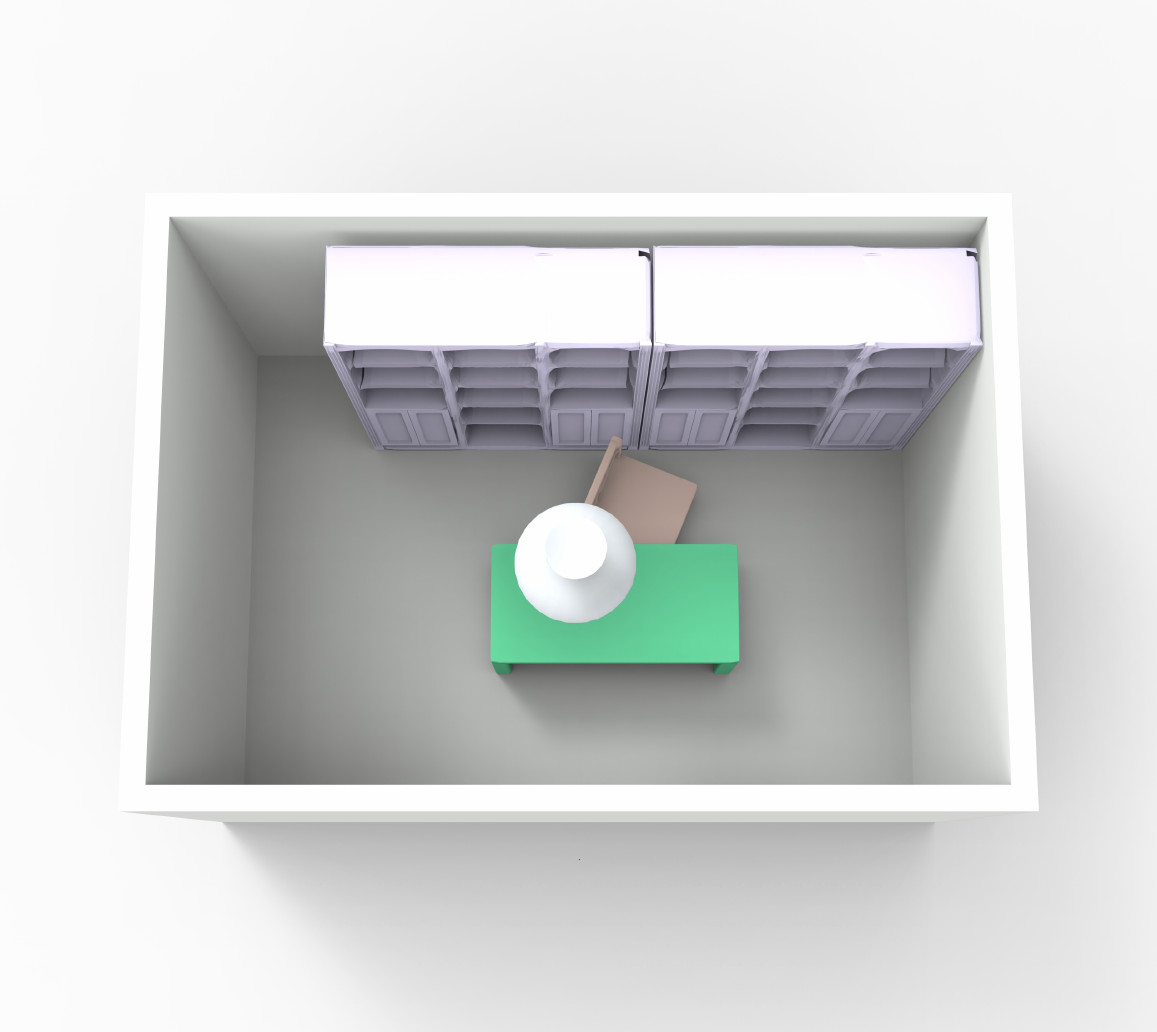}
    \includegraphics[width=0.24\linewidth]{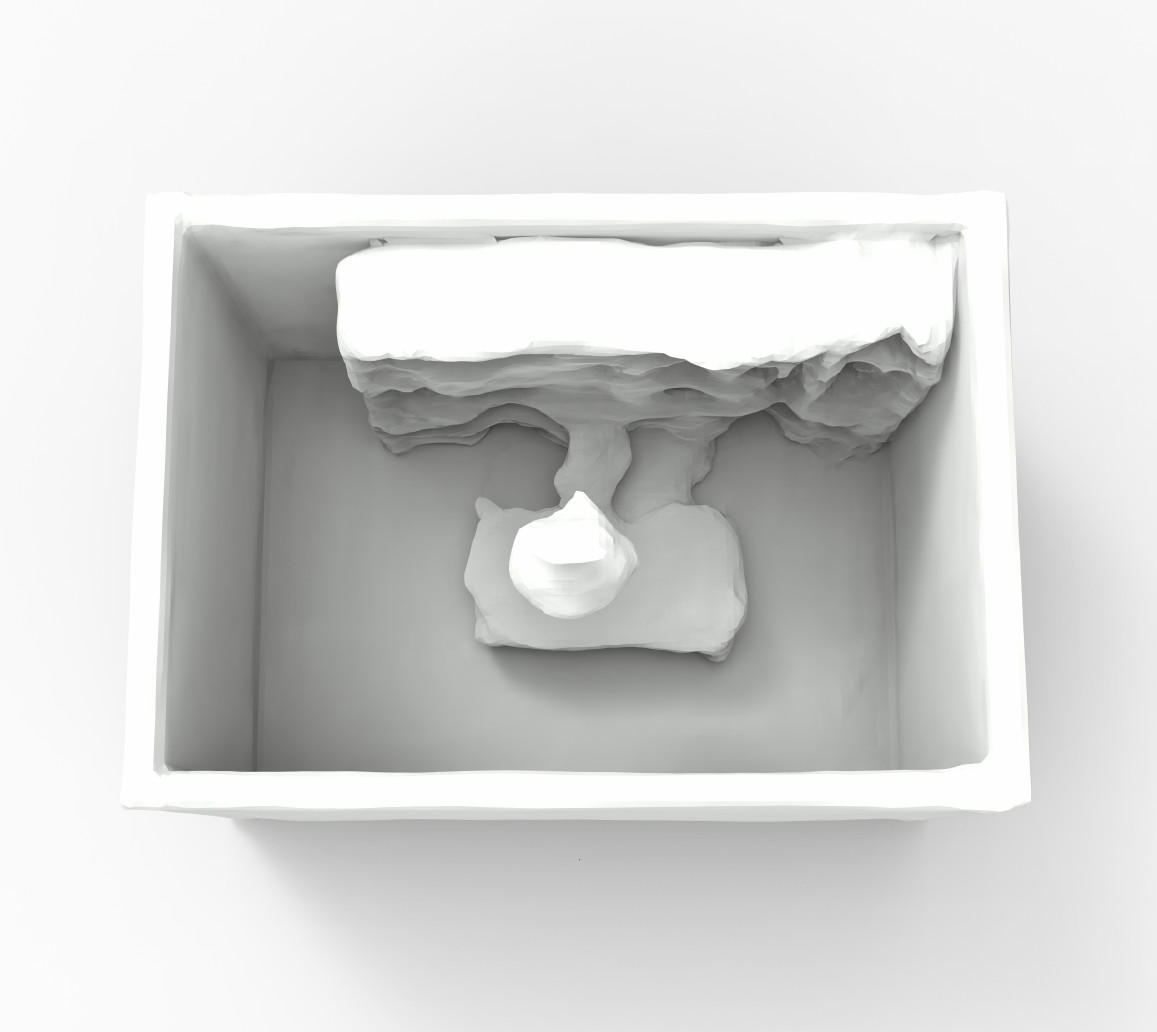}
    \includegraphics[width=0.24\linewidth]{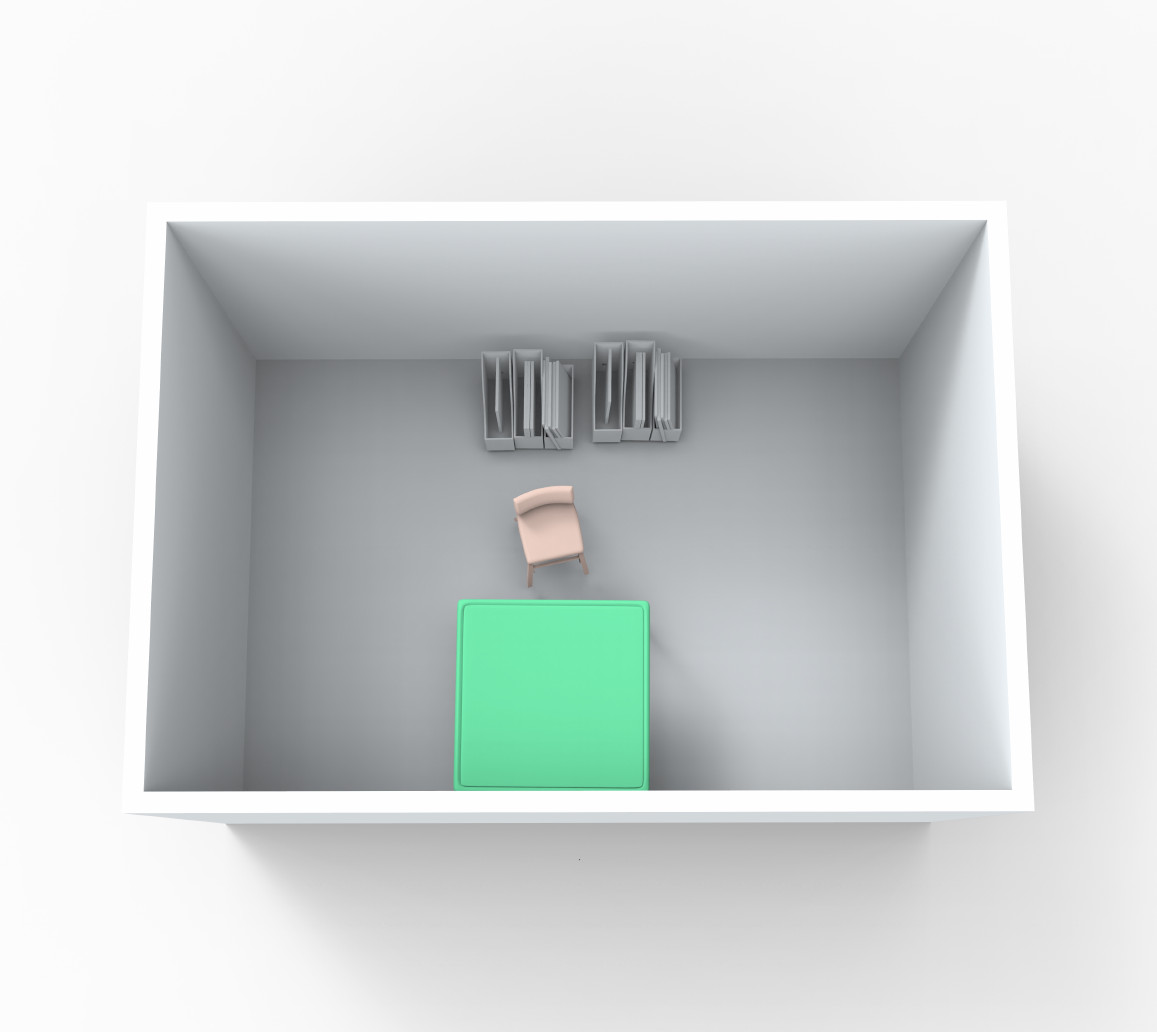}
    \includegraphics[width=0.24\linewidth]{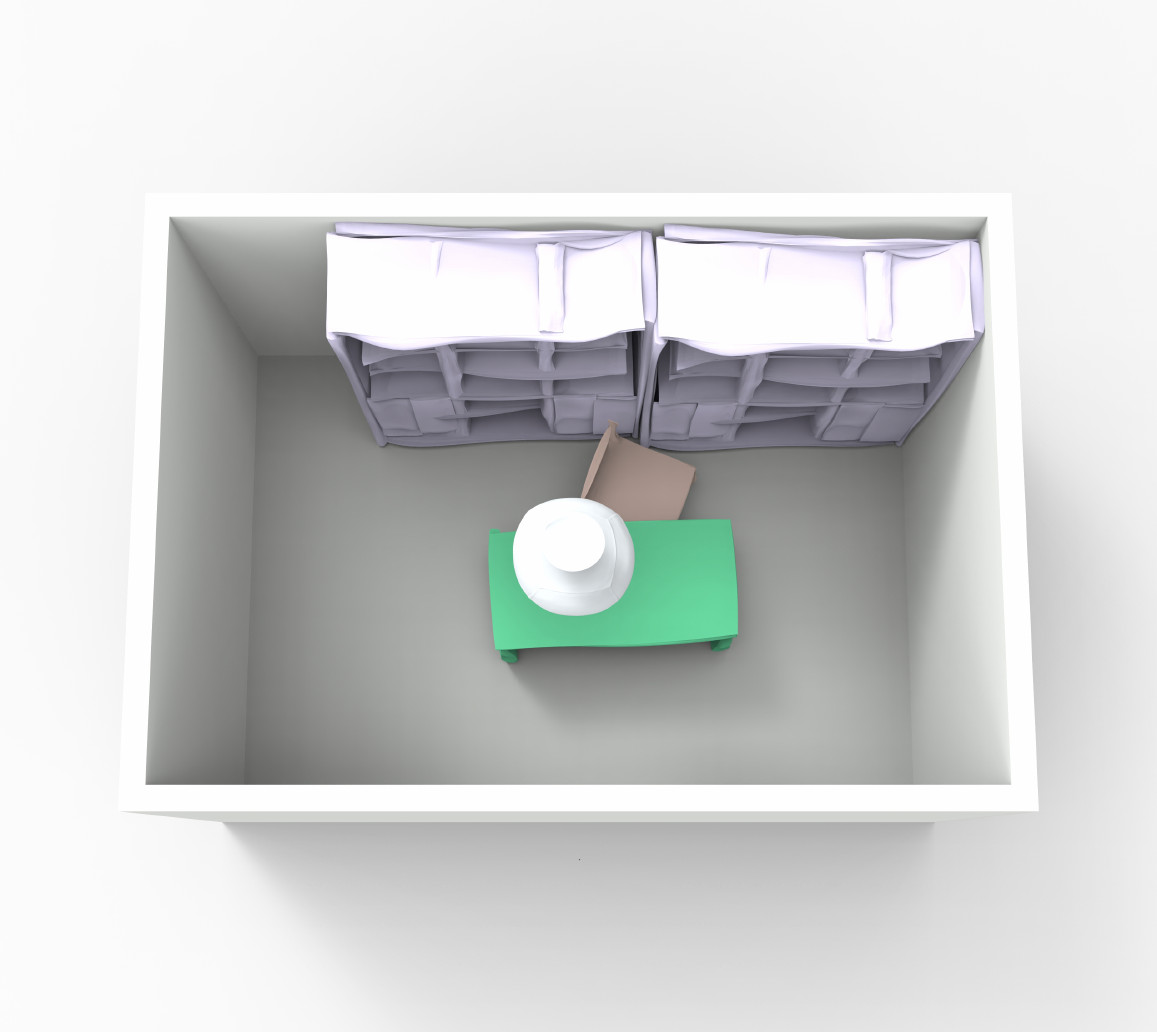}
    \\
    \includegraphics[width=0.24\linewidth]{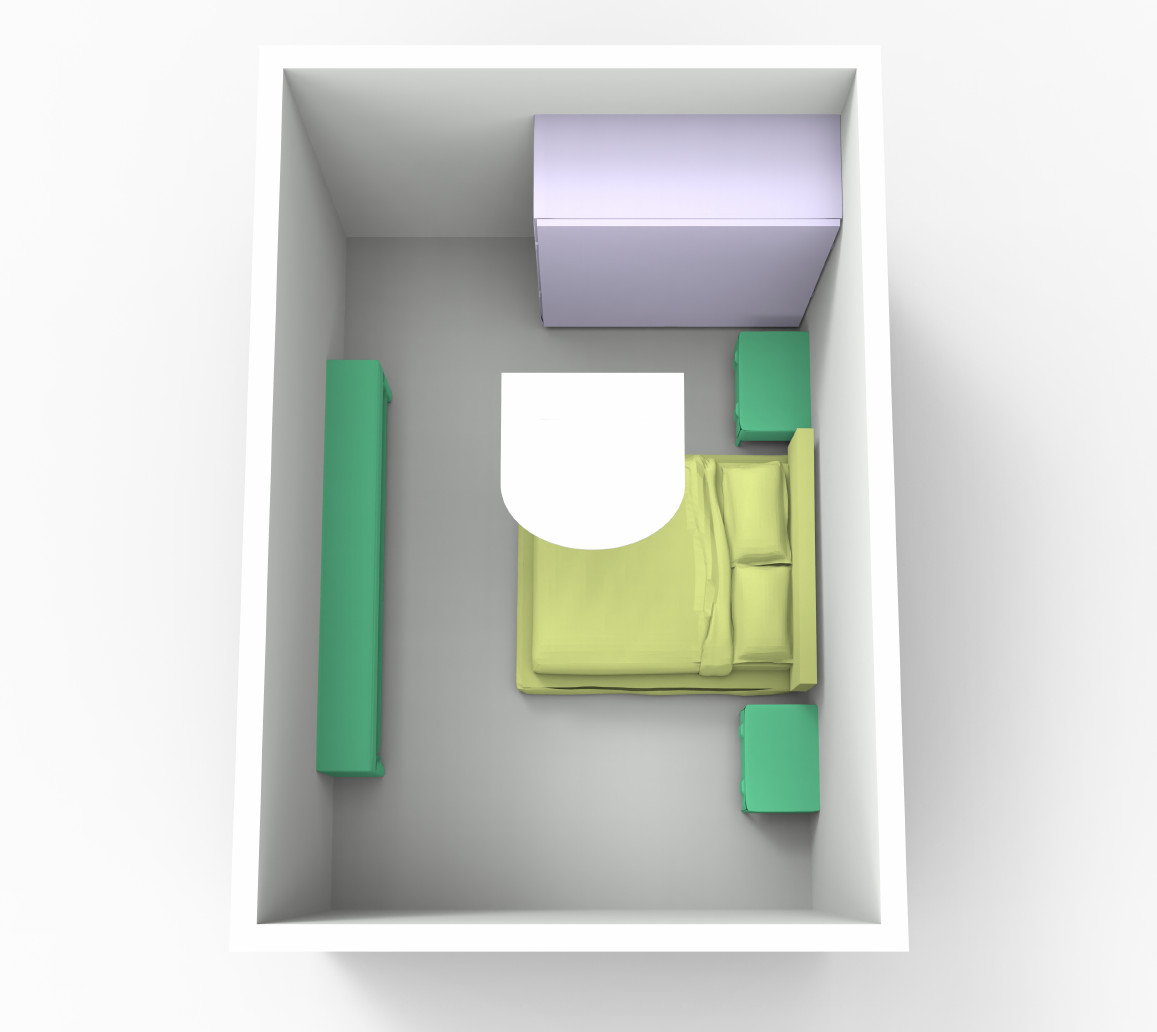}
    \includegraphics[width=0.24\linewidth]{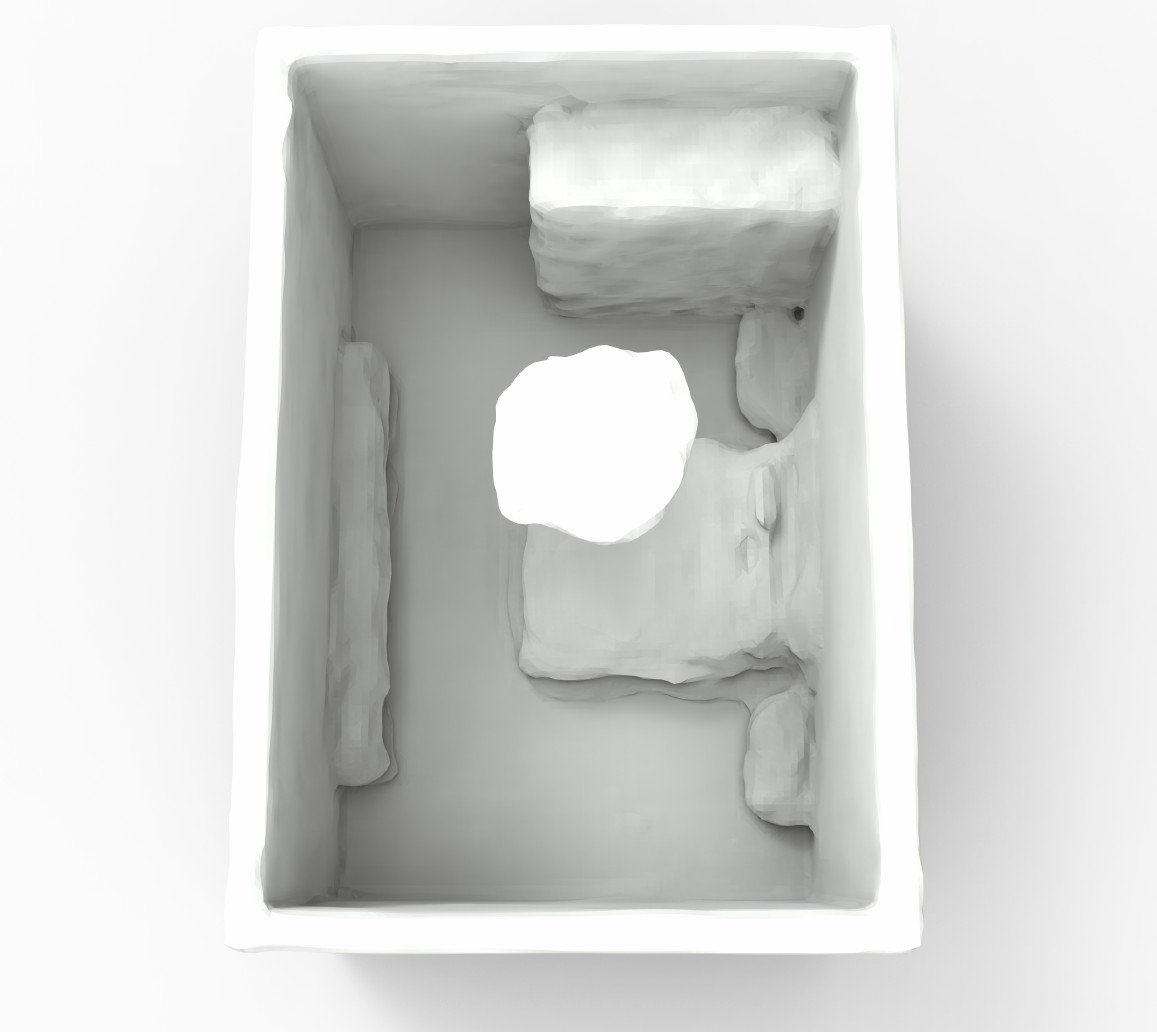}
    \includegraphics[width=0.24\linewidth]{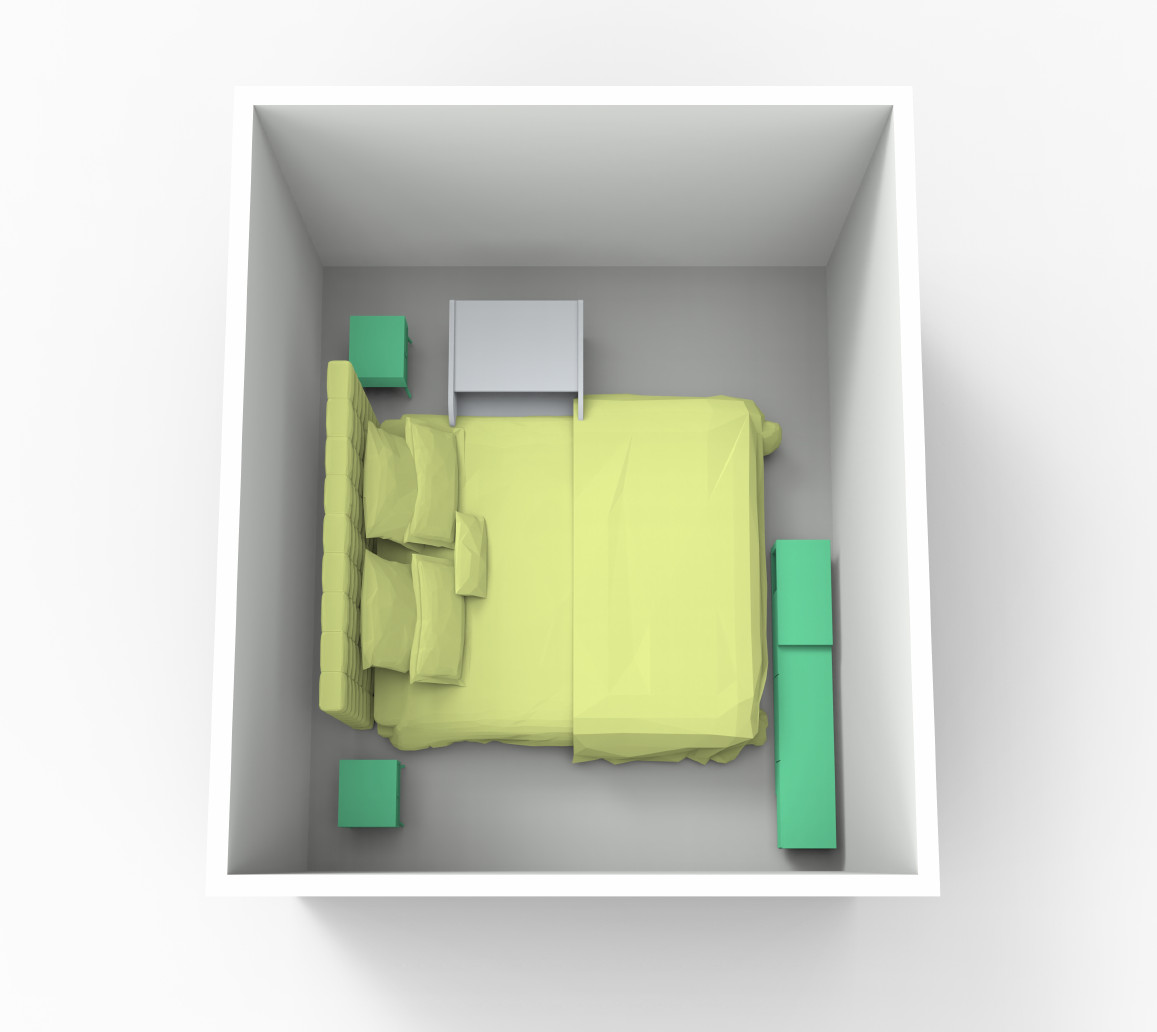}
    \includegraphics[width=0.24\linewidth]{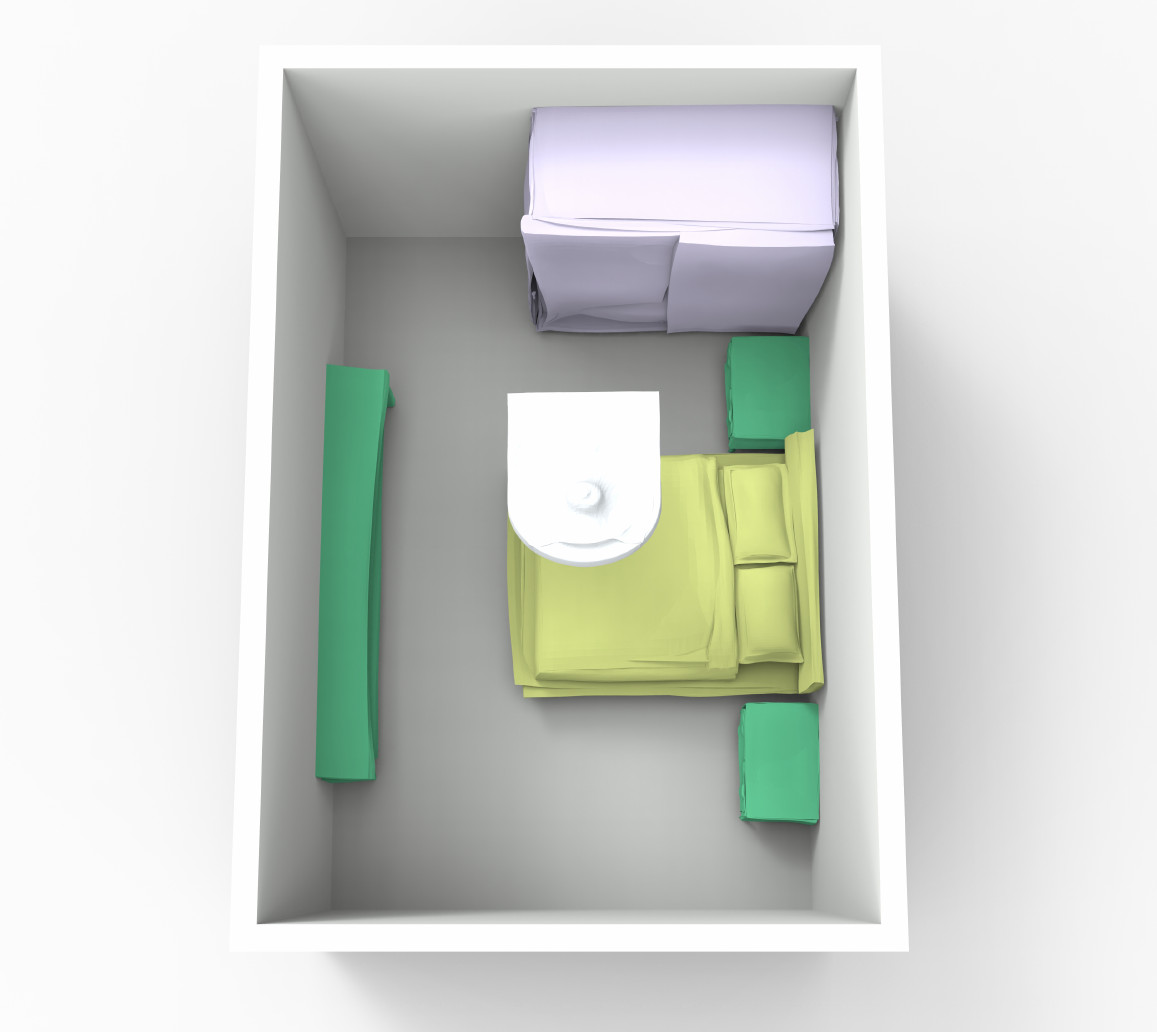}
    \\
    \includegraphics[width=0.24\linewidth]{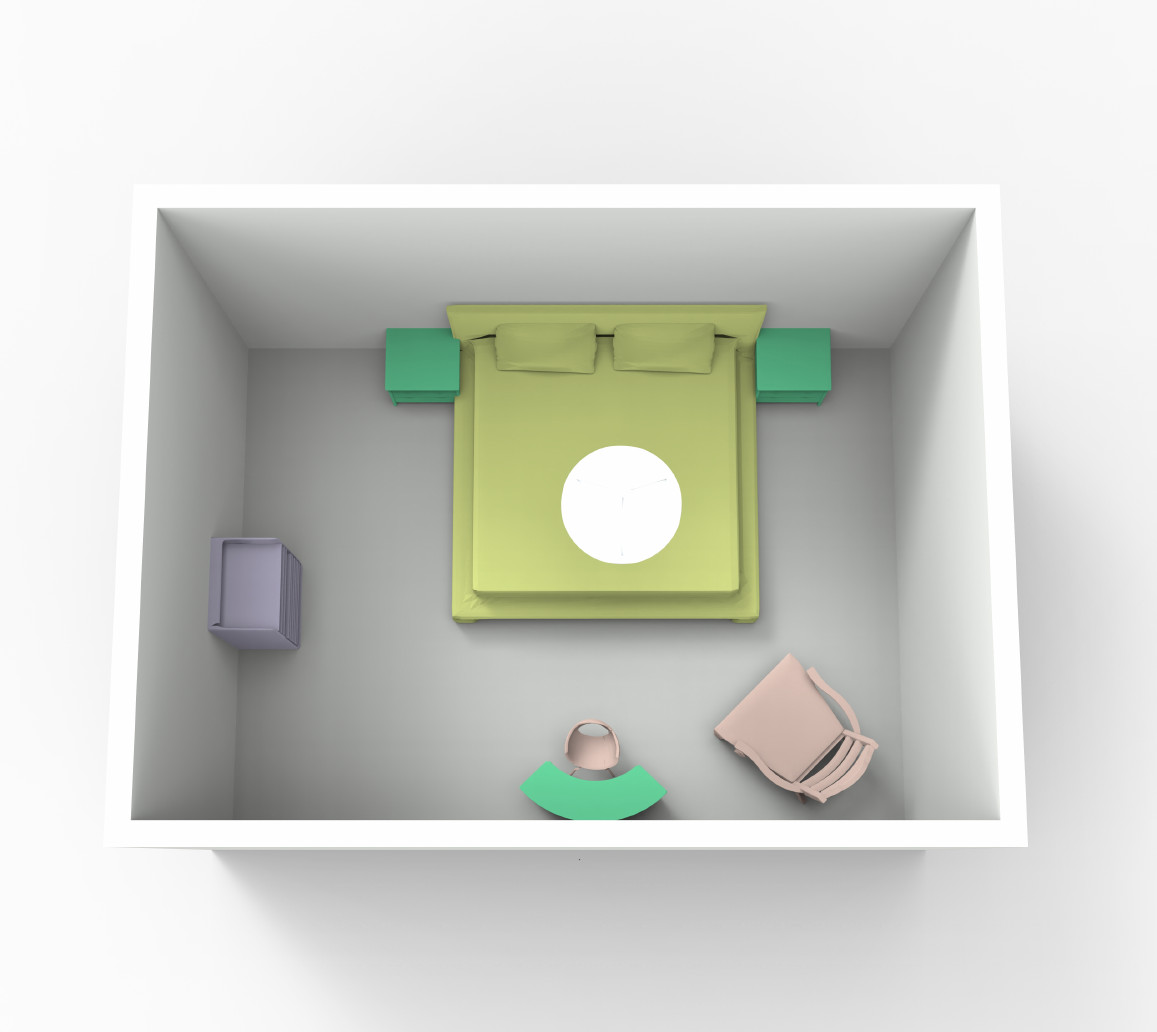}
    \includegraphics[width=0.24\linewidth]{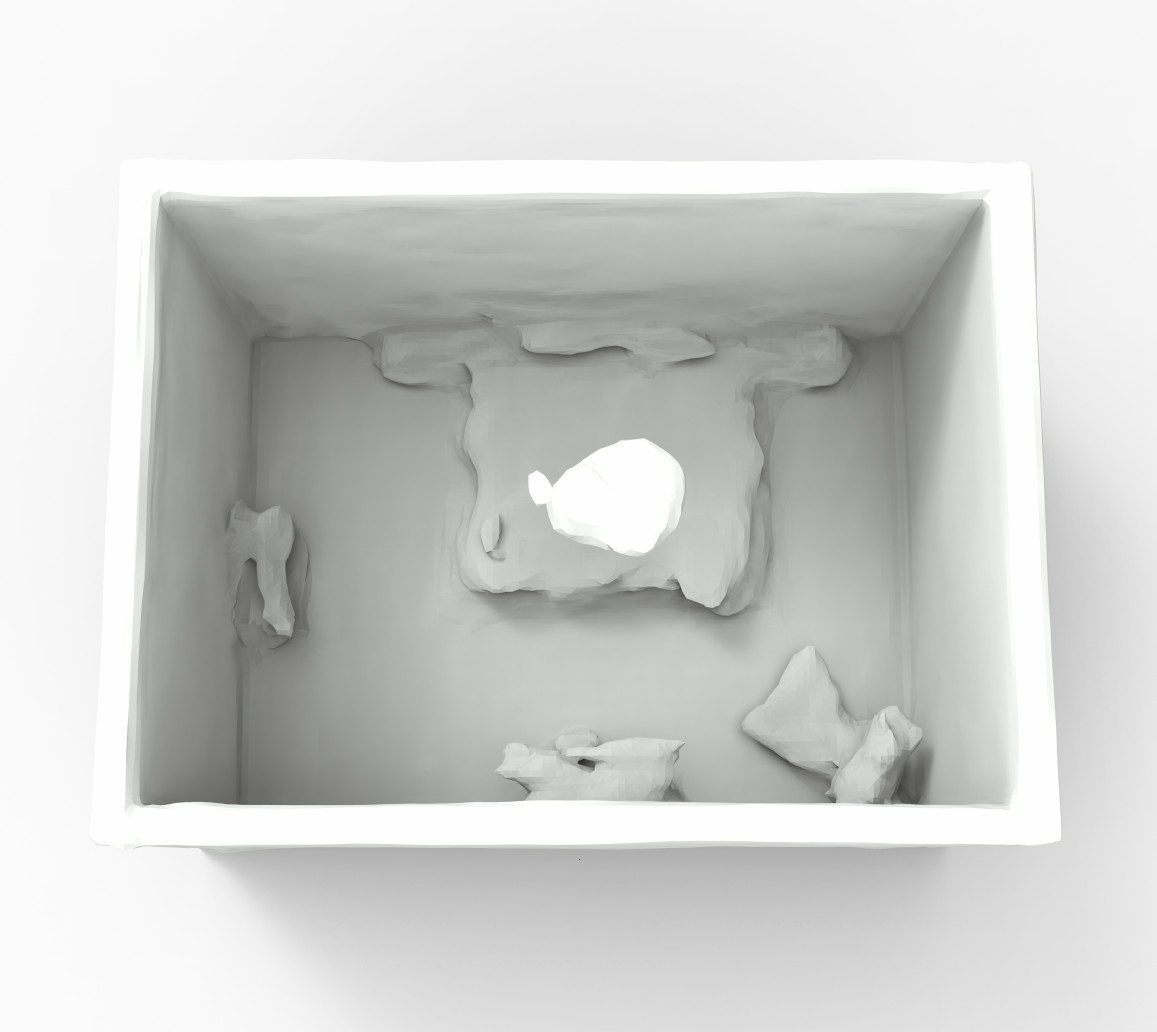}
    \includegraphics[width=0.24\linewidth]{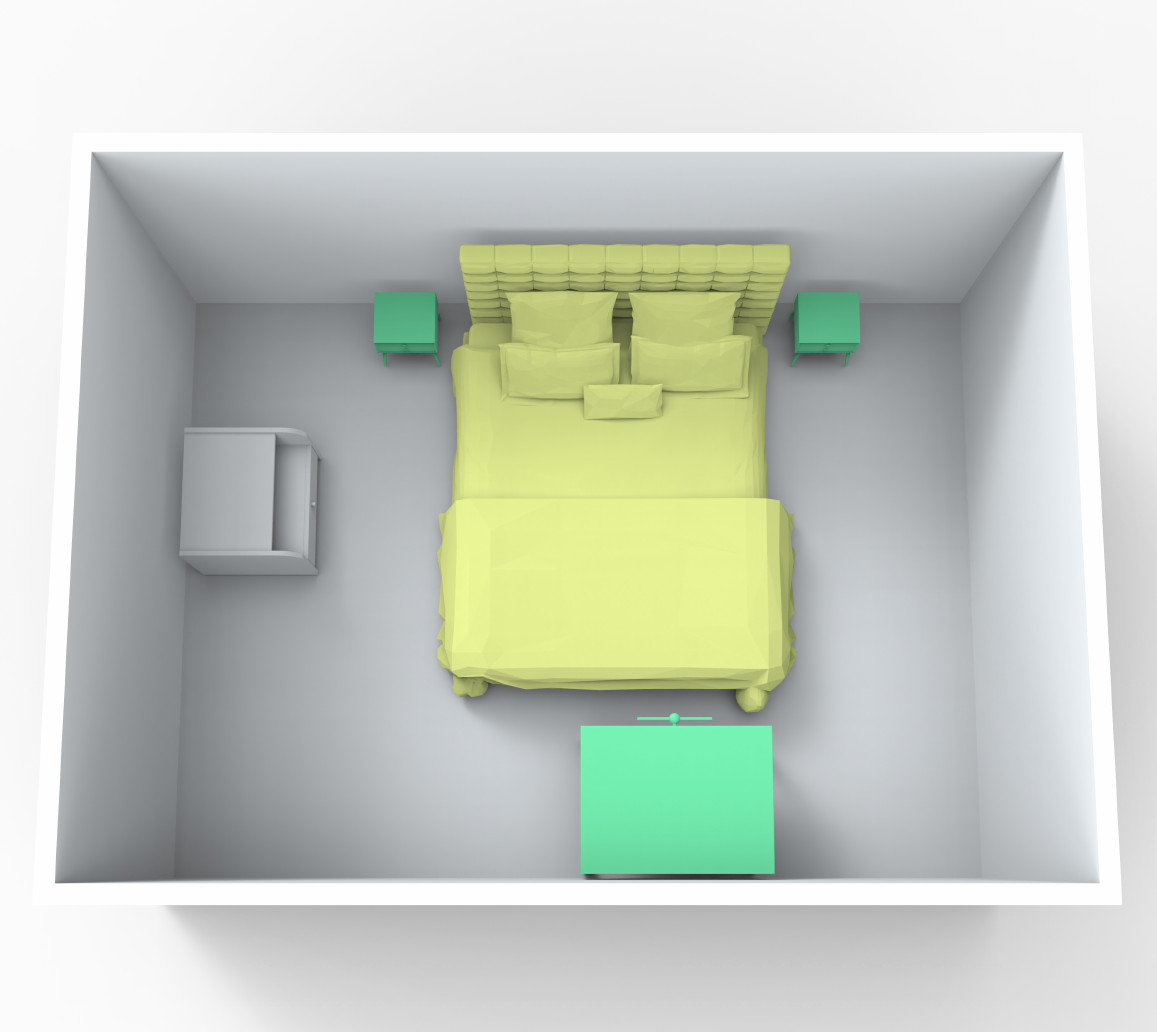}
    \includegraphics[width=0.24\linewidth]{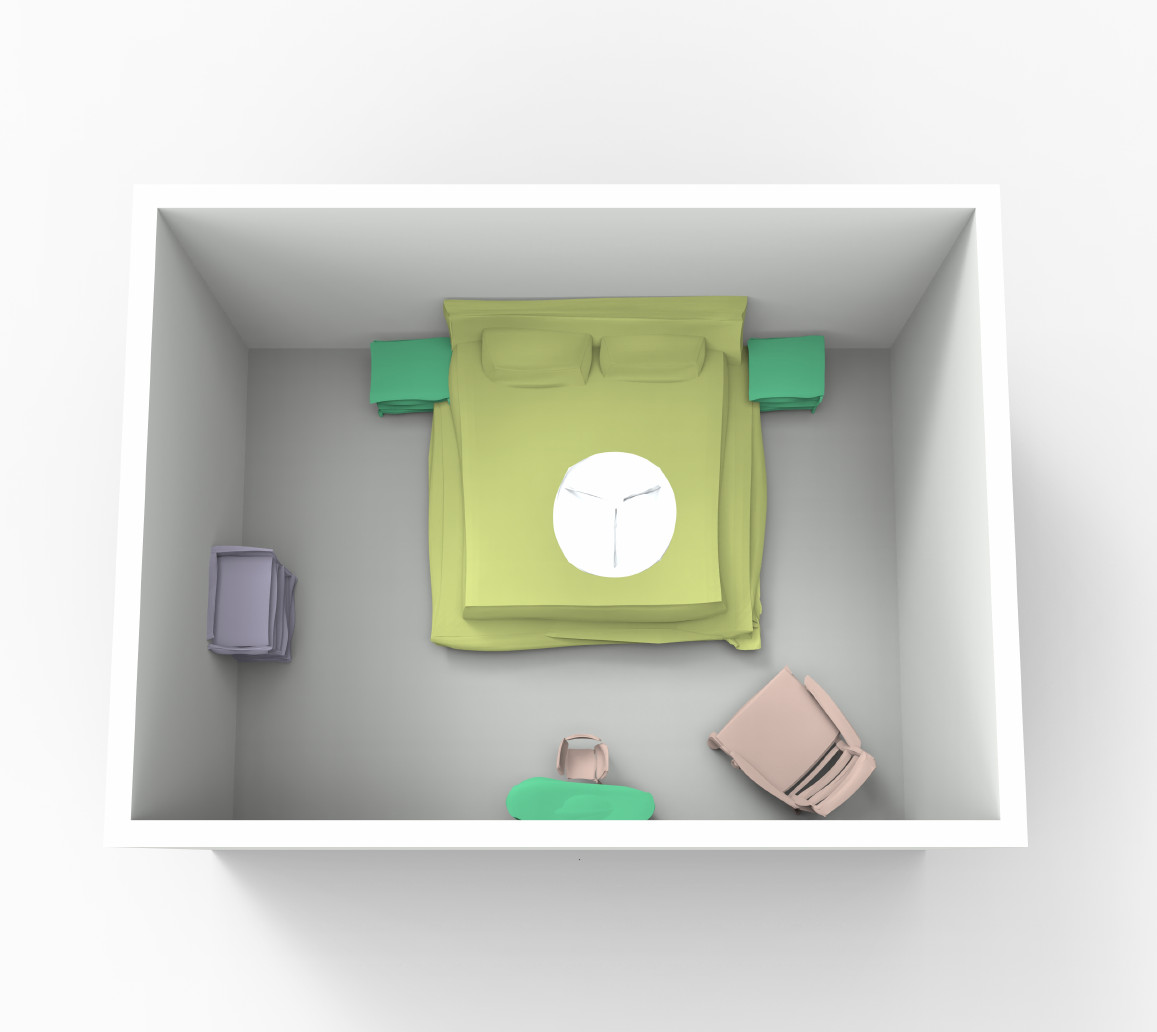}
    \\
    \subfigure[Input Room]{\includegraphics[width=0.24\linewidth]{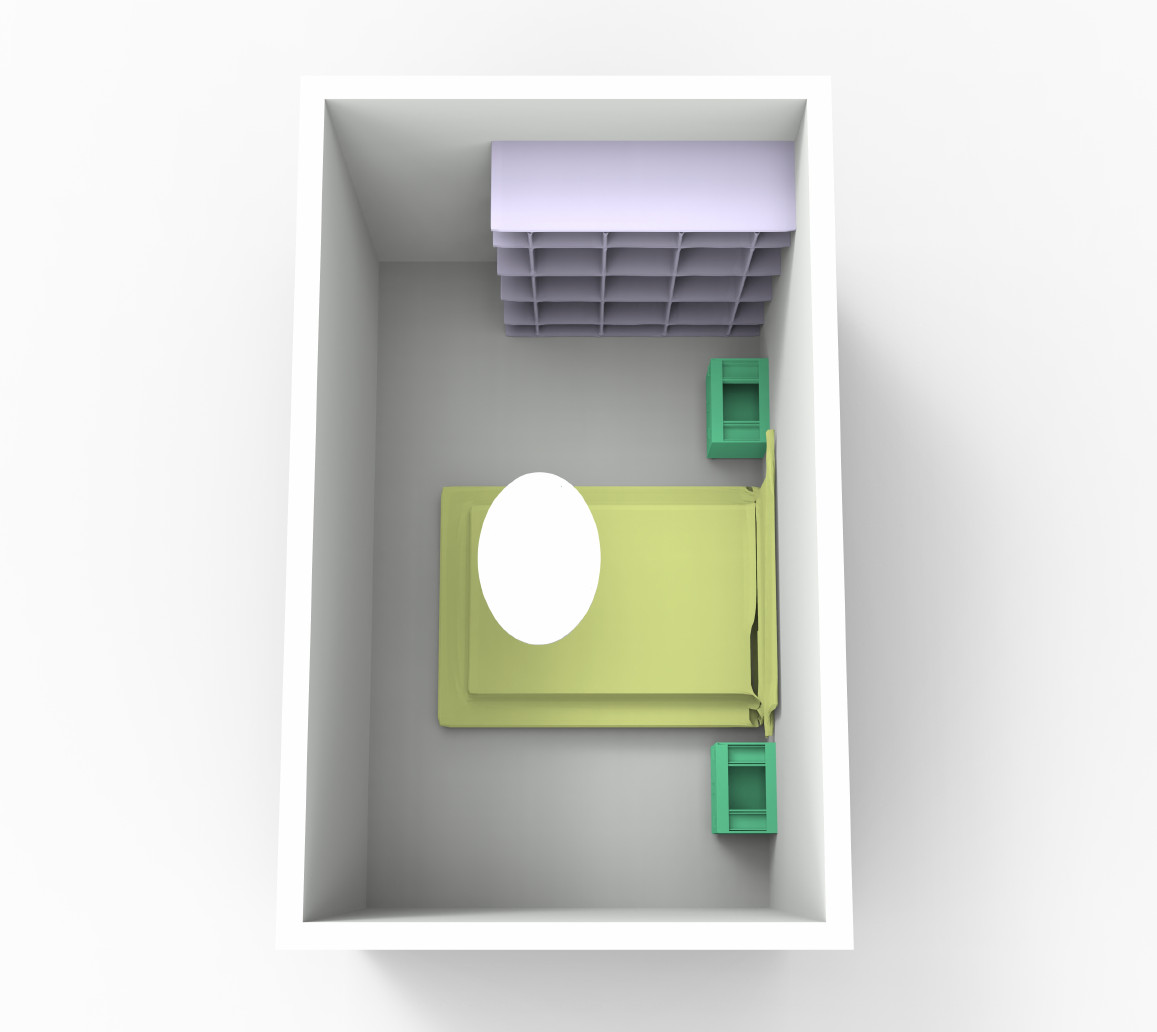}}
    \subfigure[Occ Net]{\includegraphics[width=0.24\linewidth]{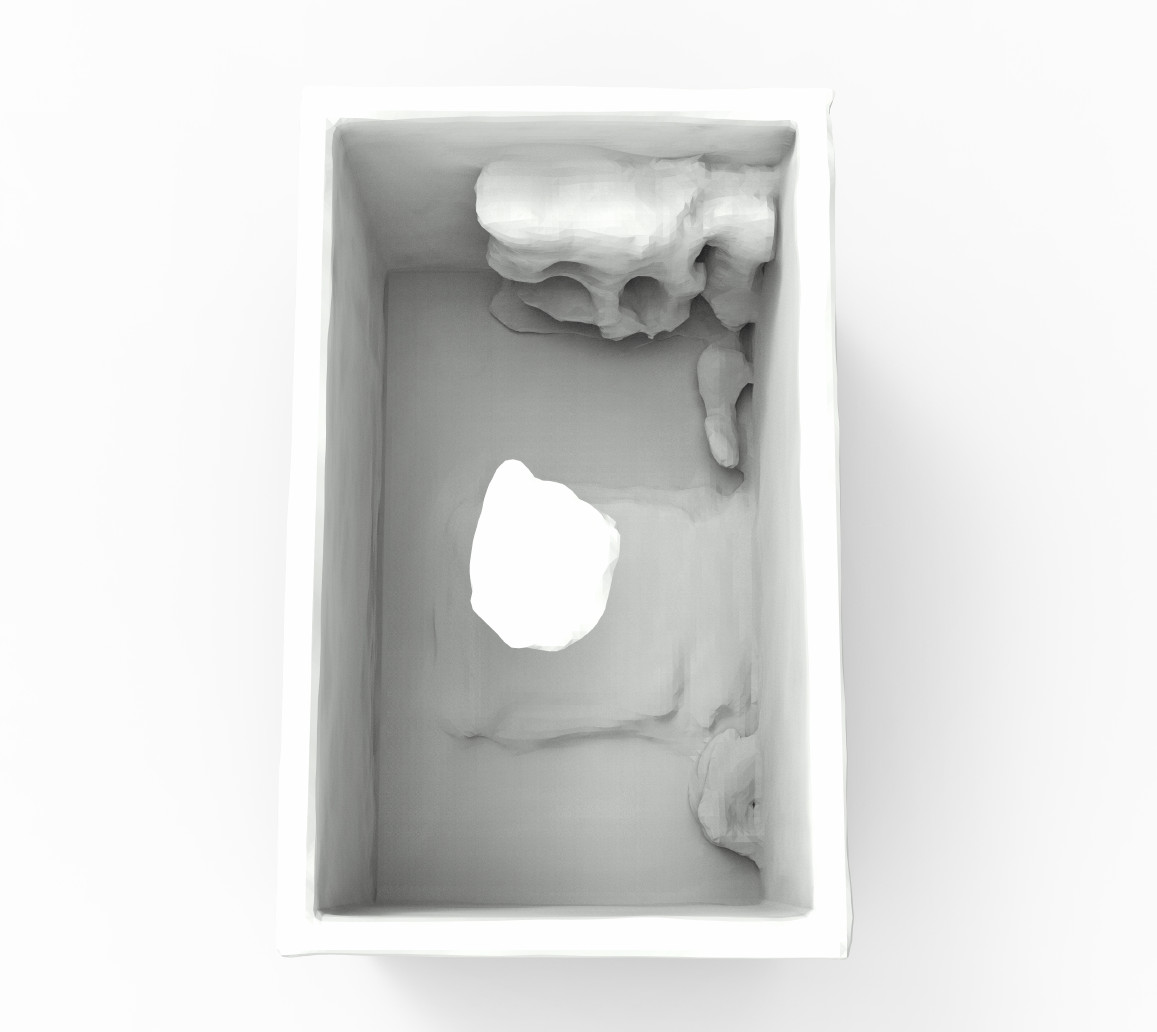}}
    \subfigure[GRAINS]{\includegraphics[width=0.24\linewidth]{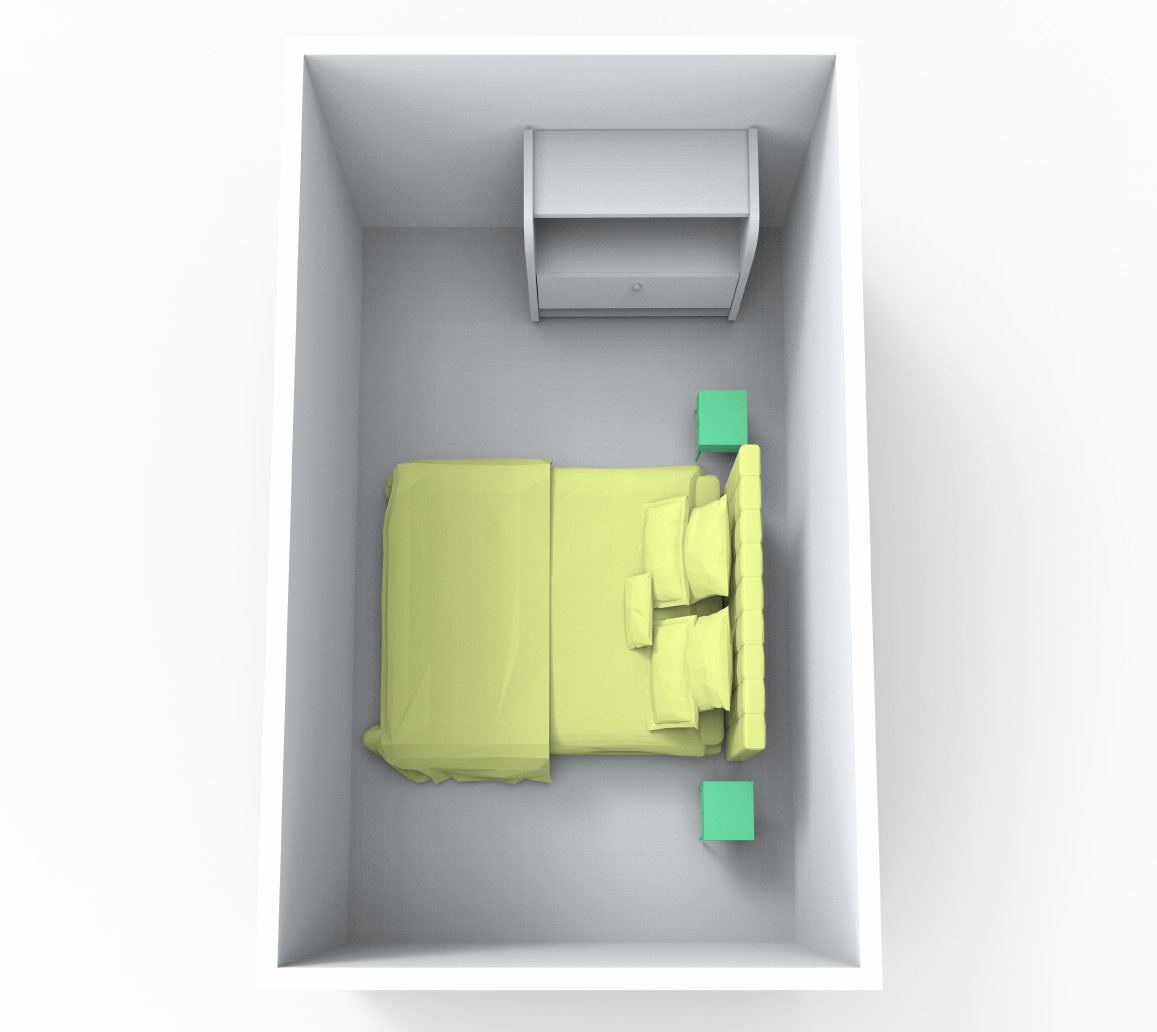}}
    \subfigure[Ours]{\includegraphics[width=0.24\linewidth]{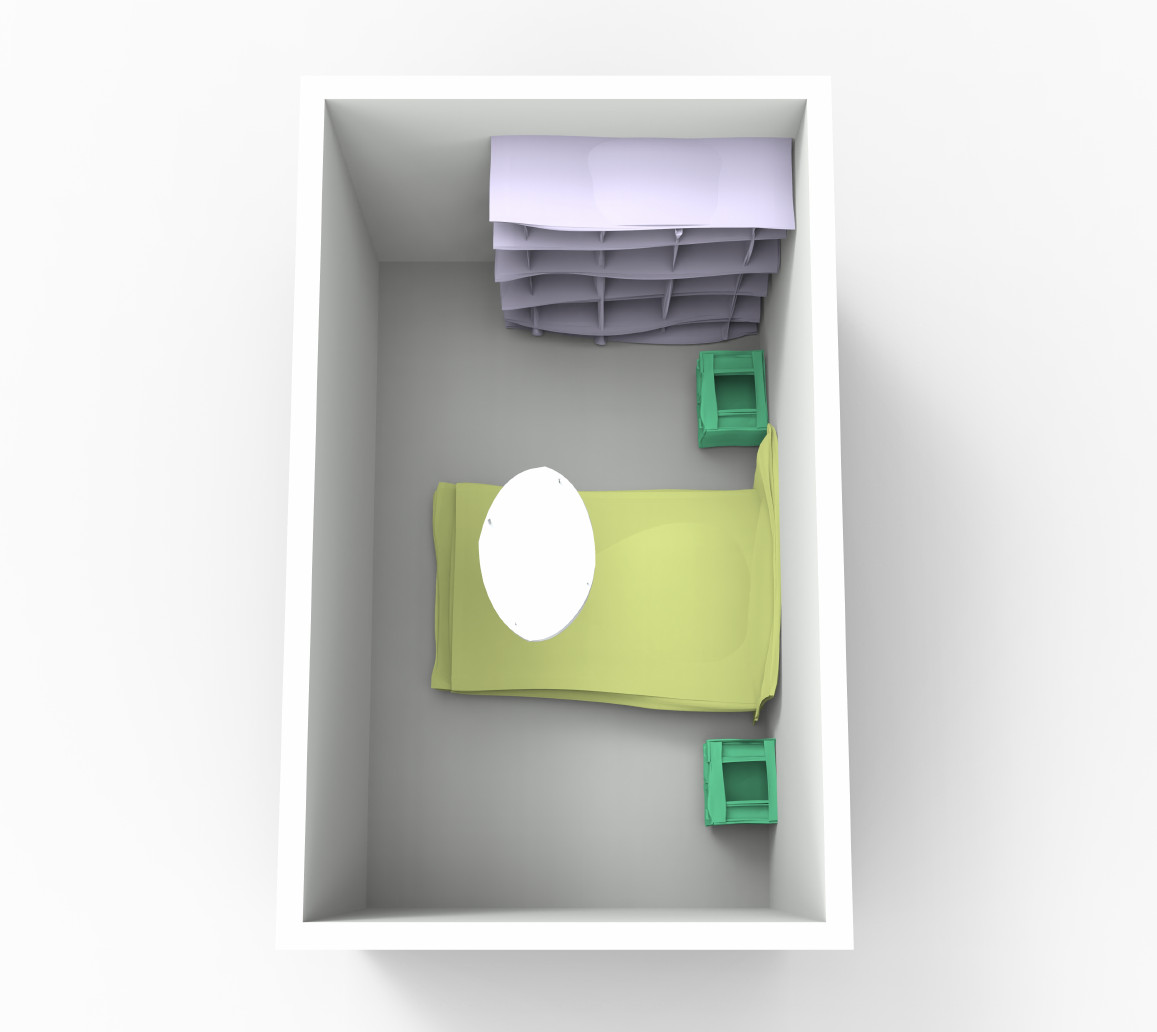}}
    \caption{\titlecap{Comparisons on room reconstruction with other methods.}{For the reconstruction task, we compare with implicit-field method--Occupancy Network (Occ Net)~\cite{mescheder2019occupancy} and GRAINS~\cite{li2019grains}. From the results, we can see that our results are capable of capturing more details (including structure and geometry) compared to the implicit-field method. GRAINS is an RvNN-VAE based generative model, which only focuses on scene generation and only reconstructs 2D boxes on the floor, so the final results of GRAINS are retrieved from the database according to 2D boxes. 
    }}
    \label{fig:reconcomparison}
\end{figure}

Table~\ref{tab:comreconstruction} shows the quantitative comparison between our method and two baseline methods on 3D-FRONT. Our method outperforms the implicit-based method Occ-Net, which demonstrates that part-level representation is very effective for  complex indoor scenes.
Figure~\ref{fig:reconcomparison} presents the qualitative comparison with other methods. It is easy to obverse that Occ-Net and GRAINS fail to capture the detailed geometry and object layouts. For example, the OCC-Net misses some object part details due to the holistic implicit representation, while our method can successfully achieve it.

%% file: main.bbl
\begin{thebibliography}{10}
\providecommand{\url}[1]{#1}
\csname url@samestyle\endcsname
\providecommand{\newblock}{\relax}
\providecommand{\bibinfo}[2]{#2}
\providecommand{\BIBentrySTDinterwordspacing}{\spaceskip=0pt\relax}
\providecommand{\BIBentryALTinterwordstretchfactor}{4}
\providecommand{\BIBentryALTinterwordspacing}{\spaceskip=\fontdimen2\font plus
\BIBentryALTinterwordstretchfactor\fontdimen3\font minus
  \fontdimen4\font\relax}
\providecommand{\BIBforeignlanguage}[2]{{%
\expandafter\ifx\csname l@#1\endcsname\relax
\typeout{** WARNING: IEEEtran.bst: No hyphenation pattern has been}%
\typeout{** loaded for the language `#1'. Using the pattern for}%
\typeout{** the default language instead.}%
\else
\language=\csname l@#1\endcsname
\fi
#2}}
\providecommand{\BIBdecl}{\relax}
\BIBdecl

\bibitem{fisher2012example}
M.~Fisher, D.~Ritchie, M.~Savva, T.~Funkhouser, and P.~Hanrahan,
  ``Example-based synthesis of 3d object arrangements,'' \emph{ACM Transactions
  on Graphics (TOG)}, vol.~31, no.~6, pp. 1--11, 2012.

\bibitem{henderson2017generative}
P.~Henderson and V.~Ferrari, ``A generative model of 3d object layouts in
  apartments,'' 2017.

\bibitem{wang2019planit}
K.~Wang, Y.-A. Lin, B.~Weissmann, M.~Savva, A.~X. Chang, and D.~Ritchie,
  ``Planit: Planning and instantiating indoor scenes with relation graph and
  spatial prior networks,'' \emph{ACM Transactions on Graphics (TOG)}, vol.~38,
  no.~4, pp. 1--15, 2019.

\bibitem{li2019grains}
M.~Li, A.~G. Patil, K.~Xu, S.~Chaudhuri, O.~Khan, A.~Shamir, C.~Tu, B.~Chen,
  D.~Cohen-Or, and H.~Zhang, ``Grains: Generative recursive autoencoders for
  indoor scenes,'' \emph{ACM Transactions on Graphics (TOG)}, vol.~38, no.~2,
  pp. 1--16, 2019.

\bibitem{li2017grass}
J.~Li, K.~Xu, S.~Chaudhuri, E.~Yumer, H.~Zhang, and L.~Guibas, ``Grass:
  Generative recursive autoencoders for shape structures,'' \emph{ACM
  Transactions on Graphics (TOG)}, vol.~36, no.~4, pp. 1--14, 2017.

\bibitem{mo2019structurenet}
K.~Mo, P.~Guerrero, L.~Yi, H.~Su, P.~Wonka, N.~J. Mitra, and L.~J. Guibas,
  ``Structurenet: hierarchical graph networks for 3d shape generation,''
  \emph{ACM Transactions on Graphics (TOG)}, vol.~38, no.~6, pp. 1--19, 2019.

\bibitem{gao2019sdm}
L.~Gao, J.~Yang, T.~Wu, Y.-J. Yuan, H.~Fu, Y.-K. Lai, and H.~Zhang, ``Sdm-net:
  Deep generative network for structured deformable mesh,'' \emph{ACM
  Transactions on Graphics (TOG)}, vol.~38, no.~6, pp. 1--15, 2019.

\bibitem{yang2020dsm}
J.~Yang, K.~Mo, Y.-K. Lai, L.~J. Guibas, and L.~Gao, ``Dsm-net: Disentangled
  structured mesh net for controllable generation of fine geometry,''
  \emph{arXiv preprint arXiv:2008.05440}, 2020.

\bibitem{fu20203dfront}
H.~Fu, B.~Cai, L.~Gao, L.-X. Zhang, J.~Wang, C.~Li, Q.~Zeng, C.~Sun, R.~Jia,
  B.~Zhao \emph{et~al.}, ``3d-front: 3d furnished rooms with layouts and
  semantics,'' in \emph{International Conference on Computer Vision (ICCV)},
  2021, pp. 10\,933--10\,942.

\bibitem{zheng2020structured3d}
J.~Zheng, J.~Zhang, J.~Li, R.~Tang, S.~Gao, and Z.~Zhou, ``Structured3d: A
  large photo-realistic dataset for structured 3d modeling,'' in \emph{European
  Conference on Computer Vision (ECCV)}.\hskip 1em plus 0.5em minus 0.4em\relax
  Springer, 2020, pp. 519--535.

\bibitem{li2021openrooms}
Z.~Li, T.-W. Yu, S.~Sang, S.~Wang, M.~Song, Y.~Liu, Y.-Y. Yeh, R.~Zhu,
  N.~Gundavarapu, J.~Shi \emph{et~al.}, ``Openrooms: An open framework for
  photorealistic indoor scene datasets,'' in \emph{IEEE Conference on Computer
  Vision and Pattern Recognition (CVPR)}, 2021, pp. 7190--7199.

\bibitem{roberts:2021}
M.~Roberts, J.~Ramapuram, A.~Ranjan, A.~Kumar, M.~A. Bautista, N.~Paczan,
  R.~Webb, and J.~M. Susskind, ``{Hypersim}: {A} photorealistic synthetic
  dataset for holistic indoor scene understanding,'' in \emph{International
  Conference on Computer Vision (ICCV)}, 2021.

\bibitem{mo2019partnet}
K.~Mo, S.~Zhu, A.~X. Chang, L.~Yi, S.~Tripathi, L.~J. Guibas, and H.~Su,
  ``Partnet: A large-scale benchmark for fine-grained and hierarchical
  part-level 3d object understanding,'' in \emph{IEEE Conference on Computer
  Vision and Pattern Recognition (CVPR)}, 2019, pp. 909--918.

\bibitem{johnson2018image}
J.~Johnson, A.~Gupta, and L.~Fei-Fei, ``Image generation from scene graphs,''
  in \emph{IEEE Conference on Computer Vision and Pattern Recognition (CVPR)},
  2018, pp. 1219--1228.

\bibitem{ashual2019specifying}
O.~Ashual and L.~Wolf, ``Specifying object attributes and relations in
  interactive scene generation,'' in \emph{International Conference on Computer
  Vision (ICCV)}, 2019, pp. 4561--4569.

\bibitem{chang2015text}
A.~Chang, W.~Monroe, M.~Savva, C.~Potts, and C.~D. Manning, ``Text to 3d scene
  generation with rich lexical grounding,'' \emph{arXiv preprint
  arXiv:1505.06289}, 2015.

\bibitem{luo2020end}
A.~Luo, Z.~Zhang, J.~Wu, and J.~B. Tenenbaum, ``End-to-end optimization of
  scene layout,'' in \emph{IEEE Conference on Computer Vision and Pattern
  Recognition (CVPR)}, 2020, pp. 3754--3763.

\bibitem{Graph2Plan20}
R.~Hu, Z.~Huang, Y.~Tang, O.~V. Kaick, H.~Zhang, and H.~Huang, ``Graph2plan:
  Learning floorplan generation from layout graphs,'' \emph{ACM Transactions on
  Graphics (TOG)}, vol.~39, no.~4, pp. 118:1--118:14, 2020.

\bibitem{Nauata2020HouseGANRG}
N.~Nauata, K.~Chang, C.-Y. Cheng, G.~Mori, and Y.~Furukawa, ``House-gan:
  Relational generative adversarial networks for graph-constrained house layout
  generation,'' in \emph{European Conference on Computer Vision (ECCV)}, 2020.

\bibitem{para2020generative}
W.~Para, P.~Guerrero, T.~Kelly, L.~Guibas, and P.~Wonka, ``Generative layout
  modeling using constraint graphs,'' \emph{arXiv preprint arXiv:2011.13417},
  2020.

\bibitem{huang2018holistic}
S.~Huang, S.~Qi, Y.~Zhu, Y.~Xiao, Y.~Xu, and S.-C. Zhu, ``Holistic 3d scene
  parsing and reconstruction from a single rgb image,'' in \emph{European
  Conference on Computer Vision (ECCV)}, 2018, pp. 187--203.

\bibitem{armeni20193d}
I.~Armeni, Z.-Y. He, J.~Gwak, A.~R. Zamir, M.~Fischer, J.~Malik, and
  S.~Savarese, ``3d scene graph: A structure for unified semantics, 3d space,
  and camera,'' in \emph{International Conference on Computer Vision (ICCV)},
  2019, pp. 5664--5673.

\bibitem{socher2011parsing}
R.~Socher, C.~C.-Y. Lin, A.~Y. Ng, and C.~D. Manning, ``Parsing natural scenes
  and natural language with recursive neural networks,'' in \emph{ICML}, 2011.

\bibitem{socher2012convolutional}
R.~Socher, B.~Huval, B.~Bath, C.~D. Manning, and A.~Ng,
  ``Convolutional-recursive deep learning for 3d object classification,''
  \emph{Advances in neural information processing systems}, vol.~25, pp.
  656--664, 2012.

\bibitem{shi2019hierarchy}
Y.~Shi, A.~X. Chang, Z.~Wu, M.~Savva, and K.~Xu, ``Hierarchy denoising
  recursive autoencoders for 3d scene layout prediction,'' in \emph{IEEE
  Conference on Computer Vision and Pattern Recognition (CVPR)}, 2019, pp.
  1771--1780.

\bibitem{zhang2019survey}
S.-H. Zhang, S.-K. Zhang, Y.~Liang, and P.~Hall, ``A survey of 3d indoor scene
  synthesis,'' \emph{Journal of Computer Science and Technology}, vol.~34,
  no.~3, pp. 594--608, 2019.

\bibitem{pintore2020state}
G.~Pintore, C.~Mura, F.~Ganovelli, L.~Fuentes-Perez, R.~Pajarola, and
  E.~Gobbetti, ``State-of-the-art in automatic 3d reconstruction of structured
  indoor environments,'' in \emph{Computer Graphics Forum}, vol.~39,
  no.~2.\hskip 1em plus 0.5em minus 0.4em\relax Wiley Online Library, 2020, pp.
  667--699.

\bibitem{yu2011make}
L.~F. Yu, S.~K. Yeung, C.~K. Tang, D.~Terzopoulos, T.~F. Chan, and S.~J. Osher,
  ``Make it home: automatic optimization of furniture arrangement,'' \emph{ACM
  Transactions on Graphics (TOG)}, vol.~30, no.~4, 2011.

\bibitem{merrell2011interactive}
P.~Merrell, E.~Schkufza, Z.~Li, M.~Agrawala, and V.~Koltun, ``Interactive
  furniture layout using interior design guidelines,'' \emph{ACM transactions
  on graphics (TOG)}, vol.~30, no.~4, pp. 1--10, 2011.

\bibitem{yeh2012synthesizing}
Y.-T. Yeh, L.~Yang, M.~Watson, N.~D. Goodman, and P.~Hanrahan, ``Synthesizing
  open worlds with constraints using locally annealed reversible jump mcmc,''
  \emph{ACM Transactions on Graphics (TOG)}, vol.~31, no.~4, pp. 1--11, 2012.

\bibitem{xu2015wall}
W.~Xu, B.~Wang, and D.-M. Yan, ``Wall grid structure for interior scene
  synthesis,'' \emph{Computers \& Graphics}, vol.~46, pp. 231--243, 2015.

\bibitem{kermani2016learning}
Z.~S. Kermani, Z.~Liao, P.~Tan, and H.~Zhang, ``Learning 3d scene synthesis
  from annotated rgb-d images,'' in \emph{Computer Graphics Forum}, vol.~35,
  no.~5.\hskip 1em plus 0.5em minus 0.4em\relax Wiley Online Library, 2016, pp.
  197--206.

\bibitem{henderson2017automatic}
P.~Henderson, K.~Subr, and V.~Ferrari, ``Automatic generation of constrained
  furniture layouts,'' \emph{arXiv preprint arXiv:1711.10939}, 2017.

\bibitem{liang2017automatic}
Y.~Liang, S.-H. Zhang, and R.~R. Martin, ``Automatic data-driven room design
  generation,'' in \emph{International Workshop on Next Generation Computer
  Animation Techniques}.\hskip 1em plus 0.5em minus 0.4em\relax Springer, 2017,
  pp. 133--148.

\bibitem{zhang2021fast}
S.-H. Zhang, S.-K. Zhang, W.-Y. Xie, C.-Y. Luo, Y.~Yang, and H.~Fu, ``Fast 3d
  indoor scene synthesis by learning spatial relation priors of objects,''
  \emph{IEEE Transactions on Visualization and Computer Graphics}, 2021.

\bibitem{xu2017autonomous}
K.~Xu, L.~Zheng, Z.~Yan, G.~Yan, E.~Zhang, M.~Niessner, O.~Deussen,
  D.~Cohen-Or, and H.~Huang, ``Autonomous reconstruction of unknown indoor
  scenes guided by time-varying tensor fields,'' \emph{ACM Transactions on
  Graphics (TOG)}, vol.~36, no.~6, pp. 1--15, 2017.

\bibitem{dong2019multi}
S.~Dong, K.~Xu, Q.~Zhou, A.~Tagliasacchi, S.~Xin, M.~Nie{\ss}ner, and B.~Chen,
  ``Multi-robot collaborative dense scene reconstruction,'' \emph{ACM
  Transactions on Graphics (TOG)}, vol.~38, no.~4, pp. 1--16, 2019.

\bibitem{xu2015autoscanning}
K.~Xu, H.~Huang, Y.~Shi, H.~Li, P.~Long, J.~Caichen, W.~Sun, and B.~Chen,
  ``Autoscanning for coupled scene reconstruction and proactive object
  analysis,'' \emph{ACM Transactions on Graphics (TOG)}, vol.~34, no.~6, pp.
  1--14, 2015.

\bibitem{ritchie2019fast}
D.~Ritchie, K.~Wang, and Y.-a. Lin, ``Fast and flexible indoor scene synthesis
  via deep convolutional generative models,'' in \emph{IEEE Conference on
  Computer Vision and Pattern Recognition (CVPR)}, 2019, pp. 6182--6190.

\bibitem{wang2018deep}
K.~Wang, M.~Savva, A.~X. Chang, and D.~Ritchie, ``Deep convolutional priors for
  indoor scene synthesis,'' \emph{ACM Transactions on Graphics (TOG)}, vol.~37,
  no.~4, pp. 1--14, 2018.

\bibitem{zhang2020deep}
Z.~Zhang, Z.~Yang, C.~Ma, L.~Luo, A.~Huth, E.~Vouga, and Q.~Huang, ``Deep
  generative modeling for scene synthesis via hybrid representations,''
  \emph{ACM Transactions on Graphics (TOG)}, vol.~39, no.~2, pp. 1--21, 2020.

\bibitem{wang2020sceneformer}
X.~Wang, C.~Yeshwanth, and M.~Nie{\ss}ner, ``Sceneformer: Indoor scene
  generation with transformers,'' in \emph{International Conference on 3D
  Vision (3DV)}.\hskip 1em plus 0.5em minus 0.4em\relax IEEE, 2021, pp.
  106--115.

\bibitem{paschalidou2021atiss}
D.~Paschalidou, A.~Kar, M.~Shugrina, K.~Kreis, A.~Geiger, and S.~Fidler,
  ``Atiss: Autoregressive transformers for indoor scene synthesis,''
  \emph{Advances in Neural Information Processing Systems}, vol.~34, 2021.

\bibitem{zhou2019scenegraphnet}
Y.~Zhou, Z.~While, and E.~Kalogerakis, ``Scenegraphnet: Neural message passing
  for 3d indoor scene augmentation,'' in \emph{International Conference on
  Computer Vision (ICCV)}, 2019, pp. 7384--7392.

\bibitem{liu2021game}
L.~Liu, Y.~Yang, Y.~Yuan, T.~Shao, H.~Wang, and K.~Zhou, ``In-game residential
  home planning via visual context-aware global relation learning,'' in
  \emph{AAAI Conference on Artificial Intelligence}, vol.~35, no.~1, 2021, pp.
  336--343.

\bibitem{yang2021scene}
H.~Yang, Z.~Zhang, S.~Yan, H.~Huang, C.~Ma, Y.~Zheng, C.~Bajaj, and Q.~Huang,
  ``Scene synthesis via uncertainty-driven attribute synchronization,'' in
  \emph{International Conference on Computer Vision (ICCV)}, 2021, pp.
  5630--5640.

\bibitem{ma2018language}
R.~Ma, A.~G. Patil, M.~Fisher, M.~Li, S.~Pirk, B.-S. Hua, S.-K. Yeung, X.~Tong,
  L.~Guibas, and H.~Zhang, ``Language-driven synthesis of 3d scenes from scene
  databases,'' \emph{ACM Transactions on Graphics (TOG)}, vol.~37, no.~6, pp.
  1--16, 2018.

\bibitem{fu2017adaptive}
Q.~Fu, X.~Chen, X.~Wang, S.~Wen, B.~Zhou, and H.~Fu, ``Adaptive synthesis of
  indoor scenes via activity-associated object relation graphs,'' \emph{ACM
  Transactions on Graphics (TOG)}, vol.~36, no.~6, pp. 1--13, 2017.

\bibitem{qi2018human}
S.~Qi, Y.~Zhu, S.~Huang, C.~Jiang, and S.-C. Zhu, ``Human-centric indoor scene
  synthesis using stochastic grammar,'' in \emph{IEEE Conference on Computer
  Vision and Pattern Recognition (CVPR)}, 2018, pp. 5899--5908.

\bibitem{ma2016action}
R.~Ma, H.~Li, C.~Zou, Z.~Liao, X.~Tong, and H.~Zhang, ``Action-driven 3d indoor
  scene evolution,'' \emph{ACM Transactions on Graphics (TOG)}, vol.~35, no.~6,
  pp. 173--1, 2016.

\bibitem{goller1996learning}
C.~Goller and A.~Kuchler, ``Learning task-dependent distributed representations
  by backpropagation through structure,'' in \emph{International Conference on
  Neural Networks (ICNN'96)}, vol.~1.\hskip 1em plus 0.5em minus 0.4em\relax
  IEEE, 1996, pp. 347--352.

\bibitem{tai2015improved}
K.~S. Tai, R.~Socher, and C.~D. Manning, ``Improved semantic representations
  from tree-structured long short-term memory networks,'' \emph{arXiv preprint
  arXiv:1503.00075}, 2015.

\bibitem{kingma2013auto}
D.~P. Kingma and M.~Welling, ``Auto-encoding variational bayes,'' \emph{arXiv
  preprint arXiv:1312.6114}, 2013.

\bibitem{niu2018im2struct}
C.~Niu, J.~Li, and K.~Xu, ``Im2struct: Recovering 3d shape structure from a
  single rgb image,'' in \emph{IEEE Conference on Computer Vision and Pattern
  Recognition (CVPR)}, 2018, pp. 4521--4529.

\bibitem{yu2019partnet}
F.~Yu, K.~Liu, Y.~Zhang, C.~Zhu, and K.~Xu, ``Partnet: A recursive part
  decomposition network for fine-grained and hierarchical shape segmentation,''
  in \emph{IEEE Conference on Computer Vision and Pattern Recognition (CVPR)},
  2019, pp. 9491--9500.

\bibitem{mo2020structedit}
K.~Mo, P.~Guerrero, L.~Yi, H.~Su, P.~Wonka, N.~J. Mitra, and L.~J. Guibas,
  ``Structedit: Learning structural shape variations,'' in \emph{IEEE
  Conference on Computer Vision and Pattern Recognition (CVPR)}, 2020, pp.
  8859--8868.

\bibitem{mo2020pt2pc}
K.~Mo, H.~Wang, X.~Yan, and L.~Guibas, ``{PT2PC}: Learning to generate 3d point
  cloud shapes from part tree conditions,'' \emph{European Conference on
  Computer Vision (ECCV)}, 2020.

\bibitem{DBScan}
M.~Ester, H.-P. Kriegel, J.~Sander, and X.~Xu, ``A density-based algorithm for
  discovering clusters in large spatial databases with noise,'' in \emph{Second
  International Conference on Knowledge Discovery and Data Mining}, ser.
  KDD'96.\hskip 1em plus 0.5em minus 0.4em\relax AAAI Press, 1996, p.
  226–231.

\bibitem{wang2011symmetry}
Y.~Wang, K.~Xu, J.~Li, H.~Zhang, A.~Shamir, L.~Liu, Z.~Cheng, and Y.~Xiong,
  ``Symmetry hierarchy of man-made objects,'' in \emph{Computer graphics
  forum}, vol.~30, no.~2.\hskip 1em plus 0.5em minus 0.4em\relax Wiley Online
  Library, 2011, pp. 287--296.

\bibitem{Gao2021SparseDD}
L.~Gao, Y.~Lai, J.~Yang, L.-X. Zhang, L.~Kobbelt, and S.~hong Xia, ``Sparse
  data driven mesh deformation,'' \emph{IEEE Transactions on Visualization and
  Computer Graphics}, vol.~27, pp. 2085--2100, 2021.

\bibitem{bouaziz2014dynamic}
S.~Bouaziz, A.~Tagliasacchi, and M.~Pauly, ``Dynamic 2d/3d registration.'' in
  \emph{Eurographics (Tutorials)}.\hskip 1em plus 0.5em minus 0.4em\relax
  Citeseer, 2014, p.~7.

\bibitem{fan2017point}
H.~Fan, H.~Su, and L.~J. Guibas, ``A point set generation network for 3d object
  reconstruction from a single image,'' in \emph{IEEE Conference on Computer
  Vision and Pattern Recognition (CVPR)}, 2017, pp. 605--613.

\bibitem{barrow1977parametric}
H.~G. Barrow, J.~M. Tenenbaum, R.~C. Bolles, and H.~C. Wolf, ``Parametric
  correspondence and chamfer matching: Two new techniques for image matching,''
  in \emph{Proceedings: Image Understanding Workshop}, 1977, pp. 21--27.

\bibitem{fu20203dfuture}
H.~Fu, R.~Jia, L.~Gao, M.~Gong, B.~Zhao, S.~Maybank, and D.~Tao, ``3d-future:
  3d furniture shape with texture,'' \emph{International Journal of Computer
  Vision}, pp. 1--25, 2021.

\bibitem{Song2017SemanticSC}
S.~Song, F.~Yu, A.~Zeng, A.~X. Chang, M.~Savva, and T.~Funkhouser, ``Semantic
  scene completion from a single depth image,'' \emph{IEEE Conference on
  Computer Vision and Pattern Recognition (CVPR)}, pp. 190--198, 2017.

\bibitem{Heusel2017GANsTB}
M.~Heusel, H.~Ramsauer, T.~Unterthiner, B.~Nessler, and S.~Hochreiter, ``Gans
  trained by a two time-scale update rule converge to a local nash
  equilibrium,'' in \emph{Advances in Neural Information Processing Systems},
  2017.

\bibitem{Rubner1998AMF}
Y.~Rubner, C.~Tomasi, and L.~Guibas, ``A metric for distributions with
  applications to image databases,'' \emph{International Conference on Computer
  Vision (ICCV)}, pp. 59--66, 1998.

\bibitem{kingma2014adam}
D.~P. Kingma and J.~Ba, ``Adam: A method for stochastic optimization,''
  \emph{arXiv preprint arXiv:1412.6980}, 2014.

\bibitem{paszke2017automatic}
A.~Paszke, S.~Gross, S.~Chintala, G.~Chanan, E.~Yang, Z.~DeVito, Z.~Lin,
  A.~Desmaison, L.~Antiga, and A.~Lerer, ``Automatic differentiation in
  pytorch,'' 2017.

\bibitem{ioffe2015batch}
S.~Ioffe and C.~Szegedy, ``Batch normalization: Accelerating deep network
  training by reducing internal covariate shift,'' p. 448–456, 2015.

\bibitem{mescheder2019occupancy}
L.~Mescheder, M.~Oechsle, M.~Niemeyer, S.~Nowozin, and A.~Geiger, ``Occupancy
  networks: Learning 3d reconstruction in function space,'' in \emph{IEEE
  Conference on Computer Vision and Pattern Recognition (CVPR)}, 2019, pp.
  4460--4470.

\bibitem{peng2020convolutional}
S.~Peng, M.~Niemeyer, L.~Mescheder, M.~Pollefeys, and A.~Geiger,
  ``Convolutional occupancy networks,'' in \emph{European Conference on
  Computer Vision (ECCV)}.\hskip 1em plus 0.5em minus 0.4em\relax Springer,
  2020, pp. 523--540.

\end{thebibliography}
